\documentclass[12pt]{article}
\usepackage{jheppub}
\usepackage{shuffle}
\usepackage{titlesec}
\usepackage{tikz}
\usepackage{stmaryrd}
\usepackage{url}
\usepackage{mathtools}
\usetikzlibrary{tikzmark,calc,,arrows,shapes,decorations.pathreplacing}
\tikzstyle{vertex}=[circle, draw, minimum size=0pt]

\usetikzlibrary{backgrounds,calc,positioning}

\usetikzlibrary{arrows.meta}

\usetikzlibrary{decorations.markings}
\tikzset{
    midarrow/.style={postaction={decorate},decoration={
            markings,
            mark=at position .5 with {\arrow{stealth}}
        }
    }
}

\tikzset{
    doublemidarrow/.style={postaction={decorate},decoration={
            markings,
            mark=at position .3 with {\arrow{stealth}},
            mark=at position .7 with {\arrow{stealth}}
        }
    }
}

\newcommand\cS{{\mathcal S}}

\newcommand\cM{{\mathcal M}}
\newcommand\cB{{\mathcal B}}
\newcommand\bZ{\mathbb{Z}}

\newcommand\fq{\mathfrak{q}}

\newcommand\bC{\mathbb{C}}
\newcommand\bR{\mathbb{R}}

\newcommand{\Kappa}{\mathrm{K}}
\newcommand{\IR}{\mathrm{IR}}
\newcommand{\RG}{\mathrm{RG}}
\newcommand{\Sh}{\mathrm{Sh}}
\newcommand{\SH}{\mathrm{sH}}

\newcommand\sqbox[1]{{
	\setbox0=\hbox{\scalebox{1}{\mbox{$\Box$}}}
	\setbox1=\hbox{\mbox{\raisebox{0.35ex}{\scriptsize #1}}}
	\mbox{\raisebox{-0.2ex}{\rlap{\hbox to \wd0{\hss{\box1}\hss}}\box0}}
}}

\makeatother

\titleclass{\subsubsubsection}{straight}[\subsection]

\newcounter{subsubsubsection}[subsubsection]
\renewcommand\thesubsubsubsection{\thesubsubsection.\arabic{subsubsubsection}}

\titleformat{\subsubsubsection}
  {\normalfont\normalsize\bfseries}{\thesubsubsubsection}{1em}{}
\titlespacing*{\subsubsubsection}
{0pt}{3.25ex plus 1ex minus .2ex}{1.5ex plus .2ex}

\makeatletter

\title{Categories of Line Defects and Cohomological Hall Algebras}
\author[1]{Davide Gaiotto,}
\author[1,2]{Nikita Grygoryev,}
\author[3]{Wei Li}
\affiliation[1]{Perimeter Institute for Theoretical Physics, Waterloo, ON N2L 2Y5, Canada}
\affiliation[2]{Department of Pure Mathematics, University of Waterloo, Waterloo, ON N2L 3G1, Canada}
\affiliation[3]{Institute of Theoretical Physics, Chinese Academy of Sciences,
100190 Beijing, P.R.\ China
}
\emailAdd{dgaiotto@perimeterinstitute.ca, ngrygoryev@perimeterinstitute.ca,\\ weili@mail.itp.ac.cn}
\abstract{Any four-dimensional Supersymmetric Quantum Field Theory with eight supercharges can be associated to a monoidal category of BPS line defects. Any Coulomb vacuum of such a theory can be conjecturally associated to an ``algebra of BPS particles'', exemplified by certain Cohomological Hall Algebras.  
We conjecture the existence of a monoidal functor from the category of line defects to a certain category of bimodules for the BPS Algebra in any Coulomb vacuum. We describe images of simple objects under the conjectural functor and study their monoidal structure in examples. We conjecture that the functor may be an equivalence of dg-categories and test the conjecture at the level of the equivariant Witten indices of the spaces of morphisms.}
\begin{document}
\maketitle
\section{Introduction}
Four-dimensional ${\cal N}=2$ Supersymmetric Quantum Field Theories typically admit Coulomb vacua whose massless spectrum consists of a free supersymmetric Abelian gauge theory \cite{Seiberg:1994rs,Seiberg:1994aj,Klemm:1994qs,Argyres:1994xh,Hanany:1995na,Argyres:1995jj,Martinec:1995by,Itoyama:1995nv,Argyres:1996eh,Katz:1996fh,Witten:1997sc,Gaiotto:2009we}. The Coulomb vacua form a complex manifold $\cB$, with singular loci where extra light degrees of freedom emerge. The gauge couplings of the IR gauge fields vary holomorphically on $\cB$, up to electric-magnetic dualities. The non-perturbative physical insights encoded in the structure of the Coulomb vacua are crucial for many mathematical applications of 4D ${\cal N}=2$ SQFTs \cite{Witten:1994cg,Donagi:1997sr,Nekrasov:2002qd,Alday:2009aq,Nekrasov:2009rc,Gaiotto:2010okc}.

The spectrum of massive 
particles in a Coulomb vacuum includes special ``half-BPS'' particles which saturate a BPS bound linking their mass $|Z_\gamma(u)|$ and their gauge and flavour charge $\gamma$. 
The ``central charge'' $Z_\gamma(u)$ is a complex linear function of the charge $\gamma$, which is an element of the lattice $\Gamma$ of gauge and flavour charges of the theory. 
The BPS bound protects BPS particle from decay away from specific co-dimension one walls in $\cB$ where the phases of two central charges align. 
The theory of wall-crossing in ${\cal N}=2$ SQFTs studies how the spectrum of BPS particles jump across such walls of marginal stability. 

In the physics literature, the BPS spectrum was initially known only in a few examples, collating information about singular loci in $\cB$ where BPS particles become massless and the free Abelian low energy description is incomplete. The Kontsevich-Soibelman wall-crossing formula \cite{Kontsevich:2008fj} greatly improved our understanding by providing a precise recipe for the $u$ dependence of the BPS spectrum.\footnote{See also \cite{Denef:2007vg} for an earlier physics story.} 

The KS wall-crossing formula was originally designed to study the properties of D-branes in Calabi-Yau compactifications of string theory via a quiver description. It sharpens pioneering work on quiver descriptions of BPS states \cite{Denef:2000nb,Denef:2002ru,Denef:2007vg,reineke2008moduli}. 
As many SQFTs can be engineered from string theory, it was found to apply to BPS particles as well. 
It can be interpreted as describing the behaviour of weakly bound configurations of BPS particles near a wall of marginal stability \cite{Andriyash:2010qv,Manschot:2010qz, Manschot:2011xc,Sen:2011aa,Alim:2011kw,Alim:2011ae,Manschot:2012rx,Cecotti:2012se,Manschot:2013sya,Mozgovoy_2013,Cordova:2015qka}.

The formula was also re-derived in the context of SQFTs \cite{Gaiotto:2010okc,Gaiotto:2010be} by looking at the effect of BPS particles on two quantities:
\begin{itemize}
\item The Seiberg-Witten geometry, a.k.a.\ the Coulomb branch of 3d vacua for a supersymmetric circle compactification of the theory.
\item The IR behaviour of half-BPS line defects in the SQFT. 
\end{itemize}
The latter perspective can be integrated with the quiver perspective by considering ``framed'' quivers \cite{Chuang:2013wt,Cordova:2013bza,Cirafici:2013bha,Cirafici:2017wlw}.
This will be an important tool for us.

The KS wall-crossing formula was derived as an equality between different presentations of the character of an ``algebra of BPS states'', defined concretely as the Cohomological Hall Algebra (CoHA) of a quiver encoding the BPS spectrum \cite{Harvey:1996gc,Kontsevich:2008fj,Galakhov:2018lta}.  

There are strong indications \cite{Li:2023zub} that framed quivers may provide a link between the IR behaviour of half-BPS line defects and the representation theory of CoHA's. The main objective of this paper is to propose and sharpen the following conjectures: 
\begin{itemize}
\item Every Coulomb vacuum $u$ gives a monoidal functor $\RG_u$ from a category of half-BPS line defects to a category of ``bimodules of framed BPS states'' for the algebra of BPS states.
\item When a quiver description is available, the bimodules are built from framed quivers by a certain variant of the CoHA construction.
\item When properly defined, the functor is a quasi-isomorphism. In particular the categories associated to different vacua are also quasi-isomorphic to each other. 
\end{itemize}
In order to formulate the conjecture more explicitly, we need to recall some definitions. 

\subsection{Half-BPS lines and the Holomorphic-Topological twist}
In physical applications \cite{Kapustin:2006hi,Kapustin:2007wm,Gaiotto:2010be,Braverman:2016wma}, the following properties of half-BPS line defects play an important role:
\begin{itemize}
\item Parallel half-BPS line defects can be fused to obtain new half-BPS line defects: the preserved SUSY prevents singularities from arising in the fusion process.
\item Half-BPS lines wrapping the KK circle are BPS order parameters for a supersymmetric circle compactification of the theory. 
They give functions on the Coulomb branch of 3d vacua which are holomorphic in a specific complex structure. 
We will denote the affine complex manifold $\cM$ parameterized by vevs of wrapped half-BPS line defects as the ``K-theoretic Coulomb branch''. 
\item The fusion of half-BPS line defects map to the product of the associated functions. 
\item The IR description of half-BPS lines in a Coulomb vacuum $u$ consist of a specific direct sum of Abelian 't Hooft-Wilson lines. The IR description is compatible with fusion.
\item Upon compactification on a circle,  wrapped half-BPS lines map under RG flow to linear combinations of wrapped Abelian 't Hooft-Wilson lines. This map often endows $\cM$ with the structure of a {\it cluster variety}. 
\end{itemize}

Following \cite{Kapustin:2006hi}, we will sharpen these statements with the help of the 
``Holomorphic-Topological'' (HT) twist of ${\cal N}=2$ SQFTs. 
The HT twist is an operation that maps an SQFT to a simpler QFT by taking the cohomology with respect to a specific super-charge. The choice of supercharge presents space-time locally as $\bR^2 \times \bC$ and makes translations in the $\bR^2$ directions and anti-holomorphic translations in the $\bC$ direction trivial in cohomology.
Many couplings of the original theory become trivial as well. 

The half-BPS line defects in the physical theory survive in the twisted theory as topological lines, say wrapping an $\bR$ factor in $\bR^2$. 
They can be freely translated in the second $\bR$ factor but their position in the $\bC$ direction matters. 
We will typically place all the lines at the origin of $\bC$ unless otherwise stated, and keep track of the quantum numbers for rotations of the complex plane around the origin. 

By general considerations, such topological line defects in the HT twist form a (dg or $A_\infty$) monoidal, $\bZ$-graded category. More precisely, topological line defects are objects in the category and local operators supported on line defects provide the morphisms. Fusion along the transversal $\bR$ direction provides the monoidal structure, and $\bC^*$ rotations of $\bC$ provide the $\bZ$ grading. The derived nature of this structure is unavoidable in the context of twisted theories, where everything is defined as the cohomology of the twisting supercharge. 

We expect half-BPS line defects to map to special topological line defects. 
Indeed, they preserve twice the amount of supersymmetry that is needed for a line defect to survive the twist. 
Although we do not know how to make this expectation concrete from first principles, we can take inspiration from a rigorous mathematical definition \cite{cautis2023canonical} of a monoidal category $\mathrm{KP}_{G,N}$, which conjecturally describes the image of half-BPS line defects in the HT twist of Lagrangian SQFTs whose gauge group is the compact form of a reductive group $G$ and which has matter fields (hypermultiplets) valued in $T^*N$ for some representation $N$ of $G$.\footnote{While a general prescription for associating a BPS quiver to an arbitrary $\mathrm{KP}_{G,N}$ is not yet known, many important cases can be handled explicitly. For instance, any pure gauge theory with a unitary gauge group has a known BPS quiver.
More recently, \cite{Ambrosino:2025qpy} gave a procedure to obtain the BPS quiver of any gauge theory with cotangent matter using the BPS quiver of the pure gauge theory as input. Together, these results cover a large and important sub-class of \cite{cautis2023canonical}, and we expect them to extend further.} 

The category $\mathrm{KP}_{G,N}$ sits in a bigger dg (triangulated) category as the {\it heart of a t-structure}. Roughly, this means that the Chan-Paton factors one encounters in the monoidal structure are vector spaces rather than complexes.\footnote{When a calculation is done in the physical theory and involves half-BPS configurations, it will involve Hilbert spaces which are representations of the  physical spin and of the R-charge $SU(2)_R$ symmetry algebra, as well as four super-charges. The twisted description only keeps track of the Cartan generator of spin, denoted as $n$, and the Cartan generator of $SU(2)_R$, which defines the ghost number. A {\it no-exotics conjecture} claims that 
the ground states which contribute to Chan-Paton factors, etc. carry trivial representations of $SU(2)_R$,
in agreement with the mathematical perspective. The no-exotics conjecture is compatible with other conjectures present in this paper, but is still rather mysterious. It would be interesting to know how the mathematical formalism could be adjusted to deal with situations where the no-exotics conjecture might be false.}

We expect this type of phenomenon to generalize to any ${\cal N}=2$ SQFT $T$ and thus denote the conjectural Abelian monoidal category of half-BPS lines as $\mathrm{Line}[T]$. If a Lagrangian description $(G,N)$ of $T$ is available, we expect $\mathrm{Line}[T] = \mathrm{KP}_{G,N}$. See \cite{Niu:2021jet} for a comparison at the level of characters of morphism spaces. It is often the case that the same SQFT admits multiple distinct Lagrangian descriptions $(G_i,N_i)$. An immediate consequence of this expectation is that $\mathrm{KP}_{G_i,N_i}$ for different descriptions of the same SQFT coincide. It would be interesting to test this statement. 

\medskip

Essentially by definition, the K-theory classes of objects in $\mathrm{Line}$, i.e.\ the elements of $K[\mathrm{Line}]$, represent the KK-compactified line defects. 
Accordingly, we expect the mathematically precise version of the second bullet-point above to be that the K-theoretic Coulomb branch is the spectrum of the K-theory ring:
\begin{equation}
\cM = \mathrm{mSpec}\,  K[\mathrm{Line}] \,.
\end{equation}
This expectation holds for the Lagrangian theories: $K[\mathrm{KP}_{G,N}]$ reproduces the BFN construction of the K-theoretic Coulomb branch, essentially by construction. 

It is also useful to consider the K-theory classes that are equivariant under the rotations of $\bC$, i.e. $K_{\bC^*}[\mathrm{Line}]$. 
Physically, these should represent the half-BPS lines inserted in Melvin space $M_\fq$ and give a non-commutative deformation of the commutative ring $K[\mathrm{Line}]$. 
This ``non-commutative K-theoretic Coulomb branch algebra'', which we will denote as $\mathcal{A}_\fq$, is an important ingredient of the story. 
The structure constants contain the spin quantum numbers of Chan-Paton factors. 

Finally, the equivariant characters of the dg morphism spaces of the $\mathrm{Line}$ category coincide with the so-called Schur indices \cite{Gadde:2011uv,Dimofte:2011py,Cordova:2016uwk} of the underlying 4d SQFT \cite{Niu:2021jet}. 
\medskip

This mathematical formalization of UV physics does not include information about the RG flow  or the notion of a Coulomb branch vacuum. 
This is similar to what happens to the B-twist of 2D $(2,2)$ SQFTs: the category of D-branes is known to 
coincide with the derived category of coherent sheaves, but the effect of RG flows on the category of branes is poorly understood and challenging to compute mathematically unless a mirror dual is available \cite{Herbst:2008jq,Hori:2013ika}.

The bare-bone formalization of the IR behaviour of line defects would be a monoidal functor (of Abelian categories)
\begin{equation}
    \RG_u: \mathrm{Line} \to \mathrm{Line}^{\IR}_u\,,
\end{equation}
landing in the category $\mathrm{Line}^{IR}_u$ of half-BPS lines in the free Abelian gauge theory that describes the IR physics in a vacuum $u$. 

The category $\mathrm{Line}^{\IR}_u$ is easily described. 
After all, it is just $\mathrm{KP}_{T_u,0}$, where $T_u$ is the IR gauge group. 
This is a very simple category, with simple objects $L^{\IR}_\gamma$ labelled by an IR electric-magnetic charge $\gamma$ valued in some lattice $\Gamma_u$. 
No IR local operators can change the gauge charge of the line, and so all morphisms are trivial except for the endomorphisms, which in turn are independent of $\gamma$ and consist of polynomials in some fermionic ghosts and their derivatives. 

Accordingly, the image $\RG_u(L)$ is essentially given by the Chan-Paton coefficients\footnote{To be precise, the Chan-Paton coefficients should carry a representation of some kind of Yangian algebra for the Abelian HT theory \cite{Costello:2013zra}, Koszul dual to the fermionic ghosts. We briefly discuss a related symmetry in Section \ref{sec:ya} but do not attempt a systematic treatment in that language.} $V_{\gamma;n}[L]$ in 
\begin{equation}
\RG_u(L) = \oplus_{\gamma \in \Gamma_u} V_{\gamma;n}(L;u) \otimes \fq^n L^{\IR}_\gamma  \,,
\end{equation}
where we denote as $\fq^n L^{\IR}_\gamma$ an elementary line assigned spin $n$. 
These are denoted as {\it framed BPS states} in physics context.\footnote{These Chan-Paton factors are a twist of vector spaces of framed BPS states which have a sharp physical definition. The discussion about the no-exotics conjecture from the previous footnote applies here as well.}

This monoidal ``fiber functor'' $\RG_u: L \to V_{\bullet;\bullet}(L;u)$ forgets much about the original category. 
Indeed, we do not lose much information by passing to the K-theoretic version of the map
\begin{equation}
\RG^0_u([L]) = \sum_{\gamma \in \Gamma_u} \mathrm{dim} V_{\gamma;n}(L;u)  \fq^n [L^{\IR}_\gamma]
\end{equation}
which only remembers the {\it framed BPS degeneracies} $\mathrm{dim} V_{\gamma;n}(L;u)$. This generating function of framed BPS degeneracies is often denoted as 
\begin{equation}
F_u([L]) = \sum_{\gamma \in \Gamma_u} \mathrm{dim} V_{\gamma;n}(L;u) \,  \fq^n \, X_\gamma
\end{equation}
where the K-theory classes $X_\gamma$ satisfy the quantum torus algebra
\begin{equation} \label{eq:qt}
X_\gamma X_{\gamma'} = \fq^{\langle \gamma, \gamma'\rangle} X_{\gamma + \gamma'}
\end{equation}
and $\langle \gamma, \gamma'\rangle$ is the anti-symmetric Dirac pairing of charges. 
The monoidal property means that 
\begin{equation} \label{eq:monoF}
F_u([L \cdot L']) =  F_u([L])  F_u([L']) 
\end{equation}
where $\cdot$ represents the monoidal structure of the category.

We should stress that the original category $\mathrm{Line}$ definitely has more information than its K-theory. 
For example, the tensor product of simple objects is typically not a direct sum of simple objects but rather some kind of extension. 

\bigskip

This paper explores a possible answer to the question:
\begin{itemize}
\item Can we find a ``better'' family of IR categories and RG functors which captures all the information of $\mathrm{Line}$?
\end{itemize}
An answer to this question would in particular give an avenue to compute the category $\mathrm{Line}$ for theories which do not admit a Lagrangian description, as well as verifying dualities.

We are inspired by the setting of gapped 2D TFTs, where the ``Algebra of the Infrared'' formalism \cite{Gaiotto:2015aoa} builds IR categories that are conjecturally quasi-isomorphic to the categories of boundary conditions or interfaces of the microscopic theory. 
BPS particles play a crucial role in the 2D story and we expect them to play a crucial role here as well. 

Accordingly, we introduce an ``algebra of BPS states'' $A_u$, perhaps in the sense of \cite{Harvey:1995fq}, exemplified by the Cohomological Hall Algebra \cite{Kontsevich:2008fj} $H(Q,W)$ of a quiver $Q$ with superpotential $W$.\footnote{The positive subalgebra of the quiver Yangian defined in \cite{Li:2020rij} can be viewed as the equivariant version of the spherical CoHA, see \cite[Sec.\ 5.2]{Galakhov:2021vbo}. For general quiver and potential, the relation between the ordinary CoHA and the equivariant version is subtle; however, as we will see later, for the 2-acyclic quivers that are the focus of the current paper, one can simply set all the equivariant parameters in the equivariant spherical CoHA to zero to obtain the ordinary one, see Appendix \ref{appsec:quiverYangian}.} 
This is the algebra of BPS states for a SQFT which arises from String Theory, 
where BPS particles are D-branes whose dynamics is encoded in $H(Q,W)$. 
In Appendix \ref{appsec:SQM} we will attempt to give a more direct physical definition of $A_u$.

\medskip

We then define a collection of $A_u$-bimodules $B_{L;u}$ and conjecture that the map
\begin{equation}
\RG_u: L \to B_{L;u}
\end{equation}
is actually the object-level part of a monoidal functor into a category of $A_u$-bimodules which are equivariant under a certain gauge action. 

\medskip

We present three pieces of evidence for our proposal:
\begin{enumerate}
\item The proposal ``works'' at the level of K-theory. Namely, we argue that the $B_{L;u}$ can be written as  
\begin{equation}
B_{L;u} = V_{L;u} \otimes A_u
\end{equation}
as right $A_u$ modules, and all our vector spaces are graded by the gauge and spin charges in such a manner that 
\begin{equation}
V_{L;u} = \oplus_{\gamma,n} V_{\gamma;n}(L;u) \,.
\end{equation}
Then 
\begin{equation}
B_{L;u} \otimes_{A_u} B_{L';u} = V_{L;u} \otimes V_{L';u} \otimes A_u \,,
\end{equation}
compatible with eq. (\ref{eq:monoF}).\footnote{Here we are glossing over an important ``Heisenberg grading'' that will be discussed in the main text.}
\item We verify in simple non-Lagrangian examples that the tensor product of $B_{L;u}$ bimodules indeed can be written as an extension of $B_{L;u}$'s, and that the structure of the extensions is compatible with the existence of functors from an $u$-independent ${\mathrm{Line}}$ category.
\item We verify that an appropriately-defined space of morphisms of bimodules has an $u$-independent graded character which coincides with the expected answer for the $\mathrm{Line}$ category (the Schur indices), compatibly with these functors being quasi-isomorphisms. 
\end{enumerate}
In the process, we will propose a sequence of refinements of the conjecture by constraining 
the types of bimodules and morphisms which can appear as images of $\RG_u$.

The formula we derive for the graded character of morphism spaces use the proposal that the Schur indices of any theory which admits a Coulomb vacuum can be reconstructed from the spectrum of BPS particles in that vacuum and from the framed BPS indices of the line defects \cite{Cordova:2015nma,Cordova:2016uwk}.

\subsection{Future directions and missing links}
A companion paper will test the existence of monoidal functors from $\mathrm{KP}_{G,N}$ for simple $G$ and $N$.

At the beginning of this project, our objective was to promote the $V_{L,u}$ spaces to modules for some algebra equipped with a co-product, perhaps a ``double'' \cite{Rapcak:2018nsl} of the algebra of BPS states. Indeed, a traditional way to ``improve'' a fiber functor is to include information about the action of the algebra of endomorphisms of the fiber functor itself \cite{private}. 
This is the idea of {\it Koszul duality} \cite{Costello:2016mgj}, which was used to great effect in other twisted theories \cite{Costello:2013zra,Paquette:2021cij}. 

We could not make this idea work and found instead the alternative route of a bimodule category. It would nevertheless be interesting to identify such a description for the category of line defects. 
Namely,  there might be a construction of modules for the ``double” of the CoHA or Yetter-Drinfeld modules of CoHA that could provide an alternative description of the category of line defects. 
As far as we know, there is currently no well-defined notion of such a double or of a “true” coproduct on CoHA suitable for our needs (despite much effort), particularly when one takes into account the additional Heisenberg grading. 
We can nevertheless speculate about the form this equivalence might take. Recall that for a Hopf algebra $A$, there is a monoidal equivalence between the category of (left–left) Yetter–Drinfeld modules over $A$ and the category of $A$-bicomodules in the category of $A$-bimodules; under this equivalence, an object $V$ is sent to $V \otimes A$. In particular, finite-dimensional $A$-modules correspond to bimodules that are free and finitely generated as right $A$-modules – precisely the form of the bimodules that appeared in our work. For CoHAs, the above correspondence would require a “chiral” version of all comodule structures. The existence and significance of such a chiral bicomodule structure on our CoHA-bimodules certainly deserve further study, and we hope to investigate this in the future.

It would be also useful to find a clear physical explanation for our proposal. In 2d analogues, the category of line defects can be built directly from the IR data by assembling BPS particles into ``fans'' which can be created by local operators \cite{Gaiotto:2015aoa}. A previous work \cite{Gaiotto:2023dvs} suggested that spaces of ``half fans'' comprised of BPS particles emanating out in an half-plane could be recovered from $A_u$ via Koszul duality. 
We are not sure how to assemble these ideas into a coherent whole. 

\section{The Cohomological Hall Algebra as a model BPS algebra}

The relation between the Cohomological Hall Algebra of \cite{Kontsevich:2010px} and 4D ${\cal N}=2$ SQFTs is surprisingly subtle. 
To the best of our knowledge, the connection can only be justified if the 4D theory $T$ can be engineered as the low energy description of a string theory ``compactification'' on a non-compact Calabi-Yau manifold $Y$ that is a fibration of ADE singularities.
Then the BPS particles in a given vacuum $u$ are associated to the cohomology classes on moduli spaces of compact stable D-branes in $Y$ and a ``BPS algebra'' $A_u$ can be tentatively defined from cohomology classes of semi-stable objects. 

We will now describe the BPS algebra in this context and formulate properties which we expect to hold in general. 

\subsection{Quivers and Superpotentials}
\label{ssec:QandW}

The definition is formalized by encoding D-branes as representations of an auxiliary quiver $Q_u$ with superpotential $W_u$. We will drop the $u$ subscript for now. 

The underlying vector space of the CoHA $H(Q,W)$ is defined as an equivariant cohomology of semi-stable quiver representations. 
In physical terms, this should be the space of supersymmetric ground states for a generic collection of D-branes described by the quiver. The product is defined via pull-backs and push-forwards of cohomology classes along certain correspondences, defined as moduli spaces of extensions of quiver representations. In Appendix \ref{appsec:SQM} we elaborate on the potential physical interpretation of the product. 

The superpotential $W$ plays an important and difficult role in the story. 
Fortunately, the detailed construction of the CoHA with superpotential will not be needed in the following, as long as the following conjecture holds true:
\begin{itemize}
\item 
For every quiver there exists at least one potential $W$ that is infinitely mutable (in the sense of \cite{Kontsevich:2008fj,derksen2008quivers}, see review below) for which the CoHA is spherically generated, namely $H(Q,W)$ coincides with the ``spherical part'' $\SH(Q)$ of $H(Q,0)$, which has a simple algebraic description.
\end{itemize}
Accordingly, whenever we refer to the ``CoHA'' in the rest of the paper, we are actually referring to $\SH(Q)$ as a conjectural stand-in for $H(Q,W)$ for the suitable infinitely mutable $(Q,W)$.

\medskip

\noindent \underline{Note added for v3.}  In earlier versions of the paper, the conjecture was stated as ``$H(Q,W)$ is spherically generated and coincides with the ``spherical part'' $\SH(Q)$ of $H(Q,0)$ whenever $W$ is infinitely-mutable. It was shown by Davison in \cite{Davison:2025ydh}  that for a given $Q$, this  might not hold for \textit{every} infinitely-mutable $W$". Take the example of  the Markov quiver $Q$, although it admits an infinitely-mutable potential $W_{\mathrm{Type-4}}$ (in \cite[Lemma 2.1.]{Davison:2025ydh})  for which $H(Q,W_{\mathrm{Type-4}})$ is spherical and $H(Q,W_{\mathrm{Type-4}})=\textrm{sH}(Q)$, it also admits another infinitely mutable superpotential $W_{\mathrm{Type-5}}$ (in \cite[Lemma 2.1.]{Davison:2025ydh}) for which $H(Q,W_{\mathrm{Type-5}})$ strictly contains its spherical subalgebra.

Note that both Type 4 and Type 5 correspond to a 4d $\mathcal{N}=2$ theory: Type 4 to a ``minimal'' theory and Type 5 to the combination of the minimal theory and a decoupled sector. (This would also be a reason to prefer Type 4 to Type 5.) The corresponding algebraic statement is that ``$H(Q,W_{\mathrm{Type-5}})$ splits as the product of $\textrm{sH}(Q)$ and a free exterior algebra generated by the non-spherical elements."

Accordingly, Davison in \cite{Davison:2025ydh} has proposed to modified the conjecture into the current form. 
This refined conjecture is sufficient for our purposes (perhaps combined with the conjecture ``any infinitely mutable potential can be related to a valid 4d N=2 theory"), since our goal is to use CoHAs to define category of line defects for ``some" 4d  $\mathcal{N}=2$ theories.



\subsection{Quivers and Mutations}
As we have now mentioned the notion of quiver mutation, it is a good time to discuss the role of quiver mutations in the physics of BPS particles. 
First, an important conjectural constraint:
\begin{itemize}
\item The ``BPS quiver'' that appears in the description of the algebra of BPS states in any 4D ${\cal N}=2$ SQFT has no loop of length $1$ or $2$, namely, it is an 2-acyclic quiver. 
\end{itemize}
With this constraint, a mutation $\mu_{i}$ of a quiver $Q$ with superpotential $W$ at a node $i$ is an operation which produces a new quiver obtained from $Q$ by a sequence of moves:\footnote{See \cite{MutationJava} for an online tool to perform quiver mutations.}
\begin{itemize}
\item Flip all arrows ending or starting at the node $i$.
\item Add a new arrow from node $j\neq i$ to node $k \neq i$ for each pair of arrows from $j$ to $i$ and from $i$ to $k$.
\item Remove any pair of arrows producing a loop of length $2$, so that the constraint that there is no loop of length 2 remains true. 
\item Modify the superpotential accordingly, in a manner we do not need to describe here.
\end{itemize}
The mutations may be obstructed if the superpotential $W$ misses certain terms. 
Reference \cite{derksen2008quivers} proves that there exist sufficiently generic superpotentials $W$ such that any arbitrary sequence of mutations can be executed, as long as $Q$ has no loops of length $1$ and $2$. 
We call such a pair $(Q,W)$ infinitely mutable. 
Whenever $W$ is not mentioned, we assume that it is infinitely mutable.\footnote{Note that we do not need to demand the stronger (and vaguer) “genericity” condition on $W$.}
Moreover, based on the conjecture in Sec.~\ref{ssec:QandW}, we choose the infinitely-mutable $W$ for which $H(Q,W)$ is spherical and coincide with the spherical subalgebra $\textrm{sH}(Q)$ of $H(Q,0)$. 

In the infinitely mutable setting, mutating twice at a node $i$  returns the original quiver. 
Mutations at adjacent nodes do not commute. 
Starting from a quiver $Q$, we can generate a whole groupoid $[Q]$ of quivers related by mutations. 

The role of the quiver mutations in the physical setting is that the BPS quiver $Q_u$ precisely jumps by mutations as we move around the space of vacua and cross certain walls of marginal stability. 
Accordingly, we will get a family of distinct BPS algebras $\SH(Q_u)$ associated to the same SQFT.   

\subsection{Stability conditions and walls}
In order to describe wall-crossing phenomena, we need to introduce the notions of the charge lattice and stability conditions. 

\medskip

At every Coulomb vacuum $u$ of a SQFT, we have a lattice $\Gamma_u$ of flavour and gauge charges, equipped with an anti-symmetric integral Dirac pairing: 
\begin{equation}
    \langle \bullet, \bullet\rangle: \Gamma_u \to \bZ \, .
\end{equation}
The lattice $\Gamma_u$ is locally constant, but undergoes monodromies as $u$ is transported around the singularities in the 4D Coulomb branch $\cB$.

\medskip

The stability conditions are controlled by the central charges $Z_\gamma(u)$, which define a linear map $\Gamma_u \to \bC$. The central charges are locally holomorphic and non-vanishing on the non-singular part of $\cB$. In the rest of the paper we will never really need to refer directly to $u$: only $\Gamma$ and $Z_\gamma$ enter our considerations. We will thus drop any reference to $u$ in the following.\footnote{
In general, not all values of the central charges occur in $\cB$. Accordingly, not all possible quiver mutations will occur along $\cB$ \cite{Cecotti:2011rv}. 
The implications of this fact are unclear.}

A ``chamber'' in the space of stability conditions is a connected region in which 
\begin{itemize}
\item $\mathrm{arg} \, Z_\gamma$ are all distinct for non-collinear charges populated by BPS particles and
\item $\mathrm{Im}\, Z_\gamma$ is non-zero for charges populated by BPS particles.
\end{itemize}
We will denote the co-dimension one loci where either condition fails as walls of marginal stability of the first or second kind, respectively  \cite{Kontsevich:2008fj}:
\begin{itemize}
\item The spectrum of BPS particles can jump across walls of the first kind, but the BPS quiver $(Q,W)$ and thus the BPS algebra $A$ remains unchanged.
\item The BPS quiver $(Q,W)$ undergoes specific mutations across walls of the second kind,
but the spectrum of BPS particles remains unchanged. On the other hand, the framed BPS degeneracies for line defects can jump. 
\end{itemize}
The jumps in the two types of BPS spectra are governed by specific wall-crossing formulae \cite{Kontsevich:2008fj,Gaiotto:2010okc,Gaiotto:2010be}. We will review below and in Section \ref{sec:CoHAbimodule} the interpretation of wall-crossing formulae in the context of CoHA's.
\medskip

In general, we expect the BPS spectrum to be encoded in a BPS quiver $(Q,W)$ when all of the BPS particles with $\mathrm{Im}\, Z_\gamma>0$ can be described as bound states of a finite collection of ``elementary'' hypermultiplet particles.
Denote the individual charges of such particles as $\gamma_i \in \Gamma$ and collectively as a vector $\vec \gamma$ of charges. In particular, all populated charges with $\mathrm{Im}\, Z_\gamma>0$ are non-negative linear combinations of the $\gamma_i$. 

The nodes of the BPS quiver correspond to the elementary particles and are thus labelled by the $\gamma_i$ charges. 
The number of arrows between nodes is uniquely determined by the Dirac pairing:\footnote{Note that the sign convention here is the opposite to that in \cite{Alim:2011kw}.}
\begin{equation}
\langle\gamma_i, \gamma_j \rangle \equiv |i \to j| - |j \to i|
\end{equation}
if we assume the absence of cycles of length $1$ and $2$. 
Equivalently, $|i \to j| =  \langle\gamma_i, \gamma_j \rangle$ when the latter is positive, or $0$ otherwise. The collection of $\gamma_i$ in $\Gamma$ thus gives us directly the BPS quiver $Q$, and we equip it with an infinitely-mutable superpotential $W$ for which $H(Q,W)$ is spherical and coincide with the spherical subalgebra $\textrm{sH}(Q)$ of $H(Q,0)$. We will denote a BPS quiver $Q$ with specific $\gamma_i$ labels as a {\it labelled quiver}.

\medskip

Next, we can describe how the BPS quiver mutates across a wall of marginal stability of the second kind. The only way a populated central charge can hit the real axis is if it is the central charge $Z_{\gamma_{i}}$ of an elementary particle.  Across the wall, the anti-particle with charge $-\gamma_{i}$ becomes an ``elementary'' particle, and the roles of elementary particles and bound states get re-shuffled. 

The new quiver across the wall is obtained from $(Q,W)$ by a mutation at $i$.
The specific reorganization of the elementary particles depends on whether $Z_{\gamma_{i}}$ is along the positive or negative real axis. 
Accordingly, the new labels $\vec \gamma'$ are related to the old ones by a ``tropical cluster transformation'', and depending on the sign of $Z_{\gamma_i}$ at the wall, we have two distinct versions $\mu^{0/\pi}_i$ of the mutations of a labelled quiver:
\begin{align}\label{eq:mupm}
\mu^{\pi}_i:\qquad \gamma'_{i} &= - \gamma_{i} \quad \textrm{and} \quad  \gamma_j' = \gamma_j + |i \to j| \gamma_{i} \qquad (Z_{\gamma_i}<0)\,,\cr
\mu^{0}_i:\qquad \gamma'_{i} &= - \gamma_{i} \quad \textrm{and} \quad  \gamma_j' = \gamma_j + |j \to i| \gamma_{i} \qquad (Z_{\gamma_i}>0)\,.
\end{align}   
These transformations are compatible with the arrow mutation rules: namely 
\begin{align}
\langle \gamma'_i, \gamma'_j \rangle 
&=-\langle \gamma_i, \gamma_j \rangle\\
\langle \gamma'_j, \gamma'_k \rangle &= \langle \gamma_j, \gamma_k \rangle + |j \to i| |i \to k| - |k \to i| |i \to j|    \,,
\end{align}
for both $\mu^{0/\pi}_i$.
For the same $i$, the two mutations $\mu^{0}_i$ and $\mu^{\pi}_i$ are inverses of each other, and the square of a mutation only changes the labelling of a quiver, by the Lefschetz monodromy:
\begin{align}
\mu^{\pi}_i\mu^{\pi}_i: \qquad \gamma_j \to \gamma_j + \langle \gamma_i, \gamma_j\rangle \gamma_i \,,\cr
\mu^{0}_i\mu^{0}_i: \qquad \gamma_j \to \gamma_j - \langle \gamma_i, \gamma_j\rangle \gamma_i \,.
\end{align}

\subsection{Heisenberg grading and Kontsevich-Soibelman spectrum generator}

The CoHA $H(Q,W)$ and many other vector spaces which we will encounter are equipped with an {\it Heisenberg grading}. 
This is a grading by $\Gamma \oplus \bZ$ such that under tensor products, the gradings $(\gamma, r)$, $(\gamma',r')$ of the factors combine 
into a grading 
\begin{equation}
(\gamma + \gamma',r+r'+\langle \gamma, \gamma'\rangle)
\end{equation}
for the tensor product.\footnote{Physically, this is the contribution of the IR electric-magnetic fields to the spin of a system of parallel dyons.} 
Equivalently,
\begin{equation}
(V \otimes V')_{\gamma, r} = \oplus_{\gamma' \in \Gamma,r' \in \bZ} V_{\gamma',r'} \otimes V'_{\gamma-\gamma',r-r'+\langle \gamma, \gamma'\rangle} \,.
\end{equation}
The Heisenberg-graded vector-spaces with finite-dimensional graded components form a monoidal category $\mathrm{Vect}^{\mathrm{Hei}}_\Gamma$.\footnote{There is a subtlety which we have ignored in this paper but may become important in future work. We are actually working with super-vectorspaces and the Grassmann parity of a twisted tensor product may involve a Grassmann parity shift by $\langle \gamma, \gamma'\rangle$ modulo $2$. Accordingly, we should perhaps define a monoidal category $\mathrm{sVect}^{\mathrm{Hei}}_\Gamma$ which includes the Grassmann shift and work with algebras over $\mathrm{sVect}^{\mathrm{Hei}}_\Gamma$.}

\smallskip

Then the CoHA $H(Q,W)$
\begin{itemize}
\item is an algebra in $\mathrm{Vect}^{\mathrm{Hei}}_\Gamma$ and
\item is supported on non-negative linear combinations of the $\gamma_i$ and non-negative spin.
\end{itemize}
An element of $\mathrm{Vect}^{\mathrm{Hei}}_\Gamma$ has an equivariant character
\begin{equation}
\chi(V) \equiv \sum_{\gamma \in \Gamma} \sum_{r \in \bZ} \mathrm{dim} V_{\gamma,n}\,  \fq^r \, X_\gamma\,.
\end{equation}
As long as the $X_\gamma$ multiply as quantum torus generators (\ref{eq:qt}), we have
\begin{equation}
\chi(V\otimes W)=\chi(V)\chi(W) \,.
\end{equation}

We define the Kontsevich-Soibelman spectrum generator $\cS$ as the character of the CoHA:\footnote{Important notation warning: in other papers on this topic we usually refer to this expression as $\cS^{-1}$, reserving the symbol $\cS$ for the expression which enters directly in the Schur index. We apologize for possible confusion, but including an inverse power in every formula grows tedious and is notationally heavy.}
\begin{equation}
\cS \equiv \chi(H(Q,W)) \,,
\end{equation}
which is a formal power series in the quantum torus algebra and in $\fq$. 
We expect that \begin{equation}
\cS = \chi(\SH(Q,W))    
\end{equation} 
for BPS quivers. This is indeed the case in the examples we checked. 

The component $H_{\gamma, \bullet}$ of the CoHA with charge 
\begin{equation}
\gamma = \vec n \cdot \vec \gamma  \equiv \sum_i n_i \gamma_i
\end{equation}
consists of the cohomology classes of quiver representations with rank $n_i$ at node $i$. 

\subsection{Harder-Narasimhan filtrations and factorization}

The reason for the name ``stability condition'' for $Z$ is that the phase
\begin{equation}
\mathrm{arg} \, Z_{\vec n \cdot \vec \gamma}
\end{equation} 
can be used to assign a ``slope'' to a quiver representation with rank $n_i$ at node $i$, and define a notion of stable and semi-stable quiver representation. 

The general theory of Harder-Narasimhan (HN) filtrations of the category of D-branes has crucial implications for the structure of the CoHA $A$ away from the walls of marginal stability: 
\begin{itemize}
\item Consider a primitive non-negative linear combination $\gamma$ of the $\gamma_i$ and its non-negative multiples $k \gamma$. 
The cohomology classes of stable objects with these ranks give a sub-algebra $S_\gamma$ of the CoHA $H$. 
If there are no stable objects, the sub-algebra consists of the identity only.
\item There is a (non-canonical) PBW-like basis of $H$ of the form 
\begin{equation}\label{eq:H}
H =  \prod^{\curvearrowright}_\gamma S_\gamma\,,
\end{equation}
where the product is taken over all populated primitive charges $\gamma$ in the clockwise order of the slopes of $Z_{\gamma}$ (i.e.\ in the decreasing order of  $\mathrm{arg} \, Z_\gamma$), namely $S_{\gamma_1}$ is to the left of $S_{\gamma_2}$ in \eqref{eq:H} if $\textrm{arg}(Z_{\gamma_1})>\textrm{arg}(Z_{\gamma_2})$.  
We also denote this as a ``PBW factorization'' of the algebra.
\end{itemize}
Accordingly, we also obtain a ``PBW factorization'' of the spectrum generator:
\begin{equation}
\cS = \prod^{\curvearrowright}_\gamma \chi(S_\gamma)
\end{equation}
The Wall Crossing Formula simply states that $\cS$ remains invariant across the first kind of walls of marginal stability, even though the order of the primitive charges and the form of the individual $\chi(S_\gamma)$ factors jump across these walls.  

\medskip

Now, suppose we approach a wall of marginal stability of the second kind, where $Z_{\gamma_i}$ becomes real for a node $i$. 
Then the factor $S_{\gamma_i}$ appears either leftmost (when $Z_{\gamma_i}<0$) or rightmost (when $Z_{\gamma_i}>0$) in the CoHA factorization.
The form of this factor is independent of the quiver: it essentially involves representations of a quiver with a single node and no arrow. 
It is an infinite exterior algebra with generators $e^{(i)}_n$ of spin $2n+1$ and charge $\gamma_i$:
\begin{equation}
S_{\gamma_i} = \Lambda^\bullet[e^{(i)}_0, e^{(i)}_1, e^{(i)}_2, \cdots] \,.
\end{equation}
The character of $S_{\gamma_i}$ is the (fermionic) quantum dilogarithm
\begin{equation}
\Phi(X_{\gamma_i}) \equiv \prod_{m=0}^\infty (1+X_{\gamma_i} \fq^{2m+1}) 
= \sum_{n=0}^\infty \frac{\fq^{n^2}}{(\fq^2;\fq^2)_n} X_{n \gamma_i}\,,
\end{equation}
where $(\fq^2;\fq^2)_n=\prod^{n}_{k=1}(1-\fq^{2k})$.
Accordingly, before the wall, we have 
\begin{equation}
\cS = 
\begin{cases}
\begin{aligned}
&\Phi(X_{\gamma_i}) \left[\prod^{\curvearrowright}_{\gamma>\gamma_i} \chi(S_\gamma)\right]  \qquad \qquad Z_{\gamma_i}<0\\
&\left[\prod^{\curvearrowright}_{\gamma<\gamma_i} \chi(S_\gamma)\right]\Phi(X_{\gamma_i}) \qquad \qquad Z_{\gamma_i}>0 \,,
\end{aligned}    
\end{cases}
\end{equation}
where $\gamma>\gamma_i$ means that $\textrm{arg}Z_{\gamma}<\textrm{arg}Z_{\gamma_i}$.
Then the main claim about quiver mutations is that after the wall we have 
\begin{equation}
\cS' = 
\begin{cases}
\begin{aligned}
&\left[\prod^{\curvearrowright}_{\gamma>\gamma_i} \chi(S_\gamma)\right] \Phi(X_{-\gamma_i}) \qquad \qquad Z_{\gamma_i}<0 \\
&\Phi(X_{-\gamma_i}) 
\left[\prod^{\curvearrowright}_{\gamma<\gamma_i} \chi(S_\gamma)\right]
\qquad \qquad Z_{\gamma_i}>0 \,.
\end{aligned}    
\end{cases}
\end{equation}

The reader might object to having two different formulae, as the CoHA is determined by the quiver, which mutates in the same manner independently of the sign of $Z_{\gamma_i}$. 
The sign, though, affects the choice of the tropical cluster transformation that gives the labels $\gamma'_i$ of the nodes in the new quiver, see \eqref{eq:mupm}.
Fortunately, the two final expressions are formally related by conjugation by a theta function $\Phi(X_{\gamma_i}) \Phi(X_{-\gamma_i})$, which precisely shifts 
\begin{equation}
X_{\gamma_j} \to X_{\gamma_j+ \langle \gamma_i, \gamma_j\rangle \gamma_i}\,,
\end{equation}
compensating for the difference in the two tropical cluster transformations. 

We will see explicit examples at the end of this Section.

\subsection{Spherical CoHA and Spherical Shuffle Algebra}

The collection of generators $e^{(i)}_n$ associated to single-node sub-algebras $S_{\gamma_i}$ of the CoHA $H$ plays an outsized role in our analysis. 
Following \cite{Kontsevich:2010px}, we define the {\it Spherical CoHA} $\SH(Q)$ as the sub-algebra of $H(Q,0)$ generated by the $e^{(i)}_n$. 

Notice the absence of the superpotential $W$ from the definition! 
As a consequence, $\SH(Q)$ only depends on the data of the $\gamma_i$ charges in the charge lattice $\Gamma$ and ignores all the intricacies of some potential string theory construction realizing the 4D theory of interest. 
It is ready-for-use for any theory for which we can identify the collection of elementary particles in some vacuum. 
In the rest of the paper we will assume a non-trivial conjecture:\begin{itemize}
\item The Spherical CoHa $\SH(Q)$ coincides with $H(Q,W)$ for some infinitely mutable $W$.
\end{itemize}
Reassuringly, in all the examples we look at,
the $\SH(Q)$ algebra enjoys the factorization and wall-crossing properties expected for $H(Q,W)$.

The CoHA $H(Q,0)$ has a straightforward combinatorial presentation as a {\it Shuffle Algebra} $\mathrm{Sh}(Q)$, which we will describe momentarily. 
Thus $\SH(Q)$ can be simply presented as the spherical part of $\mathrm{Sh}(Q)$.

\bigskip

The Shuffle Algebra description generalizes the familiar description of an infinite exterior algebra as an algebra of anti-symmetric polynomials in any number of variables and coincides with it for a one-node quiver.\footnote{This is no coincidence: the $GL(n)$-equivariant cohomology setting used to define the CoHA is equivalent to the singlet sector of a Matrix Quantum Mechanics, which is well-known to match a fermionic Fock space.} 

The shuffle algebra $\Sh(Q)$ is an algebra in $\mathrm{Vect}^{\mathrm{Hei}}_\Gamma$, supported on non-negative integer combinations $\vec n \cdot \vec \gamma$ of the $\gamma_i$ charges labelling the nodes of the quiver $Q$. As vector spaces, the graded components $\Sh(Q)_{\vec n \cdot \vec \gamma, r}$ are defined as spaces of polynomials in a collection of auxiliary variables $s_{i,a}$: the index $i$ runs over the nodes of the quiver and the second index is valued in $a=1, \cdots n_i$. The polynomials are anti-symmetric under exchange of any pair of variables $s_{i,a}$, $s_{i,b}$ associated to the same node.\footnote{The standard definition uses symmetric polynomials, related to ours by a product of Vandermonde factors for each node.} 

We denote the shuffle algebra element associated to a polynomial $p(s)$ in $n_i$ variables as 
\begin{equation}
    [p]_{\vec n}\,.
\end{equation}
Physically, each node of the quiver represents a type of D-branes/BPS particle of charge $\gamma_i$, and the $s_{i,a}$ variables can be reasonably interpreted as the positions of $n_i$ indistinguishable particles in the $\bC$ direction of the HT twist. 
The polynomial is a holomorphic analogue of a wavefunction. See Appendix \ref{appsec:SQM} for details.

The spin grading of $[p]_{\vec n}$ receives three contributions:  
\begin{equation}
2 \mathrm{deg}\, p + \sum_i n^2_i - \sum_{i,j}n_i n_j |i\to j| \,.
\end{equation}
Equivalently, $\Sh(Q)_{\vec n \cdot \vec\gamma, r}$ consists of polynomials of degree
\begin{equation}
\frac12 \left(r - \sum_i n^2_i +\sum_{i,j}n_i n_j |i\to j| \right)\,.
\end{equation}
We also assign a Grassmann parity $\sum_i n_i$ to $[p]_{\vec n}$, consistent with our physical interpretation.

The product in $\Sh(Q)$ is defined as follows. 
Consider two shuffle algebra elements described by polynomials $p_1(s^{(1)})$ and $p_2(s^{(2)})$ of $\vec n^{(1)}$ and $\vec n^{(2)}$ variables respectively. 
First define an auxiliary factor
\begin{equation}\label{eq:facDef}
\mathrm{fac}(s^{(1)},s^{(2)}) \equiv (-1)^{\sum_{i>j} n^{(1)}_i n^{(2)}_j}\cdot \prod_{{i,j}} \prod_{a_1,a_2} (s^{(1)}_{i,a_1}-s^{(2)}_{j,a_2})^{|i\to j|}\,.
\end{equation}
Then the shuffle product of $p_1(s^{(1)})$ and $p_2(s^{(2)})$ is defined as
\begin{equation}
[p_1(s^{(1)})]_{\vec n^{(1)}}\cdot [p_2(s^{(2)})]_{\vec n^{(2)}} \equiv \left[\sum_{\vec \sigma} \epsilon(\vec\sigma) \,(\mathrm{fac}\times p_1 \times \,p_2)( \vec\sigma(s))\right]_{\vec n^{(1)} + \vec n^{(2)}}\,,
\end{equation}
where the sum runs over a collection $\vec \sigma$ of shuffles 
\begin{equation}
\sigma_i \in (S_{n^{(1)}_i} \times  S_{n^{(2)}_i})\backslash S_{n^{(1)}_i + n^{(2)}_i}\,,
\end{equation}
which partition each sequence of variables $s_{i,a}$ into two sets $s^{(1)}_{i,b}$ and $s^{(2)}_{i,c}$, on which $\mathrm{fac}(s^{(1)},s^{(2)})p_1(s^{(1)})p_2(s^{(2)})$ is evaluated. 
The overall sign 
\begin{equation}
\epsilon(\vec\sigma) = \prod_i \epsilon(\sigma_i)
\end{equation}
ensures that the final answer is anti-symmetric under permutations of the $s_{i,a}$ for each node $i$. 
The definition of this shuffle product is of course compatible with the Heisenberg grading. 

We have included the overall sign $(-1)^{\sum_{i>j} n^{(1)}_i n^{(2)}_j}$ in our definition of $\mathrm{fac}$, see \eqref{eq:facDef}, to be consistent with our Grassmann parity conventions, so that in the absence of an arrow between two nodes the corresponding $e^{(1)}_n$ generators anti-commute. 
The sign requires a choice of ordering on the nodes of the quiver; different choices can be related by a sign redefinition of $p(s)$ and give the same Shuffle algebra $\Sh(Q)$. 
We can readily compute the character of the shuffle algebra: 
\begin{equation}
\chi(\mathrm{Sh}(Q)) = \sum_{n_i=0}^\infty \frac{\fq^{\sum_i n_i^2-\sum_{i,j} n_i n_j |i \to j|}}{\prod_i (\fq^2;\fq^2)_{n_i}} X_{\vec n \cdot \vec\gamma}\,.
\end{equation}

\medskip

We are actually interested in the Spherical Shuffle algebra $A \equiv \SH(Q)$, i.e.\ the sub-algebra of $\Sh(Q)$ generated by the single-node fermionic elements 
\begin{equation}
e^{(i)}_k \equiv [s_{i,1}^k]_{\vec n} \qquad \textrm{with}\quad n_j = \delta_{i,j}\,.
\end{equation}
(Recall that we denoted the fermionic sub-algebra generated by the $e^{(i)}_k$ for some $i$ as $S_{\gamma_i}$.) 
More generally, 
\begin{equation}
e_{m_1} \cdots e_{m_k} = \left[\det_{a,b} s_{a}^{m_b}\right]_{\vec n} \qquad \textrm{with}\quad 
n_j = k \delta_{i,j}\,.
\end{equation}
Finally, it is also useful to introduce generating functions 
\begin{equation}
e^{(i)}(z) \equiv \left[\frac{1}{z-s_{i,1}}\right]_{\vec n} \qquad \qquad n_j = \delta_{i,j}\,.
\end{equation}

\medskip

We can compute the products of generators for different nodes. 
If $\langle \gamma_i, \gamma_j\rangle \geq 0$, then
\begin{equation}
e^{(j)}(w) e^{(i)}(z) =\left[- \frac{1}{z-s_{i,1}}\frac{1}{w-s_{j,1}} \right]_{\vec n} \qquad \qquad n_k = \delta_{i,k} + \delta_{j,k} \,,
\end{equation}
and on the other hand 
\begin{equation}
e^{(i)}(z) e^{(j)}(w) = \left[\frac{1}{z-s_{i,1}}\frac{1}{w-s_{j,1}} (s_{i,1} - s_{j,1})^{|i \to j|} \right]_{\vec n} \qquad \qquad n_k = \delta_{i,k} + \delta_{j,k} \,.
\end{equation}
This allows us to express $e^{(i)}_n e^{(j)}_m$ as 
a linear combination of $e^{(j)}_{m'} e^{(i)}_{n'}$
with $n'+m' = n+m+|i \to j|$:
explicitly,
\begin{equation}\label{eq:ExchangeRelations}
e^{(i)}_{n} e^{(j)}_{m} = \sum_{k=0}^{|i \to j|}(-1)^{k+1} e^{(j)}_{m+k} e^{(i)}_{n+|i \to j|-k}  
\,, \qquad \qquad 
\langle \gamma_i, \gamma_j \rangle \geq 0 \,.
\end{equation}
This {\it exchange relation} is often written as
\begin{equation}\label{eq:algebraexchangerelation}
e^{(i)}(z) e^{(j)}(w) + (z - w)^{|i \to j|} e^{(j)}(w) e^{(i)}(z)\sim 0
\,, \qquad \qquad 
\langle \gamma_i, \gamma_j \rangle \geq 0 \,,
\end{equation}
where $\sim$ means up to regular terms.
This is the $h_I\rightarrow 0$ limit of the $e-e$ relations from the corresponding quiver Yangians, see \eqref{eq:eefromQY}.

\medskip

According to our conjectures, the spectrum generator should be 
reproduced by the character of the spherical shuffle algebra:
\begin{equation}
\cS = \chi(\SH(Q))\,,
\end{equation}
and should behave correctly under the mutations of $Q$. 
This statement is generically far from obvious, with a class of exceptions we discuss momentarily. 

\subsection{Ordered spectra}

A particularly simple situation is one where the nodes of the quiver can be totally ordered so that $|i \to j|=0$ if $i<j$. 
This condition prevents loops in the quiver, so $W=0$ necessarily and $H(Q,W) = H(Q,0)=\Sh(Q)$. 
We expect $\SH(Q) = \Sh(Q)$ from consistency with our general picture. 

This is indeed the case: any shuffle algebra element can be written as a product of $e^{(i)}_k$ such that no non-trivial $\mathrm{fac}$ factors appear: 
\begin{equation}
\Sh(Q) 
= \prod_{i} S_{\gamma_i}\,,
\end{equation}
where the product is taken along the order of the nodes, so that $e^{(j)}_t$ comes before $e^{(i)}_k$ if $\langle \gamma_i, \gamma_j\rangle \geq 0$. 

This is also the PBW factorization of $\SH(Q) = \Sh(Q)=H(Q,0)$, associated to a matching total order of $\mathrm{arg}\, Z_{\gamma_i}$. Accordingly,
\begin{equation}
\cS = \chi(\mathrm{Sh}(Q)) =  \prod_i \Phi(X_{\gamma_i}) \,.
\end{equation}
Intuitively, the total order condition on the quiver prevents the formation of bound states between the elementary particles when the phases of $Z_{\gamma_i}$ are ordered appropriately. 

\medskip

It is important to observe that this PBW basis consists of products of $e^{(i)}_n$'s only and thus the pairwise exchange relations \eqref{eq:ExchangeRelations} are sufficient to characterize the whole algebra and do computations without referring to the underlying shuffle algebra description. 
This will not be the case when loops are present: extra relations will appear beyond the quadratic exchange relations.

However, we emphasize that the knowledge of $\cS$ for an ordered quiver $Q$ is sufficient to reconstruct $\cS$ for all quivers in the mutation class $[Q]$, some of which may well have loops. 
This provides a candidate $\cS$ for many quivers, which can be tested against $\chi(\mathrm{Sh}(Q))$. 
We will do so in several examples. 

\subsection{Extra symmetries} \label{sec:ya}
We have discussed $\mathrm{Sh}(Q)$ as an algebra in $\mathrm{Vect}^{\mathrm{Hei}}_\Gamma$. 
We will now briefly discuss an extended symmetry algebra which is present in the problem. 

Formally, the shuffle product is compatible with a transformation:
\begin{equation}
    [p]_{\vec n} \to \left[p \prod_i \prod_{a_i=1}^{n_i} f_i(s_{i,a_i})\right]_{\vec n}
\end{equation}
for some power series
\begin{equation}
    f(s) = 1 + f_1 s + f_2 s^2 + \cdots
\end{equation}
Infinitesimally, we can define derivations $d_{i,k}$ acting as 
\begin{equation}\label{eq:derivationDef}
d_{i,k}: \quad [p]_{\vec n} \to \left[p \sum_{a_i=1}^{n_i} s_{i,a_i}^k\right]_{\vec n}
\end{equation}
These shift the spin charge by $k$.

We can denote these extra symmetry transformations as $\Gamma^\vee({\mathcal O})$ and the combination with the usual Heisenberg grading as $\Gamma_\mathrm{Hei}^\vee({\mathcal O})$.\footnote{If the $\gamma_i$ do not generate $\Gamma$, we are adding some extra symmetry generators which act trivially on the algebra.}

\section{Examples of spherical CoHA $\SH(Q)$}
In this Section we discuss quivers associated to the two simplest rank $1$ Argyres-Douglas theories \cite{Argyres:1995jj}, which are the main example employed throughout the paper. 
\subsection{The $A_2$ quiver}

This example was the main test case in \cite{Gaiotto:2023dvs}.
It describes the BPS spectrum of the simplest Argyres Douglas theory. 
The quiver has two nodes and a single arrow between them:
\begin{equation}\label{fig:A1quiver}
\begin{tikzpicture}[scale=1]
\node[vertex,minimum size=0.5mm,font=\footnotesize, 
] (a0) at (-1,0)  {$1$};
\node[vertex,minimum size=0.5mm,font=\footnotesize] (a1) at (1,0)  {$2$};
\draw[midarrow]   (a0) -- (a1) node[midway, above] {} ;
;
\end{tikzpicture}    
\end{equation}
In particular, $\langle \gamma_1, \gamma_2 \rangle = 1$.

The nodes of the quiver are totally ordered, leading to a PBW factorization in an appropriate chamber
\begin{equation}
A = \mathrm{H}(A_2,0) = \SH(A_2) = \Sh(A_2) = S_{\gamma_2} S_{\gamma_1}\,.
\end{equation}

\subsubsection{Decorated mutation orbits}
The $A_2$ quiver is mapped back to itself by a mutation at either node, 
up to a permutation of the nodes. If we include the effect of mutations (and the permutation) on charge labels, we have:
\begin{align}
\mu_1^{\pi}: \qquad \,&
\gamma'_1 = \gamma_1 + \gamma_2 
\,,\qquad \,\,\,
\gamma'_2 = - \gamma_1\,, \cr
\mu_1^{0}: \qquad \,&
\gamma'_1 = \gamma_2 
\,, \qquad \qquad \,\,\,\,\,
\gamma'_2 = - \gamma_1  \,, \cr
\mu_2^{\pi}: \qquad \,&
\gamma'_1 = - \gamma_2 
\,,\qquad \qquad 
\gamma'_2 = \gamma_1 \,, \cr
\mu_2^{0}: \qquad \,&
\gamma'_1 = - \gamma_2
\,,\qquad \qquad 
\gamma'_2 = \gamma_1 + \gamma_2 \,.
\end{align}

\subsubsection{Spectrum generator and chambers}

The above PBW factorization gives an explicit expression for the spectrum generator
\begin{equation}\label{eq:A2S2}
\cS = \Phi(X_{\gamma_2})\,\Phi(X_{\gamma_1}) \,.
\end{equation}
The famous pentagon relation for the quantum dilogarithm gives an alternative expression 
\begin{equation}\label{eq:A2S3}
\cS = \Phi(X_{\gamma_1})\Phi(X_{\gamma_1+ \gamma_2})\Phi(X_{\gamma_2}) \,.
\end{equation}
which is known to describe the BPS spectrum in a different chamber in the space of vacua, where the relative order of $\mathrm{arg} Z_{\gamma_i}$ is permuted. 

We will discuss momentarily the candidate PBW factorizations of $A$ associated to this chamber. First, we can discuss the behaviour of the spectrum generator under mutations:
\begin{itemize}
\item Recall that $\mu^{\pi}_1$ should remove a $\Phi(\gamma_1)$ factor from $\cS$ from the left. 
We should thus compare
\begin{align}
\cS &= \Phi(X_{\gamma_1})\,\Phi(X_{\gamma_1+ \gamma_2})\,\Phi(X_{\gamma_2}) \,, \cr
\cS' &= \Phi(X_{\gamma_1+ \gamma_2})\,\Phi(X_{\gamma_2})\,\Phi(X_{-\gamma_1})
\end{align}
which indeed agrees with the action of $\mu^{\pi}_1$ on the node labels.
\item Recall that $\mu^{0}_1$ should remove a $\Phi(\gamma_1)$ factor from $\cS$ from the right. 
We should thus compare
\begin{align}
\cS &= \Phi(X_{\gamma_2})\,\Phi(X_{\gamma_1})  \,, \cr
\cS' &= \Phi(X_{-\gamma_1}) \,\Phi(X_{\gamma_2})
\end{align}
which indeed agrees with the action of $\mu^{0}_1$ on the node labels.
\item Analogously, for $\mu_2^{\pi}$
\begin{align}
\cS &= \Phi(X_{\gamma_2})\,\Phi(X_{\gamma_1})  \,, \cr
\cS' &= \Phi(X_{\gamma_1}) \,\Phi(X_{-\gamma_2}) \, .
\end{align}
\item Finally, for $\mu_2^{0}$
\begin{align}
\cS &= \Phi(X_{\gamma_1})\,\Phi(X_{\gamma_1+ \gamma_2})\,\Phi(X_{\gamma_2}) \,, \cr
\cS' &= \Phi(X_{-\gamma_2}) \,\Phi(X_{\gamma_1})\,\Phi(X_{\gamma_1+ \gamma_2})
\end{align}
which indeed agrees with the action of $\mu^-_1$ on the node labels.
\end{itemize}

Concretely, the algebra $A$ is generated by the modes $e^{(1)}_n$ and $e^{(2)}_n$ associated to the two nodes of the quiver. 
The exchange relations take the form 
\begin{equation}\label{eq:exchangeA2}
    e^{(1)}_n e^{(2)}_m = e^{(2)}_{m+1} e^{(1)}_n - e^{(2)}_{m} e^{(1)}_{n+1}
\end{equation}
In the shuffle algebra language, that follows from 
\begin{equation}
[s_{1,1}^n]_{(1,0)} \cdot [s_{2,1}^m]_{(0,1)} = [(s_{1,1}-s_{2,1})s_{1,1}^n s_{2,1}^m]_{(1,1)} = e^{(2)}_{m+1} e^{(1)}_n - e^{(2)}_{m} e^{(1)}_{n+1}
\end{equation}
The PBW basis which arises from the total order takes the form
\begin{equation}\label{eq:PBWA22}
\prod_{a=1}^{k_1} e^{(2)}_{n_a} \cdot \prod_{a=1}^{k_2} e^{(1)}_{m_a} = [\pm \det_{a_1,b_1} s_{1,a_1}^{n_{b_1}}\det_{a_2,b_2} s_{2,a_2}^{m_{b_2}}]_{(k_1,k_2)}
\end{equation}

A candidate ``three factor'' PBW basis 
\begin{equation}
A =S_{\gamma_1}S_{\gamma_1+ \gamma_2}S_{\gamma_2}\,.
\end{equation}
exists but, curiously, is not unique. The product $e^{(2)}_0 e^{(1)}_0$ cannot be re-written in the opposite order and is thus a candidate for the first generator $e^{(12)}_0$ in $S_{\gamma_1+ \gamma_2}$. It can be completed to a fermionic sub-algebra $S_{\gamma_1+ \gamma_2}$ in two distinct ways:
\begin{align}
e^{(12)}_k &= e^{(2)}_0 e^{(1)}_k \,, \cr
e^{(12)}_k &= e^{(2)}_k e^{(1)}_0 \,.
\end{align}
For example, 
\begin{equation}
( e^{(2)}_0 e^{(1)}_k) (e^{(2)}_0 e^{(1)}_t) = e^{(2)}_0 e^{(2)}_1 e^{(1)}_k e^{(1)}_t \,,
\end{equation}
which is anti-symmetric in $k$ and $t$.
$e^{(12)}_n$ with the first definition can be rewritten into   
\begin{align}\label{eq:e121st}
e^{(12)}_{n+m}&=e^{(2)}_ne^{(1)}_m+\sum_{k=1}^ne^{(1)}_{n+m-k}e^{(2)}_{k-1}\,,
\end{align}
and satisfies the exchange relations with $e^{(1,2)}_n$:
\begin{align}\label{eq:e121stex}
e^{(2)}_ne^{(12)}_m&=-\sum_{k=1}^ne^{(12)}_{n+m-k}e^{(2)}_{k-1}\,,\cr
e^{(12)}_ne^{(1)}_m&=\left(\sum_{k=1}^m-\sum_{k=1}^n\right) e^{(1)}_{m+n-k}e^{(12)}_{k-1}\,.
\end{align}
Similarly, $e^{(12)}_n$ with the second definition can be rewritten into
\begin{align}\label{eq:e122nd}
e^{(12)}_{m+n}&=e^{(2)}_ne^{(1)}_m-\sum_{k=1}^m e^{(1)}_{k-1}e^{(2)}_{m+n-k}\,,
\end{align}
and satisfies
\begin{align}\label{eq:e122ndex}
e^{(2)}_ne^{(12)}_m&=-\left(\sum_{k=1}^m-\sum_{k=1}^n\right) e^{(12)}_{m+n-k}e^{(2)}_{k-1}\,, \cr
e^{(12)}_ne^{(1)}_m&=-\sum_{k=1}^m e^{(1)}_{k-1}e^{(12)}_{m+n-k}\,.
\end{align} 
Either of these collections work to define a PBW basis of the desired form:
\begin{equation}\label{eq:PBWA23}
\prod_{a=1}^{k_1} e^{(1)}_{l_a} \cdot \prod_{a=1}^{k_{12}} e^{(12)}_{m_a} \cdot \prod_{a=1}^{k_2} e^{(2)}_{n_a} \,.
\end{equation}
To check the equivalence between \eqref{eq:PBWA22} and \eqref{eq:PBWA23}, one applies iteratively the original exchange relations \eqref{eq:exchangeA2}, \eqref{eq:e121st} (or \eqref{eq:e122nd}), and the exchange relations between $e^{(12)}_n$ and $e^{(1,2)}_m$.

\subsection{The $A_3$ quiver and its relatives}
\label{ssec:sHA3}
This three-node quiver and its mutation orbit $[A_3]$ describes the BPS spectrum of the next simplest AD theory and illustrates nicely the difference between $\Sh$ and $\SH$ in the presence of loops.

\medskip

The $A_3$ quiver 
\begin{equation}\label{fig:A3Quiver}
\begin{tikzpicture}[scale=1]
\node[vertex,minimum size=0.5mm,font=\footnotesize] (a0) at (0,0)  {$1$};
\node[vertex,minimum size=0.5mm,font=\footnotesize] (a1) at (2,0)  {$2$};
\node[vertex,minimum size=0.5mm,font=\footnotesize] (a2) at (4,0)  {$3$}; 

\draw[midarrow] (a0) -- (a1) node[midway, below] {}; 
\draw[midarrow] (a1) -- (a2) node[midway, below] {}; 

\end{tikzpicture}    
\end{equation}
encodes charges $\gamma_i$ whose non-zero Dirac pairings are
\begin{equation}
\langle \gamma_1, \gamma_2 \rangle = \langle \gamma_2, \gamma_3 \rangle=1\,.
\end{equation} 
This pairing has an obvious kernel $\gamma_f \equiv \gamma_1 + \gamma_3$: a {\it flavour charge}
in the underlying physical theory. This charge is actually the Cartan generator of an $SU(2)$ symmetry of the theory, which is somewhat hidden in this chamber. 

\medskip

Since the quiver is ordered, we have a simple canonical PBW basis
\begin{equation}
\SH(A_3) = S_{\gamma_3}S_{\gamma_2}S_{\gamma_1} \,,
\end{equation}
with the corresponding spectrum generator
\begin{equation}\label{eq:A3S1}
\cS(A_3) = \Phi(X_{\gamma_3})\Phi(X_{\gamma_2})\Phi(X_{\gamma_1})\,.
\end{equation}
An application of the pentagon identity gives two other useful expressions for the spectrum generator: 
\begin{align}
\cS(A_3) &= \Phi(X_{\gamma_3})\Phi(X_{\gamma_1})\Phi(X_{\gamma_1 + \gamma_2})\Phi(X_{\gamma_2})= \Phi(X_{\gamma_2})\Phi(X_{\gamma_2+ \gamma_3})\Phi(X_{\gamma_3})\Phi(X_{\gamma_1})\,,
\end{align}
corresponding to the PBW bases
\begin{align}
\SH(A_3) &= S_{\gamma_3}S_{\gamma_1}S_{\gamma_1+\gamma_2} S_{\gamma_2}=
S_{\gamma_2}S_{\gamma_2+\gamma_3}S_{\gamma_3} S_{\gamma_1}\,.   
\end{align}
Here $S_{\gamma_1+\gamma_2}$ is generated by
\begin{equation}
e^{(12)}_n\equiv e^{(2)}_n e^{(1)}_0 \,,
\end{equation}
which satisfies \eqref{eq:e122nd}, \eqref{eq:e122ndex}, and 
\begin{equation}
e^{(12)}_ne^{(3)}_m=e^{(3)}_me^{(12)}_{n+1}-e^{(3)}_{m+1}e^{(12)}_{n}  \,.  
\end{equation}
Similarly $S_{\gamma_2+\gamma_3}$ is generated by
\begin{equation}
e^{(23)}_n\equiv e^{(3)}_0 e^{(2)}_n \,,
\end{equation}
which satisfies \eqref{eq:e121st} and \eqref{eq:e121stex} with the relabelling $1\rightarrow 2$ and $2\rightarrow 3$, and 
\begin{equation}
e^{(1)}_me^{(23)}_n=e^{(23)}_{n}e^{(1)}_{m+1}-e^{(23)}_{n+1}e^{(1)}_{m}  \,.  
\end{equation}

\medskip

Since $\Phi(\gamma_1)$ and $\Phi(\gamma_3)$ commute, these three expressions are enough for us to follow the spectrum generator across the six possible mutations of the labelled $A_3$ quiver:
\begin{itemize}
\item A mutation of $A_3$ at the first node, followed by $1 \leftrightarrow 2$ relabelling of the nodes, gives the $A_3^{(2)}$ quiver
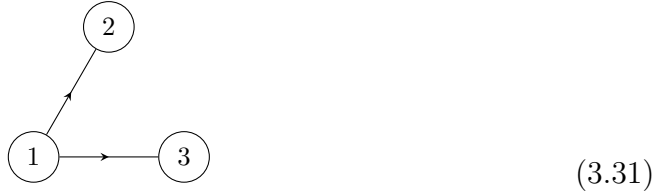
\begin{equation}\label{fig:A3Quiver2}
\begin{tikzpicture}[scale=1]
\node[vertex,minimum size=0.5mm,font=\footnotesize] (a0) at (0,0)  {$1$};
\node[vertex,minimum size=0.5mm,font=\footnotesize] (a1) at (2,0)  {$3$};
\node[vertex,minimum size=0.5mm,font=\footnotesize] (a2) at (1,1.73)  {$2$}; 

\draw[midarrow] (a0) -- (a1) node[midway, below] {}; 
\draw[midarrow] (a0) -- (a2) node[midway, left] {}; 

\end{tikzpicture}    
\end{equation} with 
\begin{equation}
\langle \gamma_1, \gamma_2 \rangle 
=\langle \gamma_1, \gamma_3 \rangle
=1\,.
\end{equation} 
The flavour charge is now $\gamma_f = \gamma_3 - \gamma_2$ and the second and third nodes play a symmetric role. This agrees with the underlying $SU(2)$ symmetry of the theory. 

This quiver is also ordered. The resulting expression for the spectrum generator
\begin{equation}
\cS(A_3^{(2)}) = \Phi(X_{\gamma_3})\Phi(X_{\gamma_2})\Phi(X_{\gamma_1})\,,
\end{equation} 
matches the mutation image of the $A_3$ spectrum generator \eqref{eq:A3S1}.

\item A mutation of $A_3$ at the third node, followed by $2 \leftrightarrow 3$ relabelling of the nodes, gives the $A_3^{(3)}$ quiver 
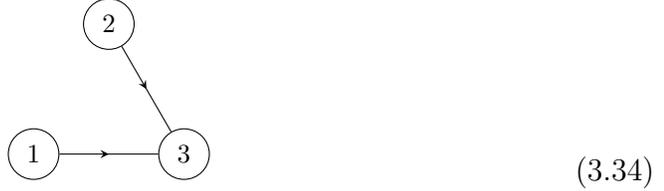
\begin{equation}\label{fig:3NodeQuiver}
\begin{tikzpicture}[scale=1]
\node[vertex,minimum size=0.5mm,font=\footnotesize] (a0) at (0,0)  {$1$};
\node[vertex,minimum size=0.5mm,font=\footnotesize] (a1) at (2,0)  {$3$};
\node[vertex,minimum size=0.5mm,font=\footnotesize] (a2) at (1,1.73)  {$2$}; 

\draw[midarrow] (a0) -- (a1) node[midway, below] {}; 
\draw[midarrow] (a2) -- (a1) node[midway, left] {}; 

\end{tikzpicture}    
\end{equation} with 
\begin{equation}
\langle \gamma_1, \gamma_3 \rangle 
= \langle \gamma_2, \gamma_3 \rangle=1\,.
\end{equation} 

Similar to the previous case, this quiver is also ordered, with the spectrum generator
\begin{equation}
\cS(A_3^{(3)}) = \Phi(X_{\gamma_3})\Phi(X_{\gamma_2})\Phi(X_{\gamma_1})\,,
\end{equation} 
again matching the mutation image of the $A_3$ spectrum generator \eqref{eq:A3S1}.
The first and second nodes play a symmetric role, corresponding to the underlying $SU(2)$ symmetry of the theory and $\gamma_f = \gamma_1 - \gamma_2$.

\item A mutation of $A_3$ at the second node gives the $\hat A_2$ quiver
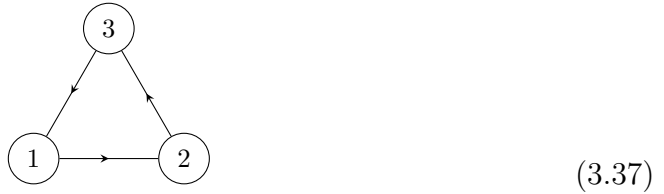
\begin{equation}\label{fig:CircularQuiver}
\begin{tikzpicture}[scale=1]
\node[vertex,minimum size=0.5mm,font=\footnotesize] (a0) at (0,0)  {$1$};
\node[vertex,minimum size=0.5mm,font=\footnotesize] (a1) at (2,0)  {$2$};
\node[vertex,minimum size=0.5mm,font=\footnotesize] (a2) at (1,1.73)  {$3$}; 

\draw[midarrow] (a0) -- (a1) node[midway, below] {}; 
\draw[midarrow] (a1) -- (a2) node[midway, right] {}; 
\draw[midarrow] (a2) -- (a0) node[midway, left]  {}; 

\end{tikzpicture}   
\end{equation}
with 
\begin{equation}
\langle \gamma_1, \gamma_2 \rangle 
= \langle \gamma_2, \gamma_3 \rangle
= \langle \gamma_3, \gamma_1 \rangle=1
\,,
\end{equation} 
which has a loop and an associated corresponding cubic superpotential.

Because of the loop, this quiver is not ordered. 
However, the spectrum generator can be computed via mutation:\footnote{It may seem surprising that the obvious $\bZ_3$ symmetry of the quiver is not manifest in the spectrum generator, but some pentagon identities can fix that. E.g. 
\begin{equation}
\cS(\hat A_2) =  \Phi(X_{\gamma_1}) \Phi(X_{\gamma_1+ \gamma_2})\Phi(X_{\gamma_2})\Phi(X_{\gamma_2 + \gamma_3})\Phi(X_{\gamma_3}) =  \Phi(X_{\gamma_1}) \Phi(X_{\gamma_1+ \gamma_2})\Phi(X_{\gamma_3})\Phi(X_{\gamma_2})\,.
\end{equation}}
\begin{equation}
\cS(\hat A_2) = \Phi(X_{\gamma_2}) \Phi(X_{\gamma_1})\Phi(X_{\gamma_2 + \gamma_3})\Phi(X_{\gamma_3})\,.
\end{equation}
We do not know how to prove directly that this is indeed the character of $\SH(\hat A_2)$, but we have computed by hand the dimensions of the graded components $\SH_{\gamma,k}(\hat A_2)$ for small values of the charges and found agreement with the above expression. 
We have also identified two candidates for the fermionic generators in $S_{\gamma_2 + \gamma_3}$:
\begin{equation}
[s_{2,1}^k]_{(0,1,1)}
\qquad \textrm{and} \qquad [s_{3,1}^k]_{(0,1,1)}\,,
\end{equation}
and tested the first few entries of the expected PBW basis 
\begin{equation}
\SH(\hat A_2) = S_{\gamma_2}S_{\gamma_1}S_{\gamma_2 + \gamma_3}S_{\gamma_3}\,.
\end{equation}
We can also compute
\begin{equation}
e^{(3)}_0 e^{(2)}_0 e^{(1)}_0 = s^{(3)}_1 - s^{(1)}_1 \,,
\end{equation}
and thus find an extra cubic relation beyond the quadratic exchange relations
\begin{equation}
e^{(3)}_0 e^{(2)}_0 e^{(1)}_0 +e^{(2)}_0 e^{(1)}_0 e^{(3)}_0 +e^{(1)}_0 e^{(3)}_0 e^{(2)}_0 =0 \, .
\end{equation}
\end{itemize}

The collection of mutations relating the quivers can be depicted schematically as 
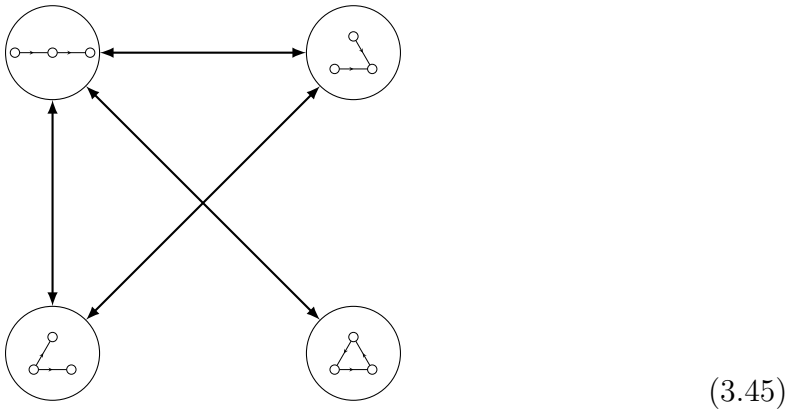
\begin{equation}
\begin{tikzpicture}[scale=1, every node/.style={scale=0.5}]
\tikzstyle{smallquiver}=[circle, draw, minimum size=2.5cm, inner sep=0pt]
\tikzstyle{vertex}=[circle, draw, inner sep=5pt, fill=white, minimum size=.5mm, font=\tiny]

\node[smallquiver] (q1) at (0,0) {
    \begin{tikzpicture}
        \node[vertex] (a0) at (0,0) {};
        \node[vertex] (a1) at (1,0) {};
        \node[vertex] (a2) at (2,0) {};
        \draw[midarrow] (a0) -- (a1);
        \draw[midarrow] (a1) -- (a2);
    \end{tikzpicture}
};
\node[smallquiver] (q2) at (4,0) {
    \begin{tikzpicture}
        \node[vertex] (b0) at (0,0) {};
        \node[vertex] (b1) at (1,0) {};
        \node[vertex] (b2) at (0.5,0.866) {};
        \draw[midarrow] (b0) -- (b1);
        \draw[midarrow] (b2) -- (b1);
    \end{tikzpicture}
};
\node[smallquiver] (q3) at (0,-4) {
    \begin{tikzpicture}
        \node[vertex] (c0) at (0,0) {};
        \node[vertex] (c1) at (1,0) {};
        \node[vertex] (c2) at (0.5,0.866) {};
        \draw[midarrow] (c0) -- (c1);
        \draw[midarrow] (c0) -- (c2);
    \end{tikzpicture}
};
\node[smallquiver] (q4) at (4,-4) {
    \begin{tikzpicture}
        \node[vertex] (d0) at (0,0) {};
        \node[vertex] (d1) at (1,0) {};
        \node[vertex] (d2) at (0.5,0.866) {};
        \draw[midarrow] (d0) -- (d1);
        \draw[midarrow] (d1) -- (d2);
        \draw[midarrow] (d2) -- (d0);
    \end{tikzpicture}
};

\tikzstyle{largequiverarrow}=[<->, >=latex, thick]

\draw[largequiverarrow] (q1) -- (q2) node[midway, above] {};
\draw[largequiverarrow] (q1) -- (q3) node[midway, left] {};
\draw[largequiverarrow] (q1) -- (q4) node[midway] {};
\draw[largequiverarrow] (q2) -- (q3) node[midway] {};

\end{tikzpicture}
\end{equation}

\section{Framed BPS states and CoHA bimodules from framed quivers}
\label{sec:CoHAbimodule}

A potential physical strategy to describe the IR behaviour of line defects is to mimic the quiver description of BPS particles and represent framed BPS states as bound states of an ``elementary'' IR 't Hooft-Wilson line defect with the ``elementary'' BPS particles. 

It is often the case that a line defect can be identified, at least formally, with a very massive particle in a larger theory that reduces to the original theory below that mass scale. 
The elementary 't Hooft-Wilson line defect can be identified as a very massive elementary BPS particle in the IR description of such an ambient theory. 
This suggests that the same mathematical technology which describes the BPS spectrum in terms of bound states between elementary BPS particles should apply to the framed BPS spectrum as well \cite{Chuang:2013wt,Cirafici:2013bha,Cordova:2013bza,Cirafici:2017wlw}.

This strategy will give candidate framed BPS degeneracies for any charge $\gamma_*$ of the elementary IR line defect. 
It is generally believed, but not proven, that for each vacuum $u$, there will be a one-to-one correspondence between half-BPS UV line defects $L$ and the ``tropical charges'' $\gamma_*(L;u)$, which mutates by tropical cluster transformations across walls of marginal stability. Our goal is to promote this 
correspondence to a monoidal functor $\RG_u$.

It was proposed that a BPS quiver description of the framed BPS spectrum will involve a ``framed'' version $Q_{\vec f}$ of the BPS quiver $Q$: we add an extra non-dynamical ``framing'' node $*$ of rank $1$ representing the extra very massive BPS particle and specify the numbers of arrows 
\begin{equation}
f_i \equiv \langle \gamma_*, \gamma_i\rangle = |*\to i| -|i\to *|
\end{equation} 
that are connected to the new node $*$. 
We also include a new superpotential $W_{\vec f}$, which we assume to be infinitely-mutable and is such that $H(Q_{\vec f},W_{\vec f})$ is spherical and is isomorphic to the spherical subalgebra $\textrm{sH}(Q_{\vec f})$ of $H(Q_{\vec f},0)$, and can thus be treated as before. 

\medskip

If we treat $(Q_{\vec f},W_{\vec f})$ as a standard quiver with potential, we can build a CoHA $H(Q_{\vec f},W_{\vec f})$, conjecturally equal to $\SH(Q_{\vec f})$.\footnote{The standard conjectural route to the framed BPS spectrum in a CoHA context is to give a separate notion of stable framed representations of $(Q,W)$ \cite{Szendroi:2007nu,Mozgovoy:2008fd,Soibelman:2014gta}. We will follow a slightly different route with a greater algebraic payoff. As a comparison, we will review in App.\ \ref{appsec:quiverYangian} how to reproduce the (unrefined) framed BPS degeneracies using the so-called quiver Yangian algorithm \cite{Li:2020rij,Galakhov:2021xum,Li:2023zub}, which produces modules (instead of bimodule) from the data of framed quivers.} This algebra is graded by an extended charge lattice $\Gamma \oplus \bZ_{\geq 0} \gamma_*$.

The $n_*=0$ subspace $H(Q_{\vec f},W_{\vec f})_\Gamma$ is expected to coincide with $H(Q,W)$. It is certainly the case that 
\begin{equation}
\SH(Q) = \SH(Q_{\vec f})_\Gamma\,.
\end{equation}
We immediately learn a remarkable fact: the $n_*=1$ subspace 
\begin{equation}
    \mathrm{H}(Q_{\vec f},W_{\vec f})_{\gamma_* + \Gamma}
\end{equation}
should be equipped with the structure of an $H(Q,W)$-bimodule! 

The extended CoHA $H(Q_{\vec f},W_{\vec f})_\Gamma$ encodes properties of BPS particles of the formal ambient theory. The $\mathrm{H}(Q_{\vec f},W_{\vec f})_{\gamma_* + \Gamma}$ subspace encodes properties of BPS particles which contain a single copy of the 
very massive elementary BPS particle. If we remove the contribution from the position $s_{*,1}$ of that particle, we should obtain a reduced space 
$\mathrm{H}(Q,W;\vec f)$ that is still a $\mathrm{H}(Q,W)$-bimodule
and encodes properties of framed BPS states for some half-BPS line defect.

\medskip

We will momentarily describe explicitly this reduced version 
\begin{equation}
\SH(Q;\vec f)\equiv    \SH(Q_{\vec f})_{\gamma_* + \Gamma}^{\mathrm{red}}
\end{equation}
as an $\SH(Q)$-bimodule. We will also adjust some signs in the definition for later convenience.

We will also illustrate how the Harder-Narasimhan filtrations of $H(Q_{\vec f},W_{\vec f})_\Gamma$ allow one to define vector spaces $V(Q,\vec f(L))$ such that 
\begin{equation}
    V(Q,\langle \gamma_*(L;u),\vec \gamma \rangle)
\end{equation} can be identified directly with the framed BPS degeneracies $V(L;u)$ of an half-BPS line defect $L$. Then 
\begin{equation}
    B_{L;u} \equiv \SH(Q;\langle \gamma_*(L;u),\vec \gamma \rangle)
\end{equation}
will be our candidate image $\RG_u(L)$ of $L$ in a monoidal category of CoHA bimodules.

\subsection{Shuffle bimodules}

In order to define the collection of $\Sh(Q)$-bimodules $\Sh(Q;\vec f)$, we simply restrict the formulae for $\Sh(Q_{\vec f})$ to the sector $n_*=1$ and set to zero the corresponding variable in the shuffle algebra description, i.e. $s_{*,1}=0$.

\medskip

The bimodule is defined in terms of polynomials in $s_{i,a}$ as for the algebra and we use the standard formulae for the shuffle product to define the left and right multiplication, say in Koszul conventions where the framing node is last. 
The factors which appear in the left and right actions are
\begin{align}
\mathrm{fac}_L(s^{(1)},s^{(2)}) &\equiv \prod_{{i}} \prod_{a_1} (s^{(1)}_{i,a_1})^{|i\to *|} \mathrm{fac}(s^{(1)},s^{(2)})\,, \cr
\mathrm{fac}_R(s^{(1)},s^{(2)}) &\equiv \prod_{{j}} \prod_{a_2} (s^{(2)}_{j,a_2})^{|*\to j|}\mathrm{fac}(s^{(1)},s^{(2)})\,,
\end{align}
respectively.\footnote{There are several possible choices of signs which could be included in this expression. Typically, they can be understood as twisting the left- or right- actions by some automorphisms of the algebra inherited from its grading. For example, one may want to have $(-s^{(2)}_{j,a_2})^{|*\to j|}$ in the second line, which twists the right action by $(-1)^{\langle \gamma_*, \gamma\rangle}$. These choices affect some details of the tensor product calculations. We discuss this in examples and leave a full analysis to future work.} 

We need to define a Heisenberg grading of $\Sh(Q;\vec f)$. This follows the usual rules but is valued in a torsor $\Gamma+\gamma_*$ with formal inner product 
\begin{equation}
\langle \gamma_*,\gamma_i\rangle = f_i = - \langle \gamma_i,\gamma_*\rangle \,.
\end{equation}
We will sometimes denote $\gamma_*$ as $\vec f$, especially if we do not yet want to choose a specific image $\gamma_*$ in $\Gamma$. 

A generic element of the bimodule is denoted as $[p]_{\vec n}$, with spin grading 
\begin{equation}
2\, \mathrm{deg}\, p + \sum_i n^2_i - \sum_{i,j}n_i n_j |i\to j| - \sum_{i} n_i(|i\to *|+|*\to i|)\,.
\end{equation}
We also assign a Grassmann parity $\sum_i n_i$ to $[p]_{\vec n}$. 
Finally, we also use the notation
\begin{equation}
b \equiv \left[1\right]_0 \in \Sh(Q;\vec f)\,.
\end{equation}

\medskip

Incidentally, simply by adding two separate framing nodes to a quiver $Q$, we obtain a map of bimodules:
\begin{equation}
\pi_{\vec f, \vec f'}: \qquad\Sh(Q;\vec f) \otimes_{\Sh(Q)} \Sh(Q;\vec f') \to \Sh(Q;\vec f+ \vec f') \,,
\end{equation}
defined by the shuffle product with factor 
\begin{equation}
\mathrm{fac}_\otimes(s^{(1)},s^{(2)}) \equiv 
 \prod_{{i}} \prod_{a_1} (s^{(1)}_{i,a_1})^{|i\to *|} \prod_{{j}} \prod_{a_2} 
(s^{(2)}_{j,a_2})^{|*'\to j|}\mathrm{fac}(s^{(1)},s^{(2)}) \,.
\end{equation}
This map is natural in a convention where $\langle \gamma_*,\gamma_*'\rangle = 0$. Otherwise, we may want to 
shift the grading of the tensor product and the map will hold up to this degree shift.

\medskip

The character of $\Sh(Q;\vec f)$ is readily computed as
\begin{equation}
\chi(\mathrm{Sh}(Q;\vec f)) = \sum_{n_i=0}^\infty \frac{\fq^{\sum_i n_i^2-\sum_{i,j} n_i n_j |i \to j|- \sum_{i} n_i(|i\to *|+|*\to i|)}}{\prod_i (\fq^2;\fq^2)_{n_i}} X_{\vec n \cdot \vec\gamma+ \gamma_*}\,.
\end{equation}

\subsection{Spherical shuffle bimodules}

We can define analogously the collection of $\SH(Q)$-bimodules $\SH(Q;\vec f)$ generated by $b \in \Sh(Q;\vec f)$ by acting with $e^{(i)}_k$'s from the left and from the right. 

\medskip

The interaction between the $e^{(i)}_k$'s and $b$ involves straightforward exchange relations: 
\begin{align} \label{eq:beexchangerelations}
b \,e^{(i)}_{k} &=e^{(i)}_{k+f_i} b\qquad \qquad \quad \, f_i>0 \,, \cr
b \, e^{(i)}_{k-f_i} &= e^{(i)}_{k} b \qquad \qquad\qquad f_i\leq 0   \,.
\end{align}
As usual, extra relations will appear in the presence of loops in the framed quiver.

Observe that the map $\pi_{\vec f, \vec f'}$ defined above may fail to map 
\begin{equation}
\SH(Q;\vec f) \otimes_{\SH(Q)} \SH(Q;\vec f') \to \SH(Q;\vec f+ \vec f')    \,,
\end{equation} as it might land outside the spherical part of the $\Sh(Q;\vec f+ \vec f')$ bimodule. Instead, we can only say that 
$\SH(Q;\vec f+ \vec f')$ is a subspace of the image of $\pi_{\vec f, \vec f'}$, i.e. a subspace of a quotient of the tensor product $\SH(Q;\vec f) \otimes_{\SH(Q)} \SH(Q;\vec f')$. We will do better with the help of PBW bases momentarily.

The tensor products of the spherical bimodules have a rich tensor structure, which is the main topic of this paper. 
We now detail the conjectural properties of $\SH(Q;\vec f)$ which follow from analogous conjectural properties of $\SH(Q_{\vec f})$.

\subsection{Mutations of framed quivers}
As we mutate the quiver $Q$ at the node $i$, we can also mutate 
$Q_{\vec f}$ at the same node. The framing arrows will be manipulated accordingly to give a new framing $Q'_{\vec f'}$ of the new quiver $Q'$:
\begin{align}
f'_i &= - f_i \,, \cr
f'_j &= f_j + |* \to i||i\to j| - |j \to i||i \to *|\,.
\end{align}

A sequence of mutations of $Q$ and vertex relabeling that gives back $Q$ will typically act non-trivially on $\vec f$. 
We give three illustrative examples here.
\begin{itemize}
\item For $Q=A_1$, the 1-node quiver with no loop, a mutation at the unique node acts as a reflection $f_1 \to - f_1$.
\item For $Q=A_2$, the 2-node quiver with one arrow from the first node to the second one, a mutation at the first node followed by relabelling of the nodes gives back $Q$ with 
\begin{equation}
f_1' = f_2+\mathrm{max}(f_1,0) \,, \qquad f_2' = - f_1 \,.
\end{equation}
This transformation, famously, has period $5$ and organizes the lattice of $\vec f$ charges into $\bZ_5$ orbits.
\item Starting from $(f_1,f_2,f_3)$ in the $A_3$ quiver, a mutation at the first and last node together with relabelling gives
\begin{equation}
f_1' = -f_3\,, \qquad 
f_2' = f_2+\mathrm{max}(f_1,0)+\mathrm{min}(f_3,0)\,,\qquad f_3' = - f_1 \,,
\end{equation}
which is a $\bZ_2$ transformation.
On the other hand, a mutation at the middle node followed by a mutation at the last node together with relabelling gives 
\begin{align}
f_1'&=-f_2+\mathrm{max}(f_3+\mathrm{max}(f_2,0),0)\,,\cr
f_2'&=-f_3-\mathrm{max}(f_2,0)\,,\cr
f_3'&=f_1+\mathrm{min}(f_2,0)+\mathrm{min}(f_3+\mathrm{max}(f_2,0),0)\,,
\end{align}
which is a $\bZ_3$ transformation. This $\bZ_6$ symmetry group generates the orbits of framed quivers. 
\end{itemize}

\subsection{PBW bases, factorizations, and mutations}
The Harder-Narasimhan filtrations of the semi-stable representations of $Q_{\vec f}$ and the factorization of the associated CoHA imply a factorization of the associated bimodules. 
If we choose the stability condition on the framed node appropriately, we can obtain the factorizations in which the bound states with the framing node are at one end or the other:
\begin{align}\label{eq:sHfactorizationLR}
\SH(Q;\vec f) &= V^{\pi}(Q;\vec f)\, \SH(Q) \,, \cr
\SH(Q;\vec f) &= \SH(Q)\,V^{0}(Q;\vec f) \,.
\end{align}
In other words: $\SH(Q;\vec f)$ is freely generated as a right- or a left- $\SH(Q)$ module. 
The graded components $V^{\pi}_{\vec f,\bullet}\SH(Q)$
and $V^{0}_{\vec f,\bullet}\SH(Q)$ ``at the origin'' of the torsor consist of $b$ only. 
More generally, for every phase $\vartheta$ in the upper half plane that is distinct from the phases of populated central charges, we have the factorization 
\begin{equation}\label{eq:sHfactorizationtheta}
\SH(Q;\vec f) = \left[\prod^{\curvearrowright}_{\gamma<\vartheta} S_\gamma \right]V^\vartheta(Q;\vec f)\left[\prod^{\curvearrowright}_{\gamma>\vartheta} S_\gamma \right]\,,
\end{equation}
where $\gamma<\theta$ means $\textrm{arg}Z_{\gamma}>\theta$ and vice versa.
Again, the graded component $V^\vartheta(Q;\vec f)_{\vec f,\bullet}$
consists of $b$ only. We expect $V^\vartheta(Q;\vec f)_{\vec f,\bullet}$ to also admit a definition in terms of ``stable framed representations of $(Q,W)$'' \cite{Soibelman:2014gta}, but we will not explore this idea here. 

As all of these expected factorizations come from the same structure in $\SH(Q_{\vec f})$, the factors $S_\gamma$ that appear in 
the $\SH(Q,\vec f)$ bimodules should coincide with those that appear in the corresponding factorization of $\SH(Q)$: although the latter is not canonical, it should at least be possible to make the {\it same} non-canonical choice for all the bimodules simultaneously. 

\medskip

The factorizations in \eqref{eq:sHfactorizationLR} and \eqref{eq:sHfactorizationtheta}  lead to the corresponding factorizations of the characters: 
\begin{align}
\chi\left(\SH(Q;\vec f)\right) &= F^{\pi}(Q;\vec f) \, \cS \,, \cr
\chi\left(\SH(Q;\vec f)\right) &= \cS \, F^{0}(Q;\vec f) \,, \cr
\chi\left(\SH(Q;\vec f)\right) & = \left[\prod^{\curvearrowright}_{\gamma<\vartheta} \chi(S_\gamma) \right]F^\vartheta(Q;\vec f)\left[\prod^{\curvearrowright}_{\gamma>\vartheta} \chi(S_\gamma) \right]\,,
\end{align}
where $F^{\pi}(Q;\vec f)$, etc. are ``monic'' power series in the quantum torus algebra with non-negative coefficients. 

Finally, if we mutate $Q$, we get a prediction for how the characters $\chi\left(\SH(Q;\vec f)\right)$ and $\chi\left(\SH(Q';\vec f')\right)$ are related: in the same manner as $\cS$ and $\cS'$!
More precisely, whatever quantum dilogarithm factors are removed from the left and added to the right of $\cS$, or vice versa, will be stripped from the left and added to the right of $\chi\left(\SH(Q;\vec f)\right)$, or vice versa. 
From now on, we will often only give the expressions for $V^{\pi}(Q;\vec f)$ and $F^{\pi}(Q;\vec f)$, and define
\begin{equation}
V(Q;\vec f) \equiv V^{\pi}(Q;\vec f)\,,\qquad 
F(Q;\vec f) \equiv F^{\pi}(Q;\vec f)\,.
\end{equation}
The expressions for $V^{0}(Q;\vec f)$ (or $V^{\theta}(Q;\vec f)$) and $F^{0}(Q;\vec f)$ (or $F^{\theta}(Q;\vec f)$) can be derived similarly.

\medskip

Chasing definitions, we reach the final statement:
\begin{itemize}
\item Depending on the sign of $Z_{\gamma_i}$, a mutation relates $F(Q;\vec f)$ and $F(Q';\vec f')$ via conjugation by $\Phi(X_{\gamma_i})$ or $\Phi(X_{-\gamma_i})^{-1}$.
\end{itemize}
Concretely, we have
 \begin{equation}
 \Phi(X_{\gamma}) X_{\gamma'} 
 = X_{\gamma'}\Phi(\fq^{2\langle \gamma, \gamma' \rangle}X_{\gamma}) \,,
 \end{equation}
which can be rewritten as
\begin{align}
\Phi(X_{\gamma})X_{\gamma'}(- \fq X_{\gamma};\fq^2)_{\langle\gamma,\gamma'\rangle}=X_{\gamma'}   \Phi(X_{\gamma}) \qquad\qquad&\langle\gamma,\gamma'\rangle\geq0 \,,\cr
\Phi(X_{\gamma})X_{\gamma'}=(- \fq X_{\gamma};\fq^2)_{\langle\gamma',\gamma\rangle}X_{\gamma'}   \Phi(X_{\gamma}) \qquad\qquad&\langle\gamma,\gamma'\rangle<0 \,,
\end{align}
which can then be used to mutate $F(Q;\vec f)$.
\medskip

The transformation rules of the $F(Q;\vec f)$ across the walls of marginal stability are the same as the wall-crossing formulae for the characters $F_L$ of framed BPS states of half-BPS line defects. Indeed, in a slightly different guise, the $V(Q;\vec f)$ have been proposed as the quiver description of framed BPS degeneracies for a line defect of tropical charge $\gamma_*$. 

\medskip

We should comment briefly on the meaning of $\vartheta$ as well.
The phase $\vartheta$ is the phase of the central charge of the very massive BPS particle we added to the system. 
When this BPS particle is identified with a line defect, $\vartheta$ determines which half of the SUSY charges is preserved by the line defect. 
It is generally expected, but not proven, that half-BPS line defects defined for some value of $\vartheta$ can be continuously deformed to define line defects for other values of $\vartheta$, so that one may discuss framed BPS degeneracies as a function of $\vartheta$. These would coincide with $V^\vartheta(Q;\vec f)$.

\subsection{Tensor products and factorization}
We can add an extra layer of structure by looking at the tensor product of bimodules: thanks to the factorization, we have 
\begin{equation}
\SH(Q;\vec f_1) \otimes_{\SH(Q)} \SH(Q;\vec f_2)= V(Q;\vec f_1)\otimes V(Q;\vec f_2)\, \SH(Q) \,,
\end{equation}
so that 
\begin{equation}
\chi(\SH(Q;\vec f_1) \otimes_{\SH(Q)} \SH(Q;\vec f_2))= F(Q;\vec f_1)F(Q;\vec f_2)\,\cS \,.
\end{equation}
This implies the analogous expressions for all $\vartheta$.

On the other hand, we know that at the level of K-theory 
\begin{equation}
    F_L F_{L'} = F_{L \cdot L'} \, .
\end{equation}
That means that the identification
\begin{equation}
F_L = F(Q,\langle \gamma_*(L),\vec \gamma \rangle)
\end{equation}
is compatible with the product in $K$-theory. This motivates a basic form of our main conjecture:
\begin{itemize}
\item There is a monoidal functor 
\begin{equation}
\RG_u: \qquad
\mathrm{Lines} \to \SH(Q)-\mathrm{Bimod}  \,,
\end{equation} mapping simple objects $L$ to $\SH(Q;\langle \gamma_*(L),\vec \gamma \rangle)$.
\end{itemize}

\medskip

If we take into account the actual common PBW factorization of the 
bimodules, we can do a bit better. 
The tensor product of two factorized expressions 
\begin{equation}
a_< v a_> \otimes a'_< v' a'_> 
\end{equation}
can be recursively simplified by bringing $a_>$ factors to the right and $a'_<$ factors to the left.
Ultimately, this allows one to always write it as $\widetilde a_< \widetilde v \otimes \widetilde v' \widetilde a'_>$. Effectively, this gives a PBW factorization of $\SH(Q;\vec f_1) \otimes_{\SH(Q)} \SH(Q;\vec f_2)$ for all $\vartheta$. 

We can thus refine our conjecture. 
We expect that it should be possible to define a monoidal category $\SH(Q)-\mathrm{Bimod}^{\mathrm{PBW}}$ of ``PBW-bimodules'', which admit compatible PBW factorizations for each $\vartheta$. It seems to be some kind of dualizability condition on the bimodules. 

We refine our conjecture to: 
\begin{itemize}
\item There is a monoidal functor 
\begin{equation}
\RG_u: \qquad \mathrm{Lines} 
\, \to \,
\SH(Q)-\mathrm{Bimod}^{\mathrm{PBW}} \,,   
\end{equation}
mapping simple objects $L$ to $\SH(Q;\langle \gamma_*(L),\vec \gamma \rangle)$.
\end{itemize}

\medskip

The PBW factorization of the tensor product simplifies the analysis of $\pi_{\vec f, \vec f'}$. Indeed, 
\begin{equation}
\pi_{\vec f, \vec f'} (\widetilde a_< \widetilde v \otimes \widetilde v' \widetilde a'_>) = \widetilde a_< \pi_{\vec f, \vec f'} (\widetilde v \otimes \widetilde v') \widetilde a'_>\,,
\end{equation}
so we would only land outside $\SH(Q;\vec f+\vec f')$ if $\pi_{\vec f, \vec f'} (\widetilde v \otimes \widetilde v')$ lands outside. 

Concretely, the framed BPS degeneracies are readily computable if the framed quiver can be ordered: then 
\begin{equation}
F^\vartheta(Q;\vec f) = X_{\gamma_*}   
\end{equation}
for $\vartheta$ corresponding to that ordering, and there is a canonical PBW basis involving ordered products of $e^{(i)}_k$'s and $b$. 
Then mutations allow one to compute $F(Q;\vec f)$ for other quivers and framings in the same orbit as an ordered one. 

\medskip

\subsection{IR gauge equivariance}
The bimodules that we have defined in this section also enjoy the obvious action of the $\Gamma^\vee({\mathcal O})$ symmetry algebra (defined in Sec.~\ref{sec:ya}), which commutes with the bimodule action. 
Our tensor product decompositions are also compatible with the $\Gamma^\vee({\mathcal O})$ action. 

We are thus led to the final form of our conjecture: 
\begin{itemize}
\item There is a monoidal functor 
\begin{equation}
\RG_u: \qquad \mathrm{Lines} 
\, \to \,
\SH(Q)-\mathrm{Bimod}^{\mathrm{PBW}}_{\Gamma^\vee({\mathcal O})}    \,,
\end{equation}
mapping simple objects $L$ to $\SH(Q;\langle \gamma_*(L),\vec \gamma \rangle)$, and involving a category of bimodules that is both $\Gamma^\vee({\mathcal O})$-equivariant and compatible with PBW factorizations.
\end{itemize}
This final refinement will turn out to be crucial to reproduce the desired categorical structures. 

\subsection{A test of invertibility for $\RG_u$}
We will now compute the graded Witten index of the $\mathrm{Hom}$ spaces in a category of $\Gamma^\vee({\mathcal O})$-equivariant bimodules. 

If we ignore momentarily the $\Gamma^\vee({\mathcal O})$ equivariance, the Witten index can be computed from a bar complex for the derived spaces of morphisms. Schematically, the complexes are the direct sum of spaces of multi-linear maps
\begin{equation}
    (A/\bC 1)^{\otimes n} \otimes B_1 \otimes (A/\bC 1)^{\otimes m} \to B_2
\end{equation}
where $B_1$ and $B_2$ are bimodules, $(A/\bC 1)$ is the algebra with the identity quotiented away and the summand is shifted in ghost number by $n+m$. The character of either of the infinite direct sums gives 
\begin{equation}
    \frac{1}{1+(\chi(A)-1)} = \chi(A)^{-1}\, .
\end{equation}
and effectively ``cancels'' a copy of $A$.

The overall structure of the calculation is already visible from the computation of $\mathrm{End}(A)$:
\begin{equation}
    \chi(\mathrm{End}_{A-\mathrm{Bimod}}(A)) = \chi(A)^{-1}\chi(A) \chi(A)^{-1}\chi(A^\vee)  =\chi(A)^{-1} \chi(A^\vee)  
\end{equation}
We impose $\Gamma^\vee({\mathcal O})$-equivariance in a relative sense: we impose $\Gamma^\vee$ invariance by hand by projecting on zero total gauge charge and add $c_{i,n}$ ghosts for the remaining generators:
\begin{equation}
\chi(\mathrm{End}_{A-\mathrm{Bimod}_{\Gamma^\vee({\mathcal O})}}(A)) = (\fq^2)^{\mathrm{rk}(\Gamma)}_\infty \mathrm{Tr} [\chi(A)^{-1} \chi(A^\vee) ]  \, ,
\end{equation}
where ``$\mathrm{Tr}$'' denotes the projection $X_\gamma \to \delta_{\gamma,0}$, and the prefactor $(\fq^2)^{\mathrm{rk}(\Gamma)}_\infty$ comes from  the contribution from the ghosts $c_{i,n}$ with $i=1,\dots,\mathrm{rk}(\Gamma)$.

Recall that $\chi(A) = \cS(\fq)$. Analogously, 
$\chi(A^\vee) = \cS(\fq^{-1})^t$, where the $t$ superscript denotes $X_{\gamma} \to X_{-\gamma}$. 
We now conjecture a marvellous relation:
\begin{equation}
\cS(\fq^{-1}) = \cS^{-1}(\fq)\,.
\end{equation}
This identity holds automatically when $\cS$ is a product of quantum dilogarithms with quantum torus arguments, as it holds for the individual factors:
\begin{equation}
    \Phi(x;\fq^{-1}) = \Phi(x;\fq)^{-1} \, ,
\end{equation}
and both inversion and $\fq \to \fq^{-1}$ invert the order of multiplication in the quantum torus algebra.  

Then 
\begin{equation}
\chi(\mathrm{End}_{A-\mathrm{Bimod}_{\Gamma^\vee({\mathcal O})}}(A)) = (\fq^2)^{\mathrm{rk}(\Gamma)}_\infty \mathrm{Tr} [\cS^{-1}(\cS^{-1})^t ] \, .
\end{equation}
This is the IR formula for the Schur index \cite{Cordova:2015nma,Cordova:2016uwk}! 

The Schur index computes the Witten index of the space of local operators of the UV theory, i.e. 
\begin{equation}
    \chi(\mathrm{End}_{A-\mathrm{Bimod}_{\Gamma^\vee({\mathcal O})}}(A)) = \chi(\mathrm{End}_{\mathrm{Lines}}(1)) \, .
\end{equation}

We can do better: 
\begin{equation}
    \chi(\mathrm{Hom}_{A-\mathrm{Bimod}}(B_1,B_2)) = \chi(A)^{-1} \chi(B_2) \chi(A)^{-1}\chi(B_1^\vee) =\chi(A)^{-1} \chi(V_2) \chi(A^\vee) \chi(V_1^\vee) 
\end{equation}
and thus for bimodules $B_{L_i}$ which are $\RG_u$ images of $L_i$ we have
\begin{equation}
\chi(\mathrm{Hom}_{A-\mathrm{Bimod}_{\Gamma^\vee({\mathcal O})}}(B_{L_1},B_{L_2})) = (\fq^2)^{\mathrm{rk}(\Gamma)}_\infty \mathrm{Tr} [( \cS^{-1} F^{\pi}_{L_2}) (\cS^{-1} F^{\pi}_{L_1})^t ]\, ,
\end{equation}
where we have assumed the Poincar\'e duality that the $F_L$ are symmetric under $\fq \to \fq^{-1}$.
This is the IR formula for the Schur index counting defect-changing local operators \cite{Cordova:2015nma,Cordova:2016uwk}, i.e.
\begin{equation}
\chi(\mathrm{Hom}_{A-\mathrm{Bimod}_{\Gamma^\vee({\mathcal O})}}(B_{L_1},B_{L_2})) = \chi(\mathrm{Hom}_{\mathrm{Lines}}(L_1,L_2)) = \chi(\mathrm{Hom}_{\mathrm{Lines}} (1,\rho(L_1)L_2))\, ,
\end{equation} 
where in the last step we have used the automorphism $\rho$ that rotates the lines by $\pi$, which satisfies $F^{\pi}_{\rho(L)}=(F^0_{L})^{t}$ and $\textrm{Tr}[ab]=\textrm{Tr}[\rho^2(b)a]$.

These formulae are fully consistent with the hypothesis that $\RG_u$ is a quasi-isomorphism at the level of 
$\mathrm{Hom}$ spaces!

\subsection{Further constraints}
At this point, we have a natural IR image for each $L_i$ 
and a plausible expectation of a quasi-isomorphism for the corresponding morphism spaces.  

It would be nice to find a stringent-enough set of constraints which could restrict the bimodule category so that the $\RG_u$ functors are surjective as well. 

We have already observed the existence of a PBW factorization for the $B_{L_i}$ and their tensor products. 
We should mention another property which appears to be common to all of these bimodules. 

Observe that the exchange relations allow one to bring generators $e_n^{(i)}$ across any element in the bimodules as long as $n$ is sufficiently large, up to a shift of the $n$ argument. One may include such property as a further constraint on the IR bimodule category.

\section{Framed BPS degeneracies, factorizations, and tensor products for the $A_1$ quiver}

The example of a BPS quiver with a single node is already rather interesting. 
The minimal option for a theory described by such a quiver is the theory of a single free hypermultiplet, with no gauge fields. 
This option has the disadvantage of not allowing any interesting half-BPS line defects. 
In order to allow for interesting defects, we can instead consider SQED$_1$: a $U(1)$ gauge theory coupled to a single hypermultiplet. 
This allows us to consider non-trivial 't Hooft-Wilson line defects.\footnote{A potential issue is that the theory is not UV free, so it should be understood as an effective description of a more complicated theory. 
This does not appear to obstruct the mathematical definition of the corresponding category $\mathrm{KP}_{\bC^*,\bC}$ and thus of the candidate $\mathrm{Line}$ category, nor the CoHA calculations.}

\medskip

The 4D Coulomb branch of the theory has two chambers. 
If we pick one of the two as a reference, denoting by $\gamma_1$ the gauge charge of the hypermultiplet with $\mathrm{Im} \, Z_{\gamma_1}>0$ in that chamber, the second chamber is reached by a mutation of the node and the elementary charge is $-\gamma_1$. 
The full charge lattice can be identified as $\Gamma =\bZ^2$, with only the second factor being populated by BPS particles: $\gamma_1 = (0,1)$.
We define a pairing $\langle (a,b), (c,d) \rangle = ad-bc$.

The 't Hooft-Wilson line defects $L_{m,e}$ of the theory are labelled by a magnetic charge $m$ and an electric charge $e$. 
The specification of an electric charge for the line defect is a bit ambiguous. 
The choice we make has the property that $\gamma_* = (m,e)$ is the tropical charge of the line defect in the reference chamber. 
If we move to the second chamber via the mutation $\mu^{\pi}_1$ or $\mu^{0}_1$, the tropical charge will change accordingly.

Notice that the original free hypermultiplet theory has a $\bZ_2$ symmetry $\gamma_1 \to - \gamma_1$. 
The lift of this symmetry to the gauge theory is a bit subtle: we can define the lift either as $\mu^{\pi}_1$ or $\mu^{0}_1$, but neither choice squares to $1$. 
Instead, it squares to a shift $e \to e \pm  m$, for $\mu^{0}_1$ and $\mu^{\pi}_1$, respectively.

\medskip

The algebra of equivariant K-theory classes $[L_{m,e}]$ is generated by the Wilson lines $[L_{0,n}] = [L_{0,1}]^n$ for integer $n$ and the 't Hooft lines $[L_{1,0}]$ and $[L_{-1,0}]$, which satisfy 
\begin{align}
[L_{\pm 1,0}][L_{0,1}] &= \fq^{\pm 2} [L_{0,1}] [L_{\pm 1,0}] \,, \cr
[L_{\pm 1,0}][L_{\mp 1,0}] &= [L_{0,0}] + \fq^{\pm} [L_{0,1}] \,.
\end{align}
In $\mathrm{Lines}$, the sum of two terms on the right hand side 
is lifted to a short exact sequence:
\begin{align}
&0 \to \fq L_{0,1} \to L_{1,0} \cdot L_{-1,0} \to L_{0,0} \to 0 \,, \cr
&0 \to L_{0,0} \to L_{-1,0} \cdot L_{1,0} \to \fq^{-1} L_{0,1} \to 0\,,
\end{align}
where we have used powers of $\fq$ to denote shifts in the equivariant spin grading. 
We have obtained these formulae from $\mathrm{KP}_{\bC^*,\bC}$, but we use opposite conventions for the order of the composition with respect to the reference \cite{cautis2023canonical} and include the equivariant spin grading. 
Finally, these relations are compatible with the lifted $\bZ_2$ symmetry
\begin{equation}\label{eq:Z2sym}
L_{1,0}\to L_{-1,-1}\,,
\qquad 
L_{-1,0} \to L_{1,0}\,,
\qquad L_{0,1}\to L_{0,-1}\,.  
\end{equation}

\subsection{Framed quivers and BPS degeneracies}
\label{ssec:A1framed}

We have a single node, with the exchange relation
\begin{equation}
\{e_n\, , e_m\}=0\,.    
\end{equation}
The framing is captured by either $|*\to 1|=f_1\geq 0$ or $|1\to *|=-f_1\geq 0$.
Both types of framed quivers are ordered. 
\begin{itemize}
\item In the latter case, 
\begin{equation}\label{eq:A1framedm}
F_{\vec f}\, \Phi(X_{\gamma_1}) 
= X_{\vec f}\, \Phi(X_{\gamma_1}) \,.
\end{equation}
Dropping some unnecessary signs, we can describe the bimodules as the span of vectors of the form 
\begin{equation}
b \left[\prod_a e_{k_a}\right] \,,
\end{equation}
with the exchange relation 
\begin{equation}
e_k \, b = b \,e_{k-f_1} \,,
\end{equation}
following from \eqref{eq:beexchangerelations}.
\item In the former case, 
\begin{equation}
F_{\vec f}\, \Phi(X_{\gamma_1}) = \Phi(X_{\gamma_1}) \, X_{\vec f} \,,
\end{equation}
i.e. 
\begin{equation} \label{eq:Ffpositive}
F_{\vec f}  = X_{\vec f} \prod_{m=0}^{f_1-1} (1+X_{\gamma_1} \fq^{2m+1- 2 f_1}) 
= (-\fq X_{\gamma_1};\fq^2)_{f_1} \, X_{\vec f} \,.
\end{equation}
Dropping some unnecessary signs, we can describe the bimodules as the span of vectors of the form 
\begin{equation}
\left[\prod_a e_{k_a}\right] b \,,
\end{equation}
with the exchange relation 
\begin{equation}
e_{k+f_1} \, b = b \,e_{k} \,,
\end{equation}
following from \eqref{eq:beexchangerelations}.
The exchange relation does not allow one to bring the generators $e_k$ with $k<f_1$ to the right of $b$. 
The best we can do is the PBW basis corresponding to 
\begin{equation}
V_{\vec f} = \left[\prod_a e_{k_a}\right]_{0\leq k_a <f_1} b \,,
\end{equation}
which indeed has character $F_{\vec f}$ \eqref{eq:Ffpositive}.
\end{itemize}
\medskip
The known framed BPS degeneracies of $L_{m,e}$ agree with 
$F_{\gamma_* = (m,e)}$, in agreement with the tropical charge. 
In particular, $F_{1,0} = X_{1,0}+ X_{1,1}$ and $F_{-1,0}= X_{-1,0}$ satisfy the correct K-theoretic relations
\begin{align}
F_{\pm 1,0} F_{\mp 1,0} &=  1+\fq^{\pm} F_{0,1} \,.
\end{align}

We are led to conjecture the existence of a monoidal functor $\RG$ mapping $L_{m,e}$ to the bimodule with $\gamma_* = (m,e)$ and thus $f_1 = m$. In order to test this conjecture, we compute tensor products of $B_{\pm 1}$, corresponding to $\gamma_* = (\pm 1,0)$. We will first ignore the $\Gamma^\vee({\mathcal O})$ equivariance and then amend that. 

The tensor products are easily computed:
\begin{itemize}
\item $B_1 \otimes_A B_{-1}$ is the most interesting. 
Let us denote the $b$'s in the two modules as $b_1$ and $b_{-1}$ to avoid confusion. 
$b_1 \otimes b_{-1}$ commutes with all the $e_k$'s except for $e_0$, which cannot be brought across it. On the other hand 
\begin{equation}
e_0 b_1 \otimes b_{-1} - b_1 \otimes b_{-1} e_0
\end{equation}
anti-commutes with all the $e_k$ generators. 
We can denote the sub-bimodule of $B_1 \otimes_A B_{-1}$ generated by this element as $\fq X_{\gamma_1} A$, as it differs from $A$ by one unit of spin and Grassmann parity as well as one unit of $\gamma_1$ charge. 
If we quotient $B_1 \otimes_A B_{-1}$ by $\fq X_{\gamma_1}A$, then we can allow $e_0$ to go across $b_1 \otimes b_{-1}$ and we get another copy of $A$:
\begin{equation}
0\to \fq X_{\gamma_1} A \to B_1 \otimes_A B_{-1} \to A \to 0 \,.
\end{equation}
This is a neat categorification of the $F_1 F_{-1}$ product. 
Crucially, the extension is non-trivial. 
\item $B_{-1} \otimes_A B_{1}$ is a bit less interesting. 
$b_{-1} \otimes b_{1}$ commutes with all the $e_k$ generators whereas $b_{-1} \otimes e_0 b_{1}$ anti-commutes with all the $e_k$ generators. 
Therefore they define sub-modules isomorphic to $A$ and $\fq^{-1} X_{\gamma_1} A$, respectively.
The tensor product is a direct sum:
\begin{equation}
B_{-1} \otimes_A B_{1} = A \oplus \fq^{-1} X_{\gamma_1} A \,.
\end{equation}
\end{itemize}
The analysis generalizes easily:
\begin{itemize}
\item If $f$ and $f'$ have the same sign, $F_f F_{f'} = F_{f+f'}$ and $B_f \otimes_A B_{f'} = B_{f+f'}$.
\item If not and $f$ is negative, we get a direct sum of shifted copies of $B_{f+f'}$. 
\item If not and $f$ is positive, we get an extension of the same collection of shifted copies of $B_{f+f'}$. 
\end{itemize}
At thus point it would appear that the functor $\RG$ remembers the extension in $L_{1,0} \cdot L_{-1,0}$ but not the extension in $L_{-1,0} \cdot L_{1,0}$. 

Luckily, the splitting of $B_{-1} \otimes_A B_1$ is not compatible with the action of $\Gamma^\vee({\mathcal O}) = {\mathcal O}^*$ on the bimodules. Recall that the non-trivial generators of ${\mathcal O}^*$
act as 
\begin{equation}
    o_k: \, e_{n} \to e_n + e_{n+k}\,,
\end{equation}
see \eqref{eq:derivationDef}.
Accordingly, 
\begin{equation}
    o_k: \, b_{-1} \otimes e_0 b_{1} \to b_{-1} \otimes e_0 b_{1} + b_{-1} \otimes b_{1}e_{k-1} 
\end{equation}
This is an extension! In other words, $A$ is an ${\mathcal O}^*$-equivariant submodule of the tensor product, but $\fq^{-1} X_{\gamma_1} A$ is not and we get a non-trivial exact sequence 
\begin{equation}
0\to A \to B_{-1} \otimes_A B_{1} \to \fq^{-1} X_{\gamma_1} A \to 0 \,.
\end{equation}

It appears that keeping track of ${\mathcal O}^*$ equivariance allows for an exact match of the monoidal structure in $\mathrm{Lines}$!

\medskip

We could repeat the analysis in the second chamber, but it is easier to employ the $\bZ_2$ symmetry \eqref{eq:Z2sym},  
which permutes the two chambers. 

\medskip

In conclusion, the monoidal structure we found is compatible with the existence of two monoidal functors 
\begin{equation}
    \mathrm{Lines} \to \SH(A_1)-\mathrm{Bimod}_{{\mathcal O}^*}\,,
\end{equation} one per chamber, which detect the extension structure of the tensor products of $L_1$ and $L_{-1}$.

As $\mathrm{Lines}= \mathrm{KP}_{\bC^*,\bC}$, it should be possible to verify the existence of the conjectured quasi-isomorphism $\RG_u$. We will do so in upcoming work.

\section{Framed BPS degeneracies, factorizations, and tensor products for the $A_2$ quiver}
\label{sec:A2}

The AD theory associated to the $A_2$ BPS quiver has a $\bZ_5$ global symmetry which is spontaneously broken on the Coulomb branch. The line defects accordingly belong to $\bZ_5$ orbits. As $\bZ_5$ acts on the Coulomb branch, the tropical charges of line defects in each $\bZ_5$ orbit are related by a judicious sequence of mutations and so are the framed BPS degeneracies. 

We expect $\bZ_5$ to be a symmetry of the monoidal category $\mathrm{Line}$ for the theory and in particular to respect tensor products. As we lack a gauge-theory description of $\mathrm{Line}$, we cannot directly predict the structure of tensor products. We can, though, 
compare $\RG$ images of $\bZ_5$ orbits of tensor products. 

\subsection{$\bZ_5$ Orbits of Framed Quivers}
\label{ssec:A2framed}
Concretely, we can generate a $\bZ_5$ orbit by the combination of a mutation at the first node and permutation of the two nodes of the $A_2$ quiver. We start the mutation sequence from an ordered framing with both framing charges negative:
\begin{align}\label{eq:A2orbitm1m2}
\vec f &= [-m_1,-m_2] \,, \cr 
\vec f &= [-m_2,m_1] \,, \cr
\vec f &= [m_1,m_2] \,, \cr
\vec f &= [m_1 + m_2,-m_1] \,, \cr
\vec f &= [m_2 ,-m_1-m_2]
\,,
\end{align}
where $m_{1,2}\geq 0$, and each mutation is w.r.t.\ node $1$ and followed by an interchange of $1\leftrightarrow 2$. 
The first three framings are ordered, while the last two have a closed loop. 
Here and below we write $\vec f$ in square brackets to avoid confusion with the corresponding charge 
\begin{equation}
\gamma_* = f_2 \gamma_1 - f_1 \gamma_2 = (f_2,-f_1)
\end{equation} in $\Gamma = \bZ \gamma_1 + \bZ \gamma_2$. 

Note that the collection of tropical charges populates the whole $\Gamma$ as we vary $m_1$ and $m_2$, except for $\gamma_* = 0$. 
They are expected to be in one-to-one correspondence with the simple UV line defects $L^{(a)}_{m_1,m_2}$, with $a=1, \cdots 5$. 
We can denote the identity line defect as $L_{0,0}$.

\medskip

We can write directly the framed BPS degeneracies for these tropical charges for the ordered cases and then obtain the remaining two via mutations:
\begin{align}
F_{[-m_1,-m_2]}\Phi(X_{\gamma_2})\Phi(X_{\gamma_1})&= X_{[-m_1,-m_2]} \Phi(X_{\gamma_2})\Phi(X_{\gamma_1}) \,, \cr
F_{[-m_2,m_1]}\Phi(X_{\gamma_2})\Phi(X_{\gamma_1})&= \Phi(X_{\gamma_2})X_{[-m_2,m_1]} \Phi(X_{\gamma_1}) \,, \cr
F_{[m_1,m_2]}\Phi(X_{\gamma_2})\Phi(X_{\gamma_1})&= \Phi(X_{\gamma_2}) \Phi(X_{\gamma_1})X_{[m_1,m_2]} \,, \cr
F_{[m_1+m_2,-m_1]}\Phi(X_{\gamma_2})\Phi(X_{\gamma_1})&= \Phi(X_{\gamma_1}) \Phi(X_{\gamma_1+ \gamma_2}) X_{[m_1+m_2,-m_1]}\Phi(X_{\gamma_2}) \,, \cr 
F_{[m_2,-m_1-m_2]}\Phi(X_{\gamma_2})\Phi(X_{\gamma_1})&=\Phi(X_{\gamma_1}) X_{[m_2,-m_1-m_2]}\Phi(X_{\gamma_1+\gamma_2}) \Phi(X_{\gamma_2}) \,.
\end{align}
We should stress that each row can be reproduced from the previous one via either $\mu_1^\pm$ mutations, as the two transformations have the same effect on $F_{\vec f}$. In practice, we find it useful to use the bulk wall-crossing identity and choose mutations so that we never have to  transport $\Phi(X_{\gamma_1})$ across $X_{\gamma_*}$. 
This gives compact expressions for the (expected) characters of $\SH(Q;\vec f)$.

The first three characters are easily understood in the language of exchange relations, as the quivers are ordered. The latter two are much more subtle and require a careful analysis of the spherical shuffle bimodules. We did some numerical test for small values of $m_i$. 

The corresponding framed BPS degeneracies are:
\begin{align}\label{eq:Fm1m2}
F_{[-m_1,-m_2]}&= X_{[-m_1,-m_2]} \,, \cr
F_{[-m_2,m_1]}&= \Phi(X_{\gamma_2})X_{[-m_2,m_1]} \Phi(X_{\gamma_2})^{-1} \,, \cr
F_{[m_1,m_2]}&= \Phi(X_{\gamma_2}) \Phi(X_{\gamma_1})X_{[m_1,m_2]}\Phi(X_{\gamma_1})^{-1} \Phi(X_{\gamma_2}) ^{-1}\,, \cr
F_{[m_1+m_2,-m_1]}&= \Phi(X_{\gamma_1}) \Phi(X_{\gamma_1+ \gamma_2}) X_{[m_1+m_2,-m_1]}\Phi(X_{\gamma_1+ \gamma_2})^{-1} \Phi(X_{\gamma_1})^{-1} \,, \cr 
F_{[m_2,-m_1-m_2]}&= \Phi(X_{\gamma_1}) X_{[m_2,-m_1-m_2]}\Phi(X_{\gamma_1})^{-1} \,.
\end{align}

\medskip

It is useful to focus on the simplest $\bZ_5$ orbit, by choosing $(m_1,m_2)=(0,1)$ in \eqref{eq:A2orbitm1m2}:
\begin{align}\label{eq:Z5orbit01}
\vec f_0 &= [0,-1] 
=-\gamma_1 \,, \cr 
\vec f_1&= [-1,0] 
=\gamma_2\,, \cr
\vec f_2 &= [0,1] \,\,\,\,\, 
=\gamma_1 \,, \cr
\vec f_3 &= [1,0] \,\,\, \,\, 
=-\gamma_2\,, \cr
\vec f_4 &= [1 ,-1] \,
=-\gamma_1-\gamma_2\,.
\end{align}
The charges on the right hand side represent the corresponding $\gamma_*$. 
We extend the definition to $\vec f_{a} = \vec f_{a+5}$ and denote the corresponding $\bZ_5$ orbit of UV line defects as $L_a$.

The first four framed quivers are actually ordered and only the last has a loop. 
We refer to the corresponding bimodules $\SH(Q;\vec f)$ simply as $B_a$, with characters 
\begin{align}
\chi(B_0) &= X_{-\gamma_1 } \Phi(X_{\gamma_2})\Phi(X_{\gamma_1}) 
\qquad \qquad \quad\,\,\,\,
= \Phi(X_{\gamma_1}) X_{-\gamma_1 } \Phi(X_{\gamma_1+\gamma_2})\Phi(X_{\gamma_2}) \,, \cr
\chi(B_1) &= \Phi(X_{\gamma_2})X_{\gamma_2} \Phi(X_{\gamma_1}) 
\qquad\qquad\qquad\,
= X_{\gamma_2}\Phi(X_{\gamma_2}) \Phi(X_{\gamma_1}) \,, \cr
\chi(B_2) &= \Phi(X_{\gamma_2}) \Phi(X_{\gamma_1})X_{\gamma_1}
\qquad \qquad \qquad\,
= \Phi(X_{\gamma_2})X_{\gamma_1} \Phi(X_{\gamma_1}) \,, \cr
\chi(B_3) &= \Phi(X_{\gamma_1}) \Phi(X_{\gamma_1+ \gamma_2}) X_{-\gamma_2}\Phi(X_{\gamma_2})  
\quad \,
=\Phi(X_{\gamma_2})\Phi(X_{\gamma_1})X_{-\gamma_2}\,, \cr 
\chi(B_4) &= \Phi(X_{\gamma_1}) X_{-\gamma_1-\gamma_2}\Phi(X_{\gamma_1+\gamma_2}) \Phi(X_{\gamma_2}) =\Phi(X_{\gamma_1}) \Phi(X_{\gamma_1+\gamma_2}) X_{-\gamma_1-\gamma_2}\Phi(X_{\gamma_2})\,,
\end{align}
where the second expressions are from the general one \eqref{eq:Fm1m2} with $(m_1,m_2)=(1,0)$. 

We gave two equivalent forms for each of these bimodules to highlight the structure of auxiliary PBW bases, which will be helpful in tensor product computations. 
We have verified experimentally that the characters are correct and the expected PBW bases
work out, with both sets of candidate $e^{(12)}_k$ generators proposed for the $S_{\gamma_1 + \gamma_2}$ sub-algebras in the previous Section. 

The corresponding framed BPS degeneracies are 
\begin{align}\label{eq:Z5orbit01Degeneracy}
F(B_0)&= X_{-\gamma_1} \,, \cr
F(B_1)&=  X_{\gamma_2}  \,, \cr
F(B_2)&= X_{\gamma_1} + X_{\gamma_1 + \gamma_2}\,, \cr
F(B_3)&= X_{-\gamma_2} + X_{\gamma_1- \gamma_2}+ X_{\gamma_1} \,, \cr 
F(B_4)&= X_{-\gamma_1 - \gamma_2} + X_{- \gamma_2} \,.
\end{align}
The pairs of expressions for the $\chi(B_a)$ make manifest that 
the framed BPS degeneracies $L^{(a)}_{m_1,m_2}$ can be 
written as $\fq^{m_1 m_2} F(B_{a-1})^{m_2} F(B_{a})^{m_1}$. 
We have also reproduced these degeneracies via a quiver Yangian computation in Appendix \ref{appsec:quiverYangian}.

\medskip

The framed BPS indices satisfy the $\bZ_5$-symmetric algebra:
\begin{align}
F(B_a) F(B_{a+1}) &=\fq^{-2} F(B_{a+1})F(B_a) \,,\cr
F(B_{a-1}) F(B_{a+1}) &= 1 + \fq^{-1} F(B_a) \,,\cr
F(B_{a+1}) F(B_{a-1}) &= 1 + \fq F(B_a)  \,.
\end{align}
This theory is not Lagrangian, so there are no clear expectations about the monoidal structure of lines except for these K-theory statements. 

A decoupling argument can provide some heuristics.
If we make $Z_{\gamma_1}$ large, the low energy effective description of the theory is a $U(1)$ gauge theory coupled to a single hypermultiplet. We can then treat, say, $L_0$ and $L_2$ as 't Hooft lines with opposite charges. 
This suggests that $L_0 \otimes L_2$ and $L_2 \otimes L_0$ should be non-trivial extensions involving the identity line and a shifted copy of $L_1$:
\begin{align}
&0 \to L_{0,0} \to L_0 \cdot L_2 \to \fq^{-1} L_1 \to 0 \,,\cr 
&0 \to \fq L_1 \to L_2 \cdot L_0 \to L_{0,0} \to 0 \,.
\end{align}
On the other hand, $L_1$ behaves as a Wilson line, so $L_0 \cdot L_1$ and $L_0 \cdot L_1$ should give the same simple object, up to spin shifts. 

If we assume that these statements in the low-energy effective theory lift to the full theory, we can apply the $\bZ_5$ action to get a collection of predictions:
\begin{align}
&0 \to L_{0,0} \to L_{a-1} \cdot L_{a+1} \to \fq^{-1} L_a \to 0 \,,\cr 
&0 \to \fq L_a \to L_{a+1} \cdot L_{a-1} \to L_{0,0} \to 0 \,,\cr
&L_a \cdot L_{a+1} = \fq^{-2} L_{a+1} \cdot L_a \,.
\end{align}

We will find that the $B_a$'s have the same monoidal structure. In particular, the monoidal structure is $\bZ_5$ symmetric even though the bimodules themselves are very different from each other. 

\subsection{Simple products}
\label{ssec:A2SimpleProducts}

First, we can verify that the tensor products of the consecutive $B_a=\SH(Q;\vec f_a)$ and $B_{a+1}=\SH(Q;\vec f_{a+1})$ give precisely $\SH(Q;\vec f_a+\vec f_{a+1})$ up to the $\fq^{\pm 1}$ shift. 
The shift comes automatically from our Heisenberg grading conventions once we embed the $\vec f$ in $\Gamma$, as long as the isomorphism maps $b \otimes b \to b$ in the relevant bimodules. 

The auxiliary bimodule map $\pi_{\vec f_a,\vec f_{a+1}}$ that we defined above indeed maps $b \otimes b \to b$. 
Then we still need to show that $\pi_{\vec f_a,\vec f_{a+1}}$ actually lands in $\SH(Q;\vec f_a+\vec f_{a+1})$ rather than $\Sh(Q;\vec f_a+\vec f_{a+1})$, i.e. that we do not encounter images of the form $b a b$ where $a$ is a non-trivial algebra element that cannot be brought to the left and/or to the right of the $b$'s.

If we can show this, then we are essentially done: by associativity of the auxiliary shuffle operations, $\pi_{\vec f_a,\vec f_{a+1}}$ maps $a_1 b \otimes b a_2 \to a_1 b a_2$ and is surjective. 
As the characters match, this is enough to prove $\pi_{\vec f_a,\vec f_{a+1}}$ is an isomorphism. 

Of the target charges  
\begin{align}
\vec f_0 +\vec f_1 &= [-1,-1] \,,\cr 
\vec f_1+\vec f_2 &= [-1,1] \,,\cr
\vec f_2+\vec f_3 &= [1,1] \,,\cr
\vec f_3+\vec f_4 &= [2,-1] \,,\cr
\vec f_4 +\vec f_0 &= [1 ,-2]\,,
\end{align}
the first three are ordered in the same manner as the summands and thus the proof is trivial: the same sets of $e^{(i)}_k$ generators can be brought to the left and to the right of the $b$'s in $b \otimes b$ and none gets ``stuck'' in between. Once we set $b \otimes b \to b$, everything follows from exchange relations. 

This leaves us with the last two cases. 
These are easily dealt with as we can write auxiliary PBW bases with the same relative order and algebra generators. Then, again, the PBW generators can be brought to the correct locations after tensor product and thus the tensor product lands surjectively on $\SH(\vec f_a+\vec f_{a+1})$.

Inspection of the characters above fixed the desired forms:
\begin{align}
&S_{\gamma_1} S_{\gamma_1 + \gamma_2} b S_{\gamma_2}\qquad 
\vec f_3 + \vec f_4 \,, \cr
&S_{\gamma_1}b S_{\gamma_1 + \gamma_2} S_{\gamma_2} \qquad 
\vec f_4 + \vec f_0\,.
\end{align}
We already have candidates $e^{(12)}_k$ generators for the $S_{\gamma_1 + \gamma_2}$ sub-algebras, so we only need to check that these bases are surjective in each bimodules. 

\subsection{Interesting tensor products}
\label{ssec:A2TensorProducts}
We expect tensor products $B_{a\pm 1} \otimes_A B_{a \mp 1}$ to give some extensions of $A$ and a shifted copy of $B_a$. 
The sum of charges are now
\begin{align}
\vec f_0 +\vec f_2 &= [0,0] \,,\cr 
\vec f_1+\vec f_3 &= [0,0] \,,\cr
\vec f_2+\vec f_4 &= [1,0] \,,\cr
\vec f_3+\vec f_0 &= [1,1] \,,\cr
\vec f_4+\vec f_1 &= [0,-1]\,,
\end{align}
so $b \otimes b$ is sometimes a candidate to generate $A$ and sometimes to generate the shifted copy of $B_a$.

Assuming the existence of compatible PBW bases, the image in the tensor product of $V^\vartheta(\vec f) \otimes V^\vartheta(\vec f')$ can give us a lot of information, especially if we minimize its size. For each pair of examples, we can find at least two $\vartheta$'s such that one $V$ is 1-dimensional and the other is 2-dimensional. Furthermore, $\pi_{\vec f, \vec f'}$ either has a kernel, giving us the surjection in the desired short exact sequence, or it lands in $\Sh$ as an isomorphism, presenting 
the injection of $\SH$ in the desired short exact sequence.

\medskip

For the computations below, we will need the exchange relations between $b$'s and $e^{(i)}_n$ given in \eqref{eq:beexchangerelations}.
For the five line defects at hand, they are:
\begin{itemize}
\item $b_{\pm \gamma_1}$:
\begin{equation}\label{eq:bpmgamma1}
[e^{(1)}_n\,,\, b_{\pm \gamma_1}]=0
\qquad \textrm{and} \qquad 
e^{(2)}_n \, b_{\pm \gamma_1}=b_{\pm \gamma_1}\, e^{(2)}_{n\mp 1}\,.
\end{equation}
\item $b_{\pm \gamma_2}$:
\begin{equation}\label{eq:bpmgamma2}
 e^{(1)}_n \, b_{\pm \gamma_2}
= b_{\pm \gamma_2}\, e^{(1)}_{n\pm 1}
\qquad \textrm{and} \qquad
[ e^{(2)}_n\,,\, b_{\pm \gamma_2}]=0\,.
\end{equation}
\item $b_{-\gamma_1-\gamma_2}$:
\begin{equation}\label{eq:bmgamma12}
 e^{(1)}_{n+1} \, b_{- \gamma_1-\gamma_2}
=b_{- \gamma_1-\gamma_2}\, e^{(1)}_{n}
\qquad \textrm{and} \qquad
e^{(2)}_n\,  b_{- \gamma_1-\gamma_2}= b_{- \gamma_1-\gamma_2} \, e^{(2)}_{n+1}\,.
\end{equation}
\end{itemize}

All the maps we describe are $\Gamma^\vee({\mathcal O})$-equivariant (sometimes non-trivially). 

\subsubsection{$B_0$ and $B_2$}
We have compatible canonical PBW bases
\begin{align}
B_0 &= b_{-\gamma_1} A \,, \cr
B_1 &= b_{\gamma_2} A \,, \cr
B_2 &= (b_{\gamma_1} \oplus e^{(2)}_0 b_{\gamma_1}) A\,, 
\end{align}
where we have added some subscripts to $b$'s for mnemonic purposes. 

To compute $B_0 \otimes_A B_2$, we can focus on
\begin{equation}
v_{02}\equiv b_{-\gamma_1}\otimes b_{\gamma_1}
\qquad\textrm{and}\qquad
u_{02}\equiv   b_{-\gamma_1}\otimes e^{(2)}_0 b_{\gamma_1}  \,.
\end{equation}
First of all, $b_{-\gamma_1}\otimes b_{\gamma_1}$ commutes with all the $e^{(i)}_k$ generators, namely any $e^{(i)}_k$ generator can be brought from the left to the right of $b_{-\gamma_1}\otimes b_{\gamma_1}$. 
Therefore $b_{-\gamma_1}\otimes b_{\gamma_1}A$ generates a copy of $A$ as a sub-bimodule 
$B_0 \otimes_A B_2$. 

On the other hand, $u_{02}$ satisfies
\begin{align}
e_n^{(1)}\, u_{02} + u_{02}\, e_{n+1}^{(1)} = v_{02}
e^{(2)}_0 e_n^{(1)}  
\qquad\textrm{and} \qquad
\{e^{(2)}_n,u_{02} \}=0\,.
\end{align}
Compare theses with the equations obeyed by (a Grassmann-shifted version of) $b_{\gamma_2}$
from \eqref{eq:bpmgamma2}, we see that $b_{-\gamma_1}\otimes b_{\gamma_1}A$ precisely contains the obstruction to extend \\
$u_{02}\to b_{\gamma_2}$ to a bimodule map to $\fq^{-1} B_1$. Here $\fq^{-1}$ denotes the spin shift together with the Grassmann parity shift $\Pi$. 
There is no way to remove the obstruction and we get 
\begin{equation}
0 \to A \to B_0 \otimes_A B_2 \to \fq^{-1} B_1 \to 0 \,.
\end{equation}

\medskip

Now consider the opposite product $B_2 \otimes_A B_0$. 
Now $v_{20}\equiv b_{\gamma_1}\otimes b_{-\gamma_1}$ commutes with all the $e^{(i)}_k$ generators except $e^{(2)}_0$. 
The offending commutator 
\begin{equation}
u_{20}\equiv [e^{(2)}_0 \, ,\, v_{20}] \,,
\end{equation}
on the other hand, satisfies 
\begin{equation}
 e^{(1)}_n \, u_{20}+ u_{20}\, e^{(1)}_{n+ 1}=0
\qquad \textrm{and} \qquad
\{ e^{(2)}_n\,,\, u_{20}\}=0\,,
\end{equation} 
which are again the same as those obeyed by the Grassmann-shifted version of $b_{\gamma_2}$, see \eqref{eq:bpmgamma2}.
This leads to the short exact sequence 
\begin{equation}
0 \to \fq B_1 \to B_2 \otimes_A B_0 \to A \to 0 \,.
\end{equation}

\subsubsection{$B_1$ and $B_3$}
We have compatible canonical PBW bases
\begin{align}
B_1 &= A(b_{\gamma_2} \oplus b_{\gamma_2}e^{(1)}_0) \,, \cr
B_2 &= A\,b_{\gamma_1} \,, \cr
B_3 &= A\,b_{-\gamma_2} \,,
\end{align}
and the calculations work exactly in the same way as the previous example. 

To compute $B_1 \otimes_A B_3$, we consider
\begin{equation}
v_{13}\equiv b_{\gamma_2}\otimes b_{-\gamma_2}
\qquad\textrm{and}\qquad
u_{13}\equiv   b_{\gamma_2} e^{(1)}_0\otimes  b_{-\gamma_2}  \,.
\end{equation} 
First of all, $b_{\gamma_2}\otimes b_{-\gamma_2}$ commutes with all the $e^{(i)}_k$ generators, and therefore $b_{\gamma_2}\otimes b_{-\gamma_2}A$ generates a copy of $A$ as a sub-bimodule $B_1 \otimes_A B_2$. 
On the other hand, 
\begin{align}
\{e^{(1)}_n,u_{13}  \}=0
\qquad\textrm{and} \qquad
e_{n+1}^{(2)}\, u_{13} - u_{13} \, e_{n}^{(2)} = v_{13} e^{(2)}_n e_0^{(1)} \,.
\end{align}
Compare this with the equation obeyed by the Grassmann-shifted version of $b_{\gamma_1}$
from \eqref{eq:bpmgamma1}, we get the short exact sequences
\begin{equation}
0 \to A \to B_1 \otimes_A B_3 \to \fq^{-1} B_2' \to 0 \,,
\end{equation}
where the prime here indicates that the definition of $\fq^\pm B_2$ has been twisted by $(-1)^{\langle \gamma_1, \gamma \rangle}$.

\medskip

For the opposite tensor product  $B_3 \otimes_A B_1$, $v_{31}\equiv b_{-\gamma_2}\otimes b_{\gamma_2}$ commutes with all the $e^{(i)}_k$ generators except $e^{(1)}_0$. 
The offending commutator 
\begin{equation}
u_{31}\equiv [e^{(1)}_0 \, ,\, v_{13}] \,,
\end{equation}
satisfies 
\begin{equation}
\{ e^{(1)}_n\,,\, u_{31}\}=0
\qquad \textrm{and} \qquad
 e^{(2)}_{n+1} \, u_{31}- u_{31}\, e^{(2)}_{n}=0\,,
\end{equation} 
which lead to the short exact sequence
\begin{equation}
0 \to \fq B_2' \to B_3 \otimes_A B_1 \to A \to 0\,,
\end{equation}
where again the prime indicates a twist by $(-1)^{\langle \gamma_1, \gamma \rangle}$.

\subsubsection{$B_2$ and $B_4$}
We expect compatible PBW bases
\begin{align}
B_2 &= A\,b_{\gamma_1} \,, \cr
B_3 &= A\,b_{-\gamma_2} \,, \cr 
B_4 &= A(b_{-\gamma_1-\gamma_2} \oplus b_{-\gamma_1-\gamma_2}e^{(2)}_0) \,.
\end{align}
The last statement, of course, concerns a non-trivial spherical bimodule and at this point we can only check it by hand for small charges. From the exchange relation perspective, there is a set of non-trivial cubic shuffle relation 
\begin{equation}\label{eq:cubicB4}
c_{0}\equiv b_{-\gamma_1-\gamma_2}e^{(2)}_0 e^{(1)}_0 + e^{(1)}_0  b_{-\gamma_1-\gamma_2}e^{(2)}_0 - e^{(2)}_0 e^{(1)}_0 b_{-\gamma_1-\gamma_2}=0\,.
\end{equation}

In the tensor product $B_2 \otimes_A B_4$, $v_{24}\equiv b_{\gamma_1} \otimes b_{-\gamma_1-\gamma_2}$ satisfies the same relations as $b_{-\gamma_2}$ (given by \eqref{eq:bmgamma12}), 
except for $e^{(2)}_{0}$, which cannot be commuted through. 
A non-trivial shuffle calculation shows that the commutator
\begin{equation}
u_{24}\equiv [e^{(2)}_{0}, v_{24}]
\end{equation}
anti-commutes with all generators in $A$.\footnote{If we had done this computation directly in terms of modes using only the exchange relations, we would have $\{e^{(1)}_{n\geq 1}, u_{24}\}=\{e^{(2)}_{n}, u_{24}\}=0$, and $\{e^{(1)}_{0}, u_{24}\} =-b_{\gamma_1}\otimes c_0$, which vanishes only upon invoking the cubic relation \eqref{eq:cubicB4}.}
Therefore we get the expected
\begin{equation}
0 \to A \to B_2\otimes_A B_4 \to \fq^{-1} B_3 \to 0\,.
\end{equation}

In the tensor product $B_4 \otimes_A B_2$, $v_{42}\equiv b_{-\gamma_1-\gamma_2} \otimes b_{\gamma_1}$ satisfies the same relations as $b_{-\gamma_2}$. 
On the other hand, 
\begin{equation}
u_{42}\equiv  b_{-\gamma_1-\gamma_2}e^{(2)}_0 \otimes b_{\gamma_1}  
\end{equation}
anti-commutes with all the $e^{(2)}_n$'s and only fails to anti-commute with $e^{(1)}_{n}$ by a descendant of $b_{-\gamma_1-\gamma_2} \otimes b_{\gamma_1}$, so that we get 
the exact sequence 
\begin{equation}
0 \to \fq B_3 \to B_4 \otimes_A B_2 \to A \to 0 \,.
\end{equation}

\subsubsection{$B_3$ and $B_0$}

Good simplifications happen in the non-canonical, compatible PBW bases:
\begin{align}
B_3 &= S_{\gamma_1}S_{\gamma_1 + \gamma_2}\,b_{-\gamma_2}S_{\gamma_2} \,, \cr
B_4 &= S_{\gamma_1}S_{\gamma_1 + \gamma_2}\,b_{-\gamma_1-\gamma_2}S_{\gamma_2} \,, \cr 
B_0 &= S_{\gamma_1}S_{\gamma_1 + \gamma_2}(b_{-\gamma_1} \oplus b_{-\gamma_1}e^{(2)}_0 e^{(1)}_0)S_{\gamma_2} \,.
\end{align}
This is a particularly interesting example, as we are tensoring bimodules defined by exchange relations but the output contains $B_4$, which satisfies extra shuffle relationships beyond the 
exchange relations. 

In $B_3 \otimes_A B_0$, the bare combination $v_{30}\equiv b_{-\gamma_2} \otimes b_{-\gamma_1}$ satisfies the same exchange relations as $b_{-\gamma_1-\gamma_2}$, but not the cubic shuffle relations such as \eqref{eq:cubicB4} that holds in $B_4$. 
Instead, the cubic combination
\begin{equation}\label{eq:A2u30}
u_{30}\equiv b_{-\gamma_2} \otimes b_{-\gamma_1}e^{(2)}_0 e^{(1)}_0+e^{(1)}_0 b_{-\gamma_2} \otimes b_{-\gamma_1}e^{(2)}_0 - e^{(2)}_0 e^{(1)}_0 b_{-\gamma_2} \otimes b_{-\gamma_1}
\end{equation}
commutes with all the $e^{(1)}_n$'s and anti-commutes with all the $e^{(2)}_n$'s and hence (almost) generates a copy of $A$. 
This gives the 
expected extension 
\begin{equation}
    0 \to A' \to B_3\otimes_A B_0 \to \fq^{-1} B_4 \to 0
\end{equation}
in a very neat way: $B_3\otimes_A B_0$ is the same as an ``exchange module'' which is a non-trivial extension of $\SH(Q;\vec f_4)$. Here again the prime denotes a twist by $(-1)^{\langle \gamma_1, \gamma\rangle}$.

\medskip

For the opposite product $B_0 \otimes_A B_3$, $v_{03}\equiv b_{-\gamma_1} \otimes b_{-\gamma_2}$ satisfies the same exchange relations as $b_{-\gamma_1-\gamma_2}$.
But remarkably it also satisfies the relations expected in $B_4$. 
On the other hand, 
\begin{equation}
u_{03}\equiv b_{-\gamma_1} e^{(2)}_0 e^{(1)}_0\otimes b_{-\gamma_2}  
\end{equation}  
commutes with all the $e^{(1)}_n$'s and anti-commutes with all the $e^{(2)}_n$'s up to elements in the image of $B_4$. 
Accordingly, 
\begin{equation}
0 \to \fq B_4 \to B_0 \otimes_A B_3 \to A' \to 0\,.
\end{equation}

\subsubsection{$B_4$ and $B_1$}
We expect compatible PBW bases
\begin{align}
B_4 &= (b_{-\gamma_1-\gamma_2} \oplus e^{(1)}_0 b_{-\gamma_1-\gamma_2}) A\,, \cr
B_0 &= b_{-\gamma_1}A \,, \cr 
B_1 &= b_{\gamma_2} A\,.
\end{align}
The analysis is completely analogous to the case of $B_2$ and $B_4$. 

In the tensor product $B_4 \otimes_A B_1$, $v_{41}\equiv b_{-\gamma_1-\gamma_2}\otimes b_{\gamma_2}$ satisfies the same relations as $b_{-\gamma_1}$, 
except for $e^{(1)}_{0}$, which cannot be commuted through. 
The shuffle calculation shows that the commutator
\begin{equation}
u_{41}\equiv [e^{(1)}_{0}, v_{41}]
\end{equation}
anti-commutes with all $e^{(1)}_n$'s and  commutes with all $e^{(2)}_n$'s.\footnote{Similar to the $B_{2}\otimes_A B_{4}$ case, if we had done this computation directly in terms of modes using only the exchange relations, we would have $\{e^{(1)}_{n}, u_{41}\}=[e^{(2)}_{n\geq 1}, u_{41}]=0$, and $[e^{(2)}_{0}, u_{41}] =-c_{0}\otimes b_{\gamma_2}$, which vanishes only upon invoking the cubic relation \eqref{eq:cubicB4}.}
Therefore we get the exact sequences
\begin{equation}
0 \to A'' \to B_4\otimes_A B_1 \to \fq^{-1} B_0 \to 0\,.
\end{equation}
where the double prime indicates a twist by $(-1)^{\langle \gamma, \gamma_2 \rangle}$.
\medskip

For $B_1 \otimes_A B_4$, $b_{\gamma_2} \otimes b_{-\gamma_1-\gamma_2}$ satisfies the same relations as $b_{-\gamma_1}$. 
On the other hand, 
\begin{equation}
u_{14}\equiv  b_{\gamma_2} \otimes e^{(1)}_0 b_{-\gamma_1-\gamma_2}  
\end{equation}
anti-commutes with all the $e^{(1)}_n$'s but fails to commute with $e^{(2)}_{n}$ by a descendant of $b_{\gamma_2} \otimes b_{-\gamma_1-\gamma_2}$, where we have used $c_0=0$ from \eqref{eq:cubicB4}. 
This finally gives 
\begin{equation}
0 \to \fq B_0 \to B_1 \otimes_A B_4 \to A'' \to 0 \,.
\end{equation}

\section{Framed BPS degeneracies, factorizations, and tensor products for quivers in the the $A_3$ family}
\label{sec:A3}

In order to study the full collection of line defects, it is useful to discuss a richer version of the AD theory associated to the $A_3$ quiver: we 
add a $U(1)$ gauge field coupled to the $\gamma_1+\gamma_3$ ``flavour'' charge of the AD theory. 
This is completely analogous to what we did for the case of the $A_1$ quiver. 

The resulting theory has a four-dimensional, non-degenerate charge lattice $\Gamma$. We can give a basis of $\Gamma$ by adjoining to the node charges 
a new charge $\gamma_4$ with $\langle \gamma_i, \gamma_4\rangle = \delta_{i,3}$.

The framing data $\vec f$ associated to a tropical charge $\gamma_*$ 
only depends on $\gamma_*$ up to shifts by $\gamma_1+ \gamma_3$. For notational convenience, we will abuse notation and also denote as $\vec f$ an actual choice of lift $\gamma_* \in \Gamma$.

The enlargement of the AD theory to the gauged AD theory has a price: the generators of the $\bZ_2 \times \bZ_3$ symmetry of the AD theory lift to the
gauged theory but may involve extra shifts of the $\gamma_4$ charge by $\gamma_1 + \gamma_3$. 

\subsection{Mutation orbits}

At first, we can ignore this subtlety and work out interesting mutation orbits of $\vec f$. 

\medskip

The $\bZ_3$ symmetry is obviously manifest in the $\hat A_2$ frame, shown in  \eqref{fig:CircularQuiver}. 
In the $\hat A_2$ frame, we can implement the $\bZ_2$ action by applying the mutation sequence with respect to the nodes $i, i+1, i+2,i$, mod 3, and followed by relabeling nodes $i+1 \leftrightarrow i+2$. 
A simple mutation orbit consists of $\bZ_2$ pairs: 
\begin{align}
i=1:\qquad\qquad &[1,0,0] \quad \leftrightarrow \quad  [0,0,-1] \,, \cr
i=2:\qquad\qquad &[0,1,0] \quad \leftrightarrow \quad  [-1,0,0]\,,\cr
i=3:\qquad \qquad &[0,0,1] \quad \leftrightarrow \quad  [0,-1,0] \,,
\end{align} 
related by $\bZ_3$ in the obvious way. 
Another important $\bZ_3$ mutation orbit consists of three $\bZ_2$-invariant charges with a loop:
\begin{equation}
[-1,0,1]\quad \rightarrow \quad 
[1,-1,0] \quad \rightarrow \quad
[0,1,-1]  
\,.    
\end{equation}
In summary, in the $\hat A_2$ frame, we have a $\bZ_2 \times \bZ_3$ orbit
\begin{align}\label{eq:Z23orbitA2hat}
\vec f_0^{+} = [1,0,0] \,,
\qquad \quad
&\vec f_0^{0} =  [-1,0,1] \,,
\qquad \quad \vec f_0^{-} = [0,0,-1] \,, \cr
\vec f_1^{+} = [0,1,0] \,,
\qquad\quad 
&\vec f_1^{0} = [1,-1,0] \,, 
\qquad \quad \vec f_1^{-} = [-1,0,0]\,,\cr
\vec f_2^{+} = [0,0,1] \,,  
\qquad\quad
&\vec  f_2^{0} = [0,1,-1]\,, 
\qquad \quad \vec f_2^{-} = [0,-1,0]  \,,  
\end{align}  
with $\bZ_3$ acting on the subscript and $\bZ_2$ acting on the superscript. 
\medskip

The subsequent computations are easier to do in the the $A_3$ frame, therefore we go back to the $A_3$ frame by mutating w.r.t.\ node $3$ and then relabeling $2\leftrightarrow$ 3. 
The charges in \eqref{eq:Z23orbitA2hat} are mapped to  
\begin{align}
\vec f_0^{+} = [1,0,0] \,,
\qquad \quad
&\vec f_0^{0} =[0,-1,0]\,, \,\,\,\,
\qquad \,\,\, \vec f_0^{-} = [0,1,-1] \,,\cr
\vec f_1^{+} = [0,0,1] \,,
\qquad\quad 
&\vec f_1^{0} = [1,0,-1]  \,,
\qquad \quad\,\vec f_1^{-} = [-1,0,0]\,,\cr
\vec f_2^{+} = [1,-1,0]   \,,
\qquad\quad\,
&\vec  f_2^{0} =  [0,1,0]  \,,
\qquad \qquad\vec f_2^{-} = [0,0,-1]\,. 
\end{align}
The characters of the corresponding bimodules are 
\begin{align}
&\chi_0^{+} = \Phi(X_{\gamma_3})\Phi(X_{\gamma_2})\Phi(X_{\gamma_1})X_{[1,0,0]} \,,\cr
&\chi_1^{+} = \Phi(X_{\gamma_3})\Phi(X_{\gamma_2})\Phi(X_{\gamma_1})X_{[0,0,1]}  \,,\cr
&\chi_2^{+} = \Phi(X_{\gamma_1})X_{[1,-1,0]}\Phi(X_{\gamma_3}) \Phi(X_{\gamma_1+\gamma_2})\Phi(X_{\gamma_2}) \,,\cr
&\chi_0^{0} =  X_{[0,-1,0]} \Phi(X_{\gamma_3})\Phi(X_{\gamma_2})\Phi(X_{\gamma_1}) \,,\cr
&\chi_1^{0} = \Phi(X_{\gamma_2})\Phi(X_{\gamma_1})X_{[1,0,-1]}\Phi(X_{\gamma_1+\gamma_2+\gamma_3})\Phi(X_{\gamma_2+\gamma_3})  \Phi(X_{\gamma_3}) \,,\cr
&\chi_2^{0} =\Phi(X_{\gamma_3})\Phi(X_{\gamma_2})\Phi(X_{\gamma_1}) X_{[0,1,0]} \,,\cr
&\chi_0^{-} = \Phi(X_{\gamma_2})X_{[0,1,-1]}\Phi(X_{\gamma_2+\gamma_3})\Phi(X_{\gamma_1}) \Phi(X_{\gamma_3})\,,\cr
&\chi_1^{-} = X_{[-1,0,0]}\Phi(X_{\gamma_3})\Phi(X_{\gamma_2})\Phi(X_{\gamma_1})  \,,\cr
&\chi_2^{-} = X_{[0,0,-1]} \Phi(X_{\gamma_3})\Phi(X_{\gamma_2})\Phi(X_{\gamma_1}) \,,
\end{align}
and thus 
\begin{align}
&F_0^{+} = X_{[1,0,0]}+X_{[1,0,0]+\gamma_1}+X_{[1,0,0]+\gamma_1+\gamma_2}+X_{[1,0,0]+\gamma_1+\gamma_2+\gamma_3} \,,\cr
&F_1^{+} = X_{[0,0,1]}+X_{[0,0,1]+\gamma_3}  \,,\cr
&F_2^{+} = X_{[1,-1,0]}+ X_{[1,-1,0]+ \gamma_1} \,,\cr
&F_0^{0} =X_{[0,-1,0]}  \,,\cr
&F_1^{0} = X_{[1,0,-1]}+X_{[1,0,-1]+\gamma_1}+X_{[1,0,-1]+\gamma_1+\gamma_2}\,,\cr
&F_2^{0} = X_{[0,1,0]}+X_{[0,1,0]+\gamma_2}+X_{[0,1,0]+\gamma_2+\gamma_3} \,,\cr
&F_0^{-} = X_{[0,1,-1]}+X_{[0,1,-1]+\gamma_2} \,,\cr
&F_1^{-} = X_{[-1,0,0]} \,, \cr
&F_2^{-} = X_{[0,0,-1]}  \,.
\end{align}

\subsection{Ring relations}
At this point, we pick an embedding in $\Gamma$. We define a collection of lifts
\begin{align}\label{eq:A3orbit}
&\vec f_0^{+} = - \gamma_2-\gamma_4 
\,,\qquad\qquad\,
\vec f_0^{0} = \gamma_3
\,,\qquad \qquad\,\,\,\,\,
\vec f_0^{-} = -\gamma_3 + \gamma_4 \,, 
\cr
&\vec f_1^{+} = - \gamma_4 
\,, \qquad \qquad\qquad\,\,
\vec f_1^{0} = -\gamma_2 
\,,\qquad\qquad
\vec f_1^{-} = \gamma_2 + \gamma_4 \,,
\cr
&\vec f_2^{+} = - \gamma_1-\gamma_2- \gamma_4 \,,\qquad 
\vec f_2^{0} =\gamma_1  
\,,\qquad\qquad\,\,\,\,\,
\vec f_2^{-} = \gamma_4\,,    
\end{align}
and extend the labelling by a modified $\bZ_3$ periodicity
\begin{align}
    \vec f^+_{i+3} &= \vec f^+_{i} -\gamma_1 - \gamma_3 \,, \cr
    \vec f^0_{i+3} &= \vec f^0_i \,, \cr
    \vec f^-_{i+3} &= \vec f^-_{i} +\gamma_1 + \gamma_3 \,, 
\end{align}
so that 
\begin{align} \label{eq:A3frameddegeneracies}
&F_0^{+} = X_{- \gamma_2-\gamma_4}+X_{- \gamma_2-\gamma_4+\gamma_1}+X_{-\gamma_4+\gamma_1}+X_{-\gamma_4+\gamma_1+\gamma_3} \,, \cr
&F_1^{+} = X_{- \gamma_4 }+X_{- \gamma_4 +\gamma_3}  \,, \cr
&F_2^{+} = X_{- \gamma_1-\gamma_2- \gamma_4}+ X_{-\gamma_2- \gamma_4}\,, \cr
&F_0^{0} = X_{\gamma_3} \,, \cr
&F_1^{0} = X_{-\gamma_2 }+X_{-\gamma_2 +\gamma_1}+X_{\gamma_1}\,, \cr
&F_2^{0} = X_{\gamma_1}+X_{\gamma_1+\gamma_2}+X_{\gamma_1+\gamma_2+\gamma_3} \,, \cr
&F_0^{-} = X_{\gamma_1 + \gamma_4}+X_{\gamma_1 + \gamma_4+\gamma_2} \,, \cr
&F_1^{-} = X_{ \gamma_2 + \gamma_4 }  \,, \cr
    &F_2^{-} = X_{\gamma_4}  
\end{align}
and
\begin{align} \label{eq:A3grshift}
F^+_{i+3} &=\fq F^+_{i} X_{-\gamma_1 - \gamma_3} \,, \cr
F^0_{i+3} &=F^0_{i} \,, \cr
F^-_{i+3} &= \fq F^-_{i} X_{\gamma_1 + \gamma_3} 
\end{align}
Powers of $X_{\gamma_1 + \gamma_3}$ here and below are framed BPS indices for $U(1)$ Wilson lines. 

We then have
\begin{align} \label{eq:A3products:shorttimesany}
    F_i^+ F_{i+1}^- &= 1 + \fq F_{i+2}^0 \,, \cr
    F_{i+1}^- F_i^+  &= 1 + \fq^{-1} F_{i+2}^0 \,, \cr
    F_{i+1}^+ F_i^- &= 1 + \fq^{-1}X_{-\gamma_1 - \gamma_3} F_{i+2}^0 \,, \cr
    F_i^- F_{i+1}^+ &= 1 + \fq X_{-\gamma_1 - \gamma_3} F_{i+2}^0 \,, \cr 
     F_i^+ F_{i}^0 &= F_{i-1}^+ + \fq F_{i-2}^+ \,, \cr
     F_{i}^0 F_i^+ &= F_{i-1}^+ + \fq^{-1} F_{i-2}^+ \,, \cr
     F_i^- F_{i}^0 &= F_{i+1}^- + \fq^{-1} F_{i+2}^- \,, \cr
     F_{i}^0 F_i^- &= F_{i+1}^- + \fq F_{i+2}^- 
\end{align}
while other products define new degeneracies and satisfy simple commutation relations:
\begin{align} \label{eq:A3products:qcommuting}
    F_i^+ F_{i+1}^+ &= F_{i+1}^+ F_i^+ \,, \cr
    F_i^+ F_{i+1}^0 &= F_{i+1}^0 F_i^+ \,, \cr
    F_i^+ F_{i-1}^0 &= \fq^2 F_{i-1}^0 F_i^+ \,, \cr
     F_i^+ F_{i}^- &= F_{i}^- F_i^+ \,, \cr
      F_i^- F_{i+1}^0 &=\fq^{-2} F_{i+1}^0 F_i^- \,, \cr
    F_i^- F_{i-1}^0 &= F_{i-1}^0 F_i^- \,, \cr
     F_i^- F_{i+1}^- &= F_{i+1}^- F_i^-
\end{align}
We also have
\begin{align} \label{eq:A3products:longtimeslong}
    F^0_i F^0_{i+1} &= X_{\gamma_1 + \gamma_3}+ \fq X_{\gamma_1 + \gamma_3} F^+_{i+2} F^-_{i+2} \,, \cr
    F^0_{i+1} F^0_i &= X_{\gamma_1 + \gamma_3}+ \fq^{-1} X_{\gamma_1 + \gamma_3} F^+_{i+2} F^-_{i+2} 
\end{align}
Finally, we can mutate at the middle node and conjugate by $\Phi(X_{\gamma_2})$ to produce the framed degeneracies in the $\hat A_2$ chamber (we also map $\gamma_3 \to \gamma_3 + \gamma_1$, $\gamma_2 \to - \gamma_1$, $\gamma_1 \to \gamma_2$):
\begin{align}\label{eq:A2hatF}
&F_0^{+} = X_{- \gamma_2- \gamma_4}+X_{\gamma_1-\gamma_2- \gamma_4}+ X_{\gamma_1- \gamma_4}\,, \cr
&F_1^{+} = X_{\gamma_1-\gamma_4}+X_{\gamma_2-\gamma_4+\gamma_1}+X_{-\gamma_4+\gamma_1+\gamma_2+\gamma_3} \,, \cr
&F_2^{+} = X_{- \gamma_4 }+X_{- \gamma_4+\gamma_1 +\gamma_3}+X_{- \gamma_4 + \gamma_3}  \,, \cr
&F_0^{0} = X_{\gamma_3}+X_{\gamma_1+\gamma_3} \,, \cr
    &F_1^{0} = X_{\gamma_1 }+X_{\gamma_1 +\gamma_2}\,, \cr
    &F_2^{0} = X_{\gamma_2}+X_{\gamma_2+\gamma_3} \,, \cr
    &F_0^{-} = X_{\gamma_2 + \gamma_4}\,, \cr
    &F_1^{-} = X_{-\gamma_1 + \gamma_4 }  \,, \cr
    &F_2^{-} = X_{\gamma_4}  
\end{align}
with periodicity involving $\gamma_1 + \gamma_2 + \gamma_3$. The $\bZ_3$ symmetry is manifest. Notice that $\gamma_4 + \gamma_2$
and $\gamma_4 - \gamma_1$ have a nice pairing with $\gamma_1$ and $\gamma_2$ respectively, so they are $\bZ_3$ images of $\gamma_4$.
We have also reproduced these degeneracies via a quiver Yangian computation in Appendix \ref{appsssec:A2hatquiver}.

The physical theory is expected to have half-BPS lines $L_i^+$, $L_i^0$ and $L_i^-$ with framed BPS degeneracies and K-theoretic 
algebra relations as above, as well as the $U(1)$ Wilson lines $L_n$. 

Simple products such as $[L_1^+]^a [L_2^+]^b [L_3^+]^c$, $[L_1^+]^a [L_2^+]^b [L_3^0]^c$, $[L_1^+]^a [L_2^0]^b [L_1^-]^c$, etc. also give linearly-independent candidates framed BPS degeneracies of half-BPS line defects with general tropical charge. 

\subsection{Bimodules}

To the collection of charges \eqref{eq:A3orbit} we associate bimodules $B_{\vec f}:=\SH(A_3,\vec f)$. 
Note that conventions \eqref{eq:A3grshift} amount to simple grading shifts \begin{equation}
B_{i+3}^a=B_i^a\otimes_A (\fq^{-1} X_{-\gamma_1-\gamma_3})^a A\,,    
\end{equation}
where $a\in \{+,0,-\}$.

To make the relation $\chi(B_i^a)=F_i^+\cS$ between bimodules and degeneracies \eqref{eq:A3frameddegeneracies} more apparent, we provide explicit PBW bases. 
\begin{align}
&B_0^{+} = (b_0^+\oplus e^{(1)}_0 b_0^+ \oplus e^{(2)}_0 e^{(1)}_0 b_0^+ \oplus e^{(3)}_0 e^{(2)}_0 e^{(1)}_0 b_0^+)\otimes_{\bC} A = A\otimes_{\bC} \,b_0^+\,, \cr
&B_1^{+} = (b_1^+\oplus e^{(3)}_0 b_1^+) \otimes_{\bC}A = A\otimes_{\bC} b_0^+ \,, \cr
&B_2^{+} = (b_2^+\oplus e^{(1)}_0 b_2^+) \otimes_{\bC}A = A\otimes_{\bC}(b_2^+\oplus b_2^+ e^{(2)}_0 ) \,, \cr
&B_0^{0} =b_0^0 \otimes_{\bC}A = A\otimes_{\bC} (b_0^0 \oplus b_0^0 e^{(2)}_0\oplus b_0^0 e^{(2)}_0 e^{(1)}_0)\,, \cr
&B_1^{0} = (b_1^0\oplus e^{(1)}_0 b_1^0\oplus e^{(2)}_0 e^{(1)}_0 b_1^0) \otimes_{\bC}A = A\otimes_{\bC} (b_1^0\oplus b_1^0 e^{(3)}_0\oplus b_1^0e^{(3)}_0 e^{(2)}_0 )\,, \cr
&B_2^{0} = (b_2^0\oplus e^{(2)}_0 b_2^0 \oplus e^{(3)}_0 e^{(2)}_0 b_2^0) \otimes_{\bC}A = A\otimes_{\bC} b_2^0 \,, \cr
&B_0^{-} = (b_0^-\oplus e^{(2)}_0 b_0^-)\otimes_{\bC} A = A \otimes_{\bC}(b_0^-\oplus b_0^- e^{(3)}_0) \,, \cr
&B_1^{-} = b_1^- \otimes_{\bC}A = A\otimes_{\bC} (b_1^-\oplus b_1^- e^{(1)}_0) \,, \cr
&B_2^{-} = b_2^- \otimes_{\bC}A = A\otimes_{\bC} (b_2^-\oplus b_2^- e^{(3)}_0 \oplus b_2^- e^{(3)}_0 e^{(2)}_0 \oplus b_2^- e^{(3)}_0 e^{(2)}_0 e^{(1)}_0).
\end{align}

As usual, each bimodule is generated by a single cyclic vector $b_{\vec f}$ satisfying exchange relations \eqref{eq:beexchangerelations}. While  $B_0^+,\,B_1^+,\,B_0^0,\,B_2^0,\,B_1^-$ and $B_2^-$ correspond to framed quivers without loops and as such have no additional relations, $B_2^+,\,B_1^0$ and $B_0^-$ each has one additional relation 
\begin{align}
-e^{(2)}_0 e^{(1)}_0 b_2^+ + e^{(1)}_0 b_2^+ e^{(2)}_0 +  b_2^+e^{(2)}_0 e^{(1)}_0 &= 0 \,, \cr
e^{(3)}_0 e^{(2)}_0 e^{(1)}_0 b_1^0 + e^{(2)}_0 e^{(1)}_0 b_1^0 e^{(3)}_0 + e^{(1)}_0 b_1^0 e^{(3)}_0  e^{(2)}_0 -  b_1^0 e^{(3)}_0 e^{(2)}_0 e^{(1)}_0 &= 0  \,, \cr
-e^{(3)}_0 e^{(2)}_0 b_0^- + e^{(2)}_0 b_0^- e^{(3)}_0 +  b_0^- e^{(3)}_0 e^{(2)}_0 &= 0.
\end{align}

\subsection{Tensor products}
Based on explicit calculations of several examples, we propose that the algebra relations should be promoted to the following statements about tensor products of bimodules. The 
relations \eqref{eq:A3products:qcommuting} are conjecturally lifted to:
\begin{align}
    B_i^+ \otimes_A B_{i+1}^+ &= B_{i+1}^+ \otimes_A B_i^+ \,, \cr
    B_i^+\otimes_A B_{i+1}^0 &= B_{i+1}^0\otimes_A B_i^+ \,, \cr
    B_i^+\otimes_A B_{i-1}^0 &= \fq^2 B_{i-1}^0\otimes_A B_i^+ \,, \cr
     B_i^+\otimes_A B_{i}^- &= B_{i}^-\otimes_A B_i^+ \,, \cr
      B_i^-\otimes_A B_{i+1}^0 &=\fq^{-2} B_{i+1}^0\otimes_A B_i^- \,, \cr
    B_i^-\otimes_A B_{i-1}^0 &= B_{i-1}^0\otimes_A B_i^- \,, \cr
     B_i^-\otimes_A B_{i+1}^- &= B_{i+1}^-\otimes_A B_i^- \,.
\end{align}
The products \eqref{eq:A3products:shorttimesany} and \eqref{eq:A3products:longtimeslong} are conjecturally lifted to exact sequences 
\begin{align}
0 &\longrightarrow \fq B_{i+2}^{0\, \prime} &\longrightarrow B_i^+\otimes_A B_{i+1}^-  &\longrightarrow A &\longrightarrow 0 \,, \cr
0 &\longrightarrow A &\longrightarrow B_{i+1}^- \otimes_A B_i^+ &\longrightarrow \fq^{-1} B_{i+2}^0 &\longrightarrow 0\,, \cr
0 &\longrightarrow A &\longrightarrow B_{i+1}^+ \otimes_A B_i^- &\longrightarrow \fq^{-1}X_{-\gamma_1 - \gamma_3} B_{i+2}^0 &\longrightarrow 0 \,, \cr
0 &\longrightarrow \fq X_{-\gamma_1 - \gamma_3} B_{i+2}^0 &\longrightarrow B_i^- \otimes_A B_{i+1}^+ &\longrightarrow A &\longrightarrow 0  \,, \cr 
0 &\longrightarrow \fq B_{i-2}^+ &\longrightarrow B_i^+ \otimes_A B_{i}^0 &\longrightarrow B_{i-1}^+ &\longrightarrow 0 \,, \cr
0 &\longrightarrow B_{i-1}^+ &\longrightarrow B_{i}^0 \otimes_A B_i^+ &\longrightarrow \fq^{-1} B_{i-2}^+ &\longrightarrow 0 \,, \cr
0 &\longrightarrow B_{i+1}^- &\longrightarrow B_i^- \otimes_A B_{i}^0 &\longrightarrow  \fq^{-1} B_{i+2}^- &\longrightarrow 0\,, \cr
0 &\longrightarrow \fq B_{i+2}^- &\longrightarrow B_{i}^0 \otimes_A B_i^- &\longrightarrow B_{i+1}^- &\longrightarrow 0 \,, \cr
0 &\longrightarrow \fq X_{\gamma_1 + \gamma_3} B^+_{i+2} \otimes_A B^-_{i+2} &\longrightarrow B^0_i \otimes_A B^0_{i+1} &\longrightarrow X_{\gamma_1 + \gamma_3}A &\longrightarrow 0\,, \cr
0 &\longrightarrow X_{\gamma_1 + \gamma_3}A &\longrightarrow B^0_{i+1} \otimes_A B^0_i &\longrightarrow \fq^{-1} X_{\gamma_1 + \gamma_3} B^+_{i+2}\otimes_A B^-_{i+2} &\longrightarrow 0
\end{align}
We have checked explicitly each entry for a convenient choice of $i$ and $\pm$ and filled in the remaining choices by using the expected $\mathbb Z_3 \times \mathbb Z_2$ symmetry.
Here we do {\it not} attempt to predict the sign twists, except for the first two, which may be required in the definition of the bimodules. 

Below we present an explicit check that $\bZ_3$ is indeed a symmetry of the problem, i.e.\ the monoidal structure of the bimodules is invariant under the corresponding sequence of mutations. We leave further checks, such as a full discussion of twists and a comparison with calculations for $\hat A_2$ and $A_3^{(1)}$ quivers, to future work.

\subsubsection{Checking $\mathbb Z_3$ symmetry}

To give an example, let's check the first two equations for all values of $i$.
\begin{itemize}
\item \textbf{Case $i=0$:} First we check $B_0^+\otimes_A B_1^-$. Moving algebra generators to the left gives a PBW basis 
\begin{align}
B_0^+\otimes_A B_1^-= A \otimes ( b_0^+b_1^- \oplus b_0^+b_1^- e^{(1)}_0 )\,.
\end{align}
There's a more convenient choice that differs from the above by a triangular redefinition: define 
\begin{equation}
v=b_0^+ \otimes b_1^- 
\qquad \textrm{and}\qquad  u=[e^{(1)}_0 ,b_0^+\otimes b_1^-] \,,
\end{equation}
then we have
\begin{align}
B_0^+\otimes_A B_1^-= A \otimes (v \oplus u ) \,.
\end{align}
Straightforward calculation using \eqref{eq:beexchangerelations} shows that $v$ commutes with all $e^{(i)}_n$'s except $e^{1}_0$, whereas $u$ satisfies 
\begin{align}\label{eq:uP0M1}
\{e^{(1)}_n,u\}=0
\,, \quad  
ue^{(2)}_{n}=e^{(2)}_{n+1}u
\,, \quad  
\{e^{(3)}_n,u\}=0\,.
\end{align}
Comparing with with the exchange relations satisfied by $b^{0}_2$ (with $\vec{f}=[0,1,0]$):
\begin{align}
[e^{(1)}_n,b^{0}_2]=0
\,, \quad  
b^{0}_2\, e^{(2)}_{n}=e^{(2)}_{n+1}\, b^{0}_2
\,, \quad  
[e^{(3)}_n,b^{0}_2]=0\,,
\end{align}
and noting that $u$ has a spin $\frac{1}{2}$,
we see that  $u$ generates a submodule isomorphic to $\fq B_2^0$ after twisted by $(-1)^{\langle\gamma_1,\gamma\rangle}$.\footnote{Recall that $\fq$ by itself caries a twist of every generator by $-1$.}
Once we quotient out the sub-module generated by $u$, $v$ is cyclic and generates $A$.\footnote{Or more precisely, the image of $v$.} 
This can be summed up in the short exact sequence
\begin{align}\label{eq:SESP0M1}
0 \to \fq B_2^{0\, \prime} \to B_0^+ \otimes_A B_1^- \to A \to 0\,,
\end{align}
where prime stands for twisting by $(-1)^{\langle\gamma_1,\gamma\rangle}$.

\medskip

For the opposite product $B_1^-\otimes B_0^+$, we can choose a left basis as 
\begin{align}
B_1^-\otimes_A B_0^+= A \otimes ( \tilde{v} \oplus  \tilde{u} )\,.
\end{align}
where \begin{equation}
\tilde{v}\equiv b_1^-\otimes b_0^+
\qquad \textrm{and}\qquad 
\tilde{u}\equiv b_1^-e^{(1)}_0\otimes b_0^+
\,.
\end{equation} 
$\tilde{v}$ commutes with all $e^{(i)}_n$'s while $\tilde{u}$ satisfies the same exchange relations \eqref{eq:uP0M1} as $u$.
The vector $\tilde{v}$ generates a submodule isomorphic to $A$ while, in the quotient, $\tilde{u}$ generates $B_2^0$. 
Finally, noting that the spin of $u$ is $-\frac{1}{2}$, we conclude
\begin{align}\label{eq:SESM1P0}
0 \to A \to B_1^- \otimes_A B_0^+ \to \fq^{-1} B_2^{0\, \prime} \to 0\,.
\end{align}

\item \textbf{Case $i=1$:} For $B_1^+\otimes B_2^-$, the right basis can be chosen as
\begin{align}
B_1^+\otimes_A B_2^-=(v\oplus u)\otimes A\,,
\end{align}
where 
\begin{equation}
v\equiv b_1^+\otimes b_2^-
\qquad \textrm{and}
\qquad
u\equiv [e^{(3)}_0,b_1^+\otimes b_2^-]\,.    
\end{equation}

The vector $v$ commutes with all $e^{i}_n$ except $e^{{3}}_0$.
The vector $u$ is of charges $(g,s)[u]=(\gamma_3, \frac{1}{2})$ and obeys
\begin{align}\label{eq:uP1M2}
\{e^{(1)}_n,u\}=0
\,, \quad  
ue^{(2)}_{n+1}=-e^{(2)}_{n}u
\,, \quad  \{e^{(3)}_n,u\}=0 \,.
\end{align}
Comparing with with the exchange relations satisfied by $b^{0}_0$ (with $\vec{f}=[0,-1,0]$):
\begin{align}
[e^{(1)}_n,b^{0}_0]=0
\,, \quad  
b^{0}_2\, e^{(2)}_{n+1}=e^{(2)}_{n}\, b^{0}_2
\,, \quad  
[e^{(3)}_n,b^{0}_2]=0\,,
\end{align}
we see that $u$ generating a submodule $B_0^0=B_3^0$ up to a shift by $\fq$. 
The class of $v$ generates $A$ in the quotient. 
\begin{align}\label{eq:SESP1M2}
0 \to \fq B_3^0 \to B_1^+ \otimes_A B_2^- \to A \to 0\,.
\end{align}

\medskip

The right basis for $B_2^-\otimes B_1^+$ is 
\begin{align}
B_2^-\otimes_A B_1^+ =(\tilde v\oplus \tilde u)\otimes A \,,
\end{align}
where 
\begin{equation}
\tilde v\equiv b_2^-\otimes b_1^+
\qquad 
\textrm{and} 
\qquad
\tilde u\equiv b_2^-e^{(3)}_0 \otimes b_1^+\,.    
\end{equation}
$\tilde{v}$ commutes with all $e^{(i)}_n$ whereas $\tilde{u}$ satisfies the same relation as $u$ in \eqref{eq:uP1M2}.
Therefore we have 
\begin{align}\label{eq:SESM2P1}
0 \to A \to B_2^- \otimes_A B_1^+ \to \fq^{-1} B_3^0 \to 0 \,.
\end{align}

Note that in \eqref{eq:SESP1M2} and \eqref{eq:SESM2P1}, $B^0_3$ is not twisted, unlike $B^0_2$ in \eqref{eq:SESP0M1} and \eqref{eq:SESM1P0}.

\item \textbf{Case $i=2$:} 
First, recall that $B_3^-=B_0^-\otimes\mathbb C \fq X_{-\gamma_1-\gamma_3}$, which comes with a Grassmann sign twist, and $B_4^0=B_1^0$. 
In $B_2^+\otimes B_3^-$, a right basis can be chosen as
\begin{align}
B_2^+\otimes_A B_3^-=(v\oplus v'\oplus v''\oplus u)\otimes A \,,
\end{align}
where we define\footnote{Note that there exists an alternative basis element $\hat v''=e^{(1)}_0b_2^+e^{(2)}_0b_3^-$ which satisfies $\hat v''=v''-ue^{(1)}_0$.}
\begin{align}
v\equiv b_2^+\otimes b_3^- \,, \quad
v'\equiv e^{(1)}_0v \,, \quad
v''\equiv e^{(2)}_0e^{(1)}_0v \,, \quad
u\equiv b_2^+e^{(2)}_0\otimes b_3^- \,.
\end{align}
Note that $b_3^-$ has a Grassmann shift relative to $b^-_0$.
One can check that $v$ has spin $\frac{1}{2}$ and satisfies
\begin{align}\label{eq:v}
&e^{(1)}_{n+1} v = - v e^{(1)}_n \,, \quad 
e^{(2)}_n v = - v e^{(2)}_n \,, \quad 
e^{(3)}_n v = - v e^{(3)}_{n+1}\,, \cr
&e^{(3)}_0e^{(2)}_0e^{(1)}_0 v - e^{(2)}_0e^{(1)}_0 v e^{(3)}_0 + e^{(1)}_0 v e^{(3)}_0e^{(2)}_0 + v e^{(3)}_0e^{(2)}_0e^{(1)}_0 =0\,,
\end{align}
which are precisely the relations satisfied by $b^{0}_4$, which has a Grassmann shift and shift by $X_{-\gamma_1-\gamma_3}$ relative to $b^0_1$ (with $\vec{f}=[1,0,-1]$).
Therefore $v$ generates a submodule isomorphic to $B_4^0$, which as a free right module is embedded as $(v\oplus v' \oplus v'')\otimes A$. 

On the other hand, $u$ commutes with all the $e^{(1)}_n$'s except that $[u,e^{(1)}_n]=v''$, commutes with all $e^{(2)}_n$'s, and  anti-commutes with all the $e^{(3)}_n$'s except that $\{u,e^{(3)}_0\}=ve^{(3)}_0e^{(2)}_0$.     
Therefore after modding out this submodule generated by $v$, the class of $u$ generates $A$ up to a sign twist, producing 
\begin{align}
0 \to \fq B_4^0 \to B_2^+ \otimes_A B_3^- \to A' \to 0 \,,
\end{align}
where the prime before $A$ stands for a twist of signs by $(-1)^{\langle \gamma_4,\gamma\rangle}$. 

\medskip

For $B_3^-\otimes B_2^+$, the right basis is 
\begin{align}
B_3^-\otimes_A B_2^+=(v\oplus v'\oplus v''\oplus u)\otimes A\,,
\end{align}
where
\begin{align}
v= b_3^-\otimes b_2^+ \,,\quad
v'= e^{(1)}_0v\,, \quad
v''= e^{(2)}_0e^{(1)}_0v \,,\quad
u=\{ e^{(2)}_0,v\}\,.
\end{align}
One can check that 
\begin{align}
e^{(1)}_n u = u e^{(1)}_n\,, \qquad e^{(2)}_n u = u e^{(2)}_n\,, \qquad e^{(3)}_n u = - u e^{(3)}_n\,,
\end{align}
implying that $u$ generates submodule isomorphic to twisted $A$. 
What's left to prove is that in the quotient, the class of $v$ behaves as $b_4^0$ and that its spin is $-\frac{1}{2}$. 
One can check that modulo $u\otimes A$, $v$ satisfies the same equation for the $v\equiv b_2^+\otimes b_3^-$ for the opposite product \eqref{eq:v}:
\begin{align}
&e^{(1)}_{n+1} [v] = -[v] e^{(1)}_n\,, \quad 
e^{(2)}_n [v] = - [v] e^{(2)}_n, \quad 
e^{(3)}_n [v] = - [v] e^{(3)}_{n+1} \,, \cr
&e^{(3)}_0e^{(2)}_0e^{(1)}_0 [v] - e^{(2)}_0e^{(1)}_0 [v] e^{(3)}_0 + e^{(1)}_0 [v] e^{(3)}_0e^{(2)}_0 + [v] e^{(3)}_0e^{(2)}_0e^{(1)}_0 =0\,.
\end{align}
Finally, we get 
\begin{align}
0 \to A' \to B_3^- \otimes_A B_2^+ \to \fq^{-1} B_4^0 \to 0 \,.
\end{align}
\end{itemize}

\section{Exploring the complex plane.}

In this Section we will explore another aspects of line defects in an HT theory: they can be moved or fused along the holomorphic direction as well, equipping the category with a structure we call a {\it meromorphic R-matrix}. 
This should be closely related to the notion of the renormalized R-matrix employed in \cite{Gautam:2019wjh}.

First of all, observe that the CoHA is equipped with a {\it translation} map:
\begin{equation}
A \to A[z]\,,
\end{equation}
which acts in the shuffle algebra description as a translation:
\begin{equation}
[p(s_{\bullet, \bullet})]_{\vec n} \to [p(s_{\bullet, \bullet}-z)]_{\vec n} \,.
\end{equation}
Here we treat $z$ as a formal variable of spin $2$ rather than as a number, to preserve the spin grading. Accordingly, the map lands on the algebra $A[z]$ of polynomials in $z$ valued in $A$. Notice that the translation maps $e^{(i)}_k$ to a linear combination of $e^{(i)}_{k-r} z^r$. 

This definition works because the algebra's $\mathrm{fac}$ only contains differences of variables. 
On the other hand, the bimodule action contains factors of $s_{i,a}$. 
We can define new $\Sh(Q)$ modules $\Sh_{\bC}(Q;\vec f)$ whose elements are polynomials in $s_{i,a}$ and $z$, by shifting the factors in $\mathrm{fac}_{L,R}$ to $(s_{i,a}-z)$. 
Intuitively, this represents bound states to a heavy particle at position $z$. 
Then $\SH_{\bC}(Q;\vec f)$ is defined in the same manner. 

\medskip

We can also define bimodules $\Sh_{0,\bC}(Q;\vec f,\vec f')$ as a modification of $\Sh_{\bC}(Q;\vec f+\vec f')$
where we use $s_{i,a}$ for $\vec f$ factors and $(s_{i,a}-z)$ for $\vec f'$ factors in the shuffle product. 
Intuitively, this represents bound states to {\it two}
heavy particles, placed at $0$ and $z$ in $\bC$. Then $\SH_{0,\bC}(Q;\vec f,\vec f')$ is defined in the same manner. It is the natural IR description of two line defects separated in the holomorphic direction. 

The structure of the HT twist then suggests the following conjecture:
\begin{itemize}
\item There are isomorphisms of $A$-bimodules 
\begin{equation}
\SH(Q;\vec f) \otimes_A \SH_{\bC}(Q;\vec f')\simeq \SH_{0,\bC}(Q;\vec f,\vec f') \simeq \SH_{\bC}(Q;\vec f') \otimes_A \SH(Q;\vec f) \,,
\end{equation}
as long as we allow inverse powers of $z$ in building the isomorphism. These are the left- and right- halves of the meromorphic R-matrix and are compatible with the analogous structure expected in $\mathrm{Line}$.
\end{itemize}
Concretely, this seems to work because once we invert $z$ we can bring every element in the new modules to the left or to the right of $b$, trivializing the calculation of the tensor product. Accordingly, 
$b \otimes b \to b$ can be naturally extended to an isomorphism. A simple example should suffice to illustrate the idea.

\subsection{The $A_1$ quiver.}
Consider for example $\SH_\bC(1)$. We have the relation 
\begin{equation}
b e_n = (e_{n+1}-z e_n) b \,.
\end{equation}

Compare $\SH(-1) \otimes_A \SH_\bC(1)$ with $\SH_{0,\bC}(1,-1)$ by mapping $b \otimes b$
to $b$ and extending the map by acting with $A$ on both sides. Can we find a suitable image for elements $b e_0\otimes b$? We can write that as 
\begin{equation}
   z^{-1} (e_0 b \otimes b-e_0 b \otimes b) 
\end{equation}
and map it to $z^{-1} [e_0, b]$ in $\SH_{0,\bC}(1,-1)$.
This was the only dangerous case, so we have an isomorphism of bi-modules 
\begin{equation}
    \SH(-1) \otimes_A \SH_\bC(1) \simeq \SH_{0,\bC}(1,-1)
\end{equation}
linear in $z^{-1}$. 

Conversely, we can immediately compare  $\SH_\bC(1)\otimes_A \SH(-1)$ with $\SH_{0,\bC}(1,-1)$ by mapping $b \otimes b$ to $b$ and extending the map by acting with $A$ on both sides. 

Combining the two isomorphisms, we get an isomorphism between $\SH(-1) \otimes_A \SH_\bC(1)$ and $\SH_\bC(1)\otimes_A \SH(-1)$ mapping $b \otimes b$ to $b \otimes b$ and $b e_0\otimes b$ to $z^{-1} [e_0, b\otimes b]$.

We see a residual structure of the exact sequences we have at $z=0$. 

\section*{Acknowledgements}
We would like to thank Wenjun Niu and Yan Soibelman for useful conversations and feedback on the draft. We would like to thank Kevin Costello and Ahsan Khan for contributions at early stages of the project. This research was supported in part by a grant from the Krembil Foundation. DG and NG are supported by the NSERC Discovery Grant program and by the Perimeter Institute for Theoretical Physics. Research at Perimeter Institute is supported in part by
the Government of Canada through the Department of Innovation, Science and Economic
Development Canada and by the Province of Ontario through the Ministry of Colleges and
Universities.
WL is supported by NSFC No.\ 11875064, No.\ 12275334, No.\ 11947302 and is 
grateful for the hospitality of Perimeter Institute, the Kavli Institute for Theoretical Physics at Santa Barbara, and the Issac Newton Institute for Mathematical Sciences at Cambridge University (during the Program ``Black holes: bridges between number theory and holographic quantum information" with EPSRC Grant ER/R014604/1) where part of this work was carried out. 

\appendix

\section{Line defects as modules of quiver Yangian}
\label{appsec:quiverYangian}

One of the main results of the current paper is the description of the line defects as the CoHA-bimodule from the corresponding framed quiver, see Section.\ \ref{sec:CoHAbimodule}.
In this appendix, we compare this with an alternative description of the line defects as modules of the so-called quiver Yangian, in terms of the ideals of the Jacobian algebra of the framed quiver \cite{Li:2023zub}.

\subsection{Framed BPS degeneracies via quiver Yangian algorithm}

Let us first briefly review the definition of the quiver Yangian. 
It was first introduced in \cite{Li:2020rij} as a formulation of the BPS algebra that describes the half-BPS D-brane bound states of IIA string theory on a toric Calabi-Yau threefold $X$, and was defined in terms of the quiver with potential associated to $X$; it was later generalized to more general quivers (with potential) \cite{Li:2023zub}.
The algebra is defined in terms of three families of operators
\begin{equation}\label{eq:modeexpansion}
e^{(i)}(z)=\sum^{\infty}_{n=0} \frac{e^{(i)}_n}{z^{n+1}} 
\qquad 
\psi^{(i)}(z)=\sum_{n\in\mathbb{Z}} \frac{\psi^{(i)}_n}{z^{n+1}}
\qquad 
f^{(i)}(z)=\sum^{\infty}_{n=0} \frac{f^{(i)}_n}{z^{n+1}}\,.
\end{equation}
The quadratic relations have a universal form for all quivers 
\begin{align}\label{eq:QYquadratic}
\psi^{(i)}(z)\, \psi^{(j)}(w)&=   \psi^{(j)}(w)\, \psi^{(i)}(z)  \;,\cr
\psi^{(i)}(z)\, e^{(j)}(w)&\simeq  \varphi^{i\Leftarrow j}(z-w)\, e^{(j)}(w)\, \psi^{(i)}(z)  \;, \cr
e^{(i)}(z)\, e^{(j)}(w)&\sim  (-1)^{|i||j|}\, \varphi^{i\Leftarrow j}(z-w) \, e^{(j)}(w)\, e^{(i)}(z)  \;, \cr
\psi^{(i)}(z)\, f^{(j)}(w)&\simeq   \varphi^{i\Leftarrow j}(z-w)^{-1} \, f^{(j)}(w)\,\psi^{(i)}(z) \;,\cr
f^{(i)}(z)\, f^{(j)}(w)&\sim  (-1)^{|i||j|}\, \varphi^{i\Leftarrow j}(z-w)^{-1}\,  f^{(j)}(w)\, f^{(i)}(z)   \;,\cr
e^{(i)}(z)f^{(j)}(w) -(-1)^{|i||j|}f^{(j)}(w)e^{(i)}(z) &\sim -  \delta_{i,j} \frac{\psi^{(i)}(z)-\psi^{(i)}(w)}{z-w}  \;,
\end{align}
where ``$\simeq$" means equality up to $z^n w^{m\geq 0}$ terms, while ``$\sim$" means equality up to $z^{n\geq 0} w^{m}$ and $z^{n} w^{m\geq 0}$ terms; $|i|=|i\rightarrow i|+1$ mod $2$.
The ``bonding factor" $\varphi^{i\Leftarrow j} (u)$ in \eqref{eq:QYquadratic} is defined as 
\begin{equation}\label{eq:BondingFactorDef}
\varphi^{i\Leftarrow j} (u) 
\equiv 
(-1)^{|j\rightarrow i|} \frac{\prod_{I\in \{i\rightarrow j\}} (u+h_I)}{\prod_{J\in \{j\rightarrow i\}} (u-h_J)}\,,
\end{equation}
where $\{h_I\}$ are the ``equivariant" weights assigned to each arrow in the quiver $Q$ and they satisfy the ``loop constraint" $\sum_{I}h_I=0$ for each corresponding term in the superpotential $W$.
The quadratic relations of the algebra were ``bootstrapped" by first fixing its action on the class of representations that describe the BPS sector, see below. 
One can then obtain the higher order relations by matching the character of the representation computed as graded dimension of ideals of the Jacobian algebra; in general the higher order relations do not have a closed form but need to be listed order by order.\footnote{When the quiver is the triple quiver of the (affine) Dynkin diagram, the quadratic relations reproduces the ordinary Yangian in the Chevalley basis, and the higher order relations obtained this way reproduce the corresponding Serre relations.} 

\medskip

To compare with the exchange relations of the CoHA \eqref{eq:algebraexchangerelation} that we used in the main text, we need to translate the quadratic relations in terms of the generating fields \eqref{eq:QYquadratic} into quadratic relations in terms of modes.
The rule is the following: first move the denominator of $\varphi^{i\Leftarrow j} (u)$ to the l.h.s.\ of the equations, then plug the mode expansion \eqref{eq:modeexpansion} into \eqref{eq:QYquadratic}, and finally take only singular terms.
For the class of 2-acyclic quivers, which are the type of quivers that we are considering in this paper, for any pair of nodes $(i,j)$, either $\langle \gamma_i, \gamma_j \rangle =|i \rightarrow j| \geq 0 $ or $\langle \gamma_i, \gamma_j \rangle =-|j \rightarrow i| \leq 0 $.
For the first case, the bonding factor defined in \eqref{eq:BondingFactorDef} is 
\begin{equation}
\varphi^{i \Leftarrow j}(u)= \prod_{I\in \{i \rightarrow j\}}(u+h_{I})
\,, \qquad \qquad 
\langle \gamma_i, \gamma_j \rangle =|i\rightarrow j| \geq 0 \,.    
\end{equation}  
These $e$'s are fermionic since $|i|=1$.
The $e-e$ mode relations from \eqref{eq:QYquadratic} is then 
\begin{equation}\label{eq:eefromQY}
e^{(i)}(z) e^{(j)}(w) +  \left(\prod_{I\in \{i \rightarrow j\}}(z-w+h_{I})\right) e^{(j)}(w) e^{(i)}(z)\sim 0
\,, \qquad 
\langle \gamma_i, \gamma_j \rangle =|i\rightarrow j|\geq 0 \,,    
\end{equation}
reproducing the exchange relations of the CoHA \eqref{eq:algebraexchangerelation} upon setting all the equivariant weights $h_I=0$.\footnote{Note that in general it is important to turn off $h_I$ only at the last step, in particular, after moving the denominator of $\varphi^{i\Leftarrow j} (u)$ to the other side of the equation; however, for the 2-acyclic quivers, which are the focus of this paper, this subtlety is absent since there the bonding factors $\varphi^{i\Leftarrow j}$ have either only numerators or only denominators. } 
(The other situation $\langle \gamma_i, \gamma_j \rangle =-|j \rightarrow i| \leq 0 $ can be treated in the same way.) 
Therefore, the subalgebra of the quiver Yangian that is generated by the $e$'s can be viewed as the equivariant version of the spherical CoHA.\footnote{ 
Conversely, for quivers (with potentials) that arise from IIA on certain toric Calabi-Yau threefolds without compact four cycles --- not the type that we are considering in this paper --- the Drinfeld double of the equivariant spherical CoHA (after quotienting out certain relations on the Cartan generators) was shown to be isomorphic to the corresponding quiver Yangian \cite{Rapcak:2020ueh}.}

\medskip

For the class of quivers considered in this paper, the Cartan elements $\psi^{(i)}_n$ can have negative modes, which make it hard to interpret them physically.
Nevertheless, the quiver Yangian provides an algorithm to compute the framed BPS degeneracies that has a somewhat different flavor from the method used in the main text. 
To explain this, we first need to describe the action of the quiver Yangian.
(Note that we only need the subalgebra generated by the $e$'s and $\psi$'s to apply this algorithm.)

\medskip

The representations of the quiver Yangian can be described by the ideals of the Jacobian algebra, with action
\begin{align}\label{eq:psiefAnsatz}
\psi^{(i)}(z)|\Pi\rangle&= \Psi_{\Pi}^{(i)}(z)|\Pi\rangle \;,\cr
e^{(i)}(z)|\Pi\rangle &=\sum_{\sqbox{$i$} \,\in \,\textrm{Add}(\Pi)} 
 \frac{\llbracket \Pi\rightarrow \Pi+\sqbox{$i$}\rrbracket}{z-h(\sqbox{$i$})}|\Pi+\sqbox{$i$}\rangle \;,\cr
f^{(i)}(z)|\Pi\rangle &=\sum_{\sqbox{$i$}\, \in\, \textrm{Rem}(\Pi)}
\frac{\llbracket \Pi\rightarrow \Pi-\sqbox{$i$}\rrbracket}{z-h(\sqbox{$i$})}|\Pi-\sqbox{$i$}\rangle \;.
\end{align}
In this notation, each $\sqbox{$i$}$ is a path from the framing node $*$ to a vertex $i\in Q_0$, and its equivariant weight is the sum of the weights of all arrows along the corresponding path $h(\sqbox{$i$})=\sum_{I\in p^{*\rightarrow i}}h_I$.
The eigenvalues of the Cartans are
\begin{equation}\label{eq:ChargeFunctionDef}
\Psi^{(i)}_{\Pi}(z) ={}^{\sharp}\psi^{(i)}_{0}(z) \prod_{j \in Q_0} \prod_{\sqbox{$j$} \in \Pi } \varphi^{i\Leftarrow j}(z-h(\sqbox{$j$}))   \,,
\end{equation}
with the framing captured by the rational function \cite{Galakhov:2021xum,Li:2023zub}:
\begin{equation}\label{eq:varphi0Def}
{}^{\sharp}\psi_{0}^{(i)}(z) 
=\frac{\prod_{\tilde{I}
\in \{\mathfrak{i} \rightarrow \infty\}} (z+h(\tilde{I}))}{\prod_{\tilde{J}
\in\{\infty \rightarrow \mathfrak{i}\}} (z-h(\tilde{J}))}\,.
\end{equation}
The coefficients for the $e/f$ generators are
\begin{equation}\label{eq:EFCoefficientsSim}
\llbracket \Pi\rightarrow \Pi\pm \sqbox{$i$}\rrbracket \sim \left(\textrm{Res}_{u=h(\sqbox{$i$}) }\Psi^{(i)}_{\Pi}(u)\right)^{\frac{1}{2}}\,,
\end{equation}
where ``$\sim$" denotes signs that will not be relevant in the current discussion. 
One can check that the action \eqref{eq:psiefAnsatz} respects the quadratic relations \eqref{eq:QYquadratic}.

The action \eqref{eq:psiefAnsatz} provides a simple algorithm to compute the (unrefined) framed BPS degeneracies. 
One simply computes the eigenvalue function $\Psi^{(i)}_{\Pi}(z)$ iteratively and counts the number of poles that correspond to the action of $e$. 
To restore $\mathfrak{q}$, one can either match the degeneracies with the $\prod e^{(i)}_n$ acting on the vacuum or use the prescription of \cite{Li:2023zub}.
We now apply this algorithm on a few examples that appeared in the main text.

\medskip

\subsubsection{Example: $A_1$ quiver}

To apply the algorithm on a given framed quiver, we only need two pieces of information: the bonding factors $\varphi^{i\Leftarrow j} (u) $ defined in \eqref{eq:BondingFactorDef}, and the factor ${}^{\sharp}\psi_{0}^{(i)}(z) $ that characterizes the framing \eqref{eq:varphi0Def}.

For the $A_1$ quiver considered in Section.\ \ref{ssec:A1framed}, the bonding factor is trivial:
\begin{equation}
\varphi^{1\Leftarrow 1}(u)=1   \,. 
\end{equation}
As for the ${}^{\sharp}\psi_{0}^{(i)}(z)$ factor, recall that there are two types of framed quivers, depending on the sign of $|*\rightarrow 1|=f_1$.
When $f_1\leq 0$, 
\begin{equation}
{}^{\sharp}\psi_{0}^{(1)}(z) =  \prod^{-f_1}_{i=1}(z-\mathfrak{z}_i) \,,  
\end{equation}
where $\mathfrak{z}_{i}$ is the weight of the $i^{\textrm{th}}$ arrow from $1$ to $*$.
Since it has no pole, the character is $Z=1$, reproducing the frame BPS degeneracies $F_{\vec{f}}=X_{\vec{f}}$
see \eqref{eq:A1framedm}.
On the other hand, when $f_1\geq 0$,
\begin{equation}
{}^{\sharp}\psi_{0}^{(1)}(z)=\prod^{f_1}_{i=1}\frac{1}{z-\mathfrak{p}_i}   \,,
\end{equation}
where $\mathfrak{p}_{i}$ is the weight of the $i^{\textrm{th}}$ arrow from $*$ to $1$.
Now ${}^{\sharp}\psi_{0}^{(i)}(z)$ has $f_1$ poles, and applying the algorithm gives \begin{equation}\label{eq:ZA1f}
Z=(1+X_{\gamma_1})^{f_1} \,,
\end{equation}
reproducing the $\mathfrak{q}\rightarrow 1$ limit of \eqref{eq:Ffpositive}.
The $2^{f_1}$ terms in \eqref{eq:ZA1f} correspond to all possible combinations
$e_{n_k}e_{n_{k-1}}\cdots e_{0}$ acting on the vacuum, with $f_1-1\geq n_{j}>n_{j-1}\geq 0$, given the refined version 
\begin{equation}\label{eq:ZA1fref}
Z=(-\mathfrak{q}X_{\gamma_1};\mathfrak{q}^2)_{f_1} \,. 
\end{equation}

\subsubsection{Example: $A_2$ quiver}
\label{appsssec:A2}

For the $A_2$ quiver considered in Section.\ \ref{ssec:A2framed}, the bonding factors are
\begin{equation}
\varphi^{1\Leftarrow 2}(z)=z
\qquad \textrm{and} \qquad
\varphi^{2\Leftarrow 1}(z)=-\frac{1}{z} \,.
\end{equation}
Let us first focus on the simplest $Z_5$ orbit, given in \eqref{eq:Z5orbit01}. 
We list their corresponding ${}^{\sharp}\psi_{0}^{(i)}(z)$ and the resulting characters.
\begin{enumerate}
\item $\vec f_0 = [0,-1] 
=-\gamma_1$:
\begin{equation}
\psi^{(1)}_0(z)=1
\qquad \textrm{and} \qquad 
\psi^{(2)}_0(z)=z-\mathfrak{z}    \,, 
\end{equation}
giving 
\begin{equation}
Z(X)=1    \,.
\end{equation}

\item $\vec f_1= [-1,0] 
=\gamma_2$:
\begin{equation}
\psi^{(1)}_0(z)=z-\mathfrak{z}
\qquad \textrm{and} \qquad 
\psi^{(2)}_0(z)=1   \,,
\end{equation}
giving 
\begin{equation}
Z(X)=1    \,.
\end{equation}

\item $\vec f_2 = [0,1]
=\gamma_1$:
\begin{equation}
\psi^{(1)}_0(z)=1
\qquad \textrm{and} \qquad 
\psi^{(2)}_0(z)=\frac{1}{z-\mathfrak{p}}  \,,
\end{equation}
giving 
\begin{equation}
Z(X)=1+X_{\gamma_2} 
\quad \rightarrow \quad 1+\fq  X_{\gamma_2}  \,.
\end{equation}
where we have restored $\fq$ by noting that the second term corresponds to $e^{(2)}_0$ acting on the vacuum.

\item $\vec f_3 = [1,0] 
=-\gamma_2$:
\begin{equation}
\psi^{(1)}_0(z)=\frac{1}{z-\mathfrak{p}}
\qquad \textrm{and} \qquad 
\psi^{(2)}_0(z)=1    \,,
\end{equation}
giving 
\begin{equation}
Z(X)=1+X_{\gamma_1}  +X_{\gamma_1+\gamma_2}  
\quad\rightarrow\quad
1+\fq X_{\gamma_1}  + \fq X_{\gamma_1+\gamma_2}
\,,
\end{equation} 
where we have restored $\fq$ by noting that the second and third term correspond to $e^{(1)}_0 $ and $e^{(2)}_0 e^{(1)}_0 $ acting on the vacuum, respectively.

\item $\vec f_4 = [1 ,-1] 
=-\gamma_1-\gamma_2$:
\begin{equation}
\psi^{(1)}_0(z)=\frac{1}{z-\mathfrak{p}}
\qquad \textrm{and} \qquad 
\psi^{(2)}_0(z)=z-\mathfrak{p} \,,    
\end{equation}
giving 
\begin{equation}
Z(X)=1+X_{\gamma_1}  
\quad \rightarrow \quad 1+\fq  X_{\gamma_1}  \,,
\end{equation}
with the second term corresponding to $e^{(1)}_0$ acting on vacuum.
\end{enumerate}
After multiplied by their respective $X_{\vec{f}}$ from the right, they reproduce the framed degeneracies \eqref{eq:Z5orbit01Degeneracy}.
The generalization to more complicated framings is straightforward.

\subsubsection{Example: $\hat{A}_2$ quiver}
\label{appsssec:A2hatquiver}

For the $\hat{A}_2$ quiver considered in \eqref{fig:CircularQuiver}, the bonding factors are
\begin{align}
\varphi^{1\Leftarrow 1}(z)=1 \,, 
\qquad \qquad \,\,\,
& \varphi^{1\Leftarrow 2}(z)=z+h_1\,,
\qquad \,
\varphi^{1\Leftarrow 3}(z)=\frac{1}{z-h_3}\,,
\cr
\varphi^{2\Leftarrow 1}(z)=\frac{1}{z-h_1}\,,
\qquad 
&\varphi^{2\Leftarrow 2}(z)=1\,,
\qquad\qquad\,\,\,
\varphi^{2\Leftarrow 3}(z)=z+h_2\,,
\cr
\varphi^{3\Leftarrow 1}(z)=z+h_3\,,
\qquad\,
&
\varphi^{3\Leftarrow 3}(z)=\frac{1}{z-h_2}\,,
\qquad 
\varphi^{3\Leftarrow 3}(z)=1\,,
\end{align}
with the constraint
\begin{equation}
h_1+h_2+h_3=0\,.    
\end{equation}

Let's consider the simplest $\bZ_2\times
 \bZ_3$ orbit, given in \eqref{eq:Z23orbitA2hat}.
 We list their corresponding ${}^{\sharp}\psi_{0}^{(i)}(z)$ and the resulting characters.
\begin{itemize}
\item $\vec{f}^{+}$.
First consider $\vec{f}^{+}_0=[1,0,0]$, we have
\begin{equation}
\psi^{(1)}_0(z)=\frac{1}{z} \,,
\qquad 
\psi^{(2)}_0(z)=1   \,,     
\qquad 
\psi^{(3)}_0(z)=1\,,
\end{equation}
giving
\begin{equation}
\vec{f}^{+}_1=[1,0,0]: \qquad Z(X)=1+X_{\gamma_1}+X_{\gamma_1+\gamma_2} 
\quad \rightarrow\quad
1+\fq X_{\gamma_1}+\fq X_{\gamma_1+\gamma_2}   \,,
\end{equation}
where the representation terminates before $X_{\gamma_1+\gamma_2+\gamma_3}$ because the zero from \\ $\varphi^{3\Leftarrow 1}(z)=z+h_3$ cancels the required pole, and the second and third terms correspond to $e^{(1)}_0$ and $e^{(2)}_0e^{(1)}_0$ acting on the vacuum.
After multiplied by $X_{-\gamma_{2}-\gamma_4}$ from the right, it reproduces the corresponding result in \eqref{eq:A2hatF}.
Similarly for the other two cases in this $\bZ_3$ orbit:
\begin{align}
&\vec{f}^{+}_1=[0,1,0]: \qquad Z(X)=1+X_{\gamma_2}+X_{\gamma_2+\gamma_3} 
\quad \rightarrow\quad
1+\fq X_{\gamma_2}+\fq X_{\gamma_2+\gamma_3}   \,,\cr
&\vec{f}^{+}_2=[0,0,1]: \qquad Z(X)=1+X_{\gamma_3}+X_{\gamma_3+\gamma_1} 
\quad \rightarrow\quad
1+\fq X_{\gamma_3}+\fq X_{\gamma_3+\gamma_1}    \,.
\end{align}
After multiplied by $X_{\gamma_{1}-\gamma_4}$ and $X_{-\gamma_4}$, respectively, from the right, they reproduce the corresponding results in \eqref{eq:A2hatF}.
\item $\vec{f}^{0}$.
First consider $\vec{f}^{0}_0=[1,-1,0]$, we have
\begin{equation}
\psi^{(1)}_0(z)=\frac{1}{z}   \,,     
\qquad 
\psi^{(2)}_0(z)=z-h_2\,,
\qquad
\psi^{(3)}_0(z)=1 \,,
\end{equation}
where the zero at $z=h_2$ in $\psi^{(2)}_0(z)=\varphi^{2\Leftarrow *}(z)$ is due to the loop $*\rightarrow 1 \rightarrow 2 \rightarrow *$.
This gives
\begin{equation}
\vec{f}^{0}_0=[1,-1,0]: \qquad Z(X)=1+X_{\gamma_1}
\quad \rightarrow\quad
	1+\fq X_{\gamma_1} \,.
\end{equation}
where the representation terminates before $X_{\gamma_1+\gamma_2}$ because the zero from 
$\psi^{(2)}_0(z)=z-h_2$ cancels the required pole, and the $\fq$ is restored by noting that the second term corresponds to $e^{(1)}_0$ acting on the vacuum. 
After multiplied by $X_{\gamma_{3}}$ from the right, it reproduces the corresponding result in \eqref{eq:A2hatF}.
Similarly for the other two cases in this $\bZ_3$ orbit:
\begin{align}
&\vec{f}^{0}_1=[0,1,-1]: \qquad Z(X)=1+X_{\gamma_2}
\quad \rightarrow\quad
1+\fq X_{\gamma_2}   \,,\\
&\vec{f}^{0}_2=[-1,0,1]: \qquad Z(X)=1+X_{\gamma_3}
\quad \rightarrow\quad
1+\fq X_{\gamma_3} \,.
\end{align} 
After multiplied by $X_{\gamma_{1}}$ and $X_{\gamma_{2}}$, respectively,  from the right, they reproduce the corresponding results in \eqref{eq:A2hatF}.
\item $\vec{f}^{-}$.
First consider $\vec{f}^{-}_1=[0,0,-1]$, we have
\begin{equation}\label{eq:psi0A2hatfm1}
\psi^{(1)}_0(z)=1 \,,
\qquad 
\psi^{(2)}_0(z)=1   \,,     
\qquad 
\psi^{(3)}_0(z)=z\,,
\end{equation}
giving
\begin{equation}
\vec{f}^{-}_1=[0,0,-1]: \qquad Z(X)=1 \,,
\end{equation}
since there is no pole in the factors in \eqref{eq:psi0A2hatfm1}.
Similarly for the other two cases in this $\bZ_3$ orbit:
\begin{align}
&\vec{f}^{-}_2=[-1,0,0]: \qquad Z(X)=1 \,,\cr
&\vec{f}^{-}_3=[0,-1,0]: \qquad Z(X)=1 \,.
\end{align} 
These reproduces the framed degeneracies in \eqref{eq:A2hatF}.
\end{itemize}

\subsection{Higher order relations from quiver Yangian arguments}

\subsubsection{Three-node-one-loop quiver}

In the definition of the quiver Yangian, although the quadratic relations are easy to write down explicitly in terms of the quiver and superpotential $(Q,W)$, it takes more work to obtain the higher order relations.

It was conjectured that one way to obtain all the higher order relations of these algebras is to demand that the character of the vacuum representation of the algebra computed by applying the $e^{(i)}_{n}$'s on the ground states should agree with the counting of the ideals of the Jacobian algebra $J(Q,W)$.
For example, one can reproduce the Serre relations of the affine Yangian of $\mathfrak{gl}_{n|m}$ this way, order by order.

\medskip

Let's now apply this argument on the three-node-one-loop quiver.
The generators are
\begin{equation}
e^{(i)}(z)=\sum^{\infty}_{n=0}\frac{e^{(i)}_n}{z^{n+1}}    
\,, \qquad 
a=1,2,3\,.
\end{equation}
We assign the weights to the arrow
\begin{equation}
h_{I_{i\rightarrow i+1}} =    h_i\,,
\end{equation}
with the constraint
\begin{equation}
h_1+h_2+h_3=0    \,,
\end{equation}
due to the superpotential.

The bonding factor was defined in \eqref{eq:BondingFactorDef}.
For the case at hand, they are
\begin{equation}
\begin{array}{ccc}
\varphi^{1\Leftarrow 1}(u)=1  \qquad   
& \varphi^{1\Leftarrow 2}(u)=u+h_1  \qquad 
&\varphi^{1\Leftarrow 3}(u)=\frac{1}{u-h_3} \\
\varphi^{2\Leftarrow 1}(u)=\frac{1}{u-h_1} \qquad     
& \varphi^{2\Leftarrow 2}(u)=1 \qquad 
&\varphi^{2\Leftarrow 3}(u)=u+h_2\\
\varphi^{3\Leftarrow 1}(u)=u+h_3\qquad
& \varphi^{3\Leftarrow 2}(u)=\frac{1}{u-h_2}\qquad
&\varphi^{3\Leftarrow 3}(u)=1 \,.
\end{array}    
\end{equation}

Now consider the framing $|*\rightarrow 1|$=1.
Let's compute the state and its eigenvalue of $\psi^{(i)}(z)$ order by order. 
By construction, these eigenvalues are
\begin{equation}
\psi^{(i)}(u)|\Kappa\rangle =\Psi_{\Kappa}^{(i)}(u) |\Kappa\rangle \
\,, \qquad
\Psi_{\Kappa}^{(i)}(u)    =\psi^{(i)}_0(u)\prod_{j \in Q_0}\prod_{\sqbox{$j$}\in \Kappa}\varphi^{i\Leftarrow j}(u-h(\sqbox{$j$})) \,.
\end{equation}
\begin{enumerate}
\item Ground state.
The charge functions of the ground state are
\begin{equation}
\Psi^{(1)}(u)=\frac{1}{u}
\,, \qquad
\Psi^{(2)}(u)=1
\,,\qquad
\Psi^{(3)}(u)=1  \,,
\end{equation}
which means that we can add one ``atom" of ``color" $1$ at $z=0$.
\item Level-$1$.
After adding the this atom, the state has charge function
\begin{equation}
\Psi^{(1)}(u)=\frac{1}{u}
\,, \qquad
\Psi^{(2)}(u)=\frac{1}{u-h_1}
\,,\qquad
\Psi^{(3)}(u)=u+h_3  \,,
\end{equation}
which means that we can add an atom of color $2$ at $z=h_1$.
\item Level-$2$.
After adding the this atom, the state has charge function
\begin{equation}
\Psi^{(1)}(u)=1
\,, \qquad
\Psi^{(2)}(u)=\frac{1}{u-h_1}
\,,\qquad
\Psi^{(3)}(u)=1 \,,
\end{equation}
which means that the representation terminates at this level.
\end{enumerate}
The character is then
\begin{equation}
Z(X)=1+X_1+X_1X_2 \,.   
\end{equation}
Now let's try to reproduce this character by applying the $e^{(i)}_n$ on $|0\rangle_1$.
\begin{enumerate}
\item Ground state: 
\begin{equation}
|0\rangle_1    \,.
\end{equation}
\item At level-$1$, we need to  impose
\begin{equation}\label{eq:3loop1level1}
e^{(1)}_{n \geq 1}|0\rangle_1 
=e^{(2)}_{n }|0\rangle_1  
=e^{(3)}_{n }|0\rangle_1 
=0 \,,
\end{equation}
in order to have the one state 
\begin{equation}
e^{(1)}_0 |0\rangle_1    
\end{equation}
at this level.
\item Level-$2$.
Applying $e^{(i)}_n$ on $e^{(1)}_0 |0\rangle_1$, after imposing the quadratic relations and \eqref{eq:3loop1level1}, we have
\begin{equation}
e^{(2)}_0 e^{(1)}_0    |0\rangle_1 \,,
\end{equation}
reproducing the $x_1 x_2$ term in the character.
\item Level-3.
Now apply $e^{(i)}_n$ on $e^{(2)}_0 e^{(1)}_0 |0\rangle_1$.
After imposing the quadratic relations and \eqref{eq:3loop1level1}, but before imposing any higher order relations, we would have
\begin{equation}\label{eq:3loop1level3}
e^{(3)}_0e^{(2)}_0 e^{(1)}_0 |0\rangle_1    \,.
\end{equation}
Since the character terminates at level-2, there must be an cubic relations such that 
\begin{equation}
e^{(3)}_0e^{(2)}_0 e^{(1)}_0 |0\rangle_1  =0    \,.
\end{equation}
\end{enumerate}

Now repeat this argument for the other two framings of this kind, i.e.\ $|*\rightarrow 2|=1$ or $|*\rightarrow 3|=1$, we obtain
\begin{equation}
e^{(3)}_0e^{(2)}_0 e^{(1)}_0 |0\rangle_1  
=e^{(1)}_0e^{(3)}_0 e^{(2)}_0 |0\rangle_2 
=e^{(2)}_0e^{(1)}_0 e^{(3)}_0 |0\rangle_3 
=0 \,.
\end{equation}
This is compatible with the cubic relations that is $\mathbb{Z}_3$ symmetric
\begin{equation}
e^{(3)}_0e^{(2)}_0 e^{(1)}_0 +e^{(1)}_0e^{(3)}_0 e^{(2)}_0 +e^{(2)}_0e^{(1)}_0 e^{(3)}_0=0  \,.   
\end{equation}

\section{BPS particles and $(2,2)$ Supersymmetric Quantum Mechanics}
\label{appsec:SQM}

Recall the 4D ${\cal N}=2$ SUSY algebra with central charges:
\begin{align}
\{ Q_{\alpha}^A, \bar{Q}^B_{\dot{\beta}} \} &= P_{\alpha\dot{\beta}} \epsilon^{AB} \,, \\
\quad
\{ Q_{\alpha}^A, Q_{\beta}^B \} &= i \epsilon_{\alpha\beta}\epsilon^{AB} Z \,, 
\\
\quad 
\{ \bar{Q}^A_{\dot{\alpha}}, \bar{Q}^B_{\dot{\beta}} \} &= -i \epsilon_{\dot{\alpha}\dot{\beta}}\epsilon^{AB} \bar{Z} \,.
\end{align}
Here the capital Latin indices denote the doublets of the $SU(2)_R$ R-symmetry, the dotted and undotted lowercase Greek indices denote the spinor indices of the two chiralities.
The $Z$ term is a central charge which can appear in a sector with charged particles.
\medskip

The HT twist of the theory breaks both symmetry groups to the Cartan sub-algebra. The twisting super-charge can be taken to be 
\begin{equation}
Q= Q_-^- + \bar Q_{\dot -}^-  \,,
\end{equation}
which obeys
\begin{align}
\{ Q, \bar{Q}^+_{\dot{-}} \} &= -\{Q,Q_{-}^+\} = P_{-\dot{-}} \,, \\
\{ Q, \bar{Q}^+_{\dot{+}} \} &= P_{-\dot{+}}+ i\bar{Z} \,, \\
\qquad \qquad
\{Q,Q_{+}^+\} &= -P_{+\dot{-}} -i Z\,.
\end{align}
We identify $P_{-\dot{-}}$ with the anti-holomorphic derivative $\partial_{\bar z}$ in the holomorphic plane. On the other hand, $P_{-\dot{+}}$ and $P_{+\dot{-}}$ are holomorphic and anti-holomorphic momenta in the topological plane. We see that the world-lines of particles of central charge $Z$ align with the phase of the central charge in the topological plane. 

\medskip

A BPS particle such as an hypermultiplet breaks half of the supersymmetries. 
We will now focus on stationary or slow-moving objects. We break the Lorentz symmetry to the $SU(2)_s$ rotation group by identifying dotted and undotted indices:
\begin{align}
\{ Q_{\alpha}^A, \bar{Q}^B_{\beta} \} &=  P_0 \epsilon_{\alpha\beta} \epsilon^{AB}+ P_{\alpha\beta} \epsilon^{AB} \,, \\
\{ Q_{\alpha}^A, Q_{\beta}^B \} &= i \epsilon_{\alpha\beta}\epsilon^{AB} Z \,, \\
\{ \bar{Q}^A_{\alpha}, \bar{Q}^B_{\beta} \} &= - i\epsilon_{\alpha\beta}\epsilon^{AB} \bar{Z} \,,
\end{align}
where now $P_{\alpha \beta}= P_{\beta \alpha}$ denotes the momentum and $P_0$ the energy. 

A stationary BPS particle with positive imaginary central charge preserves 
\begin{equation}
q^A_\alpha \equiv Q_{\alpha}^A + \bar{Q}^A_{\alpha}\,,
\end{equation}
which satisfy 
\begin{equation}
\{ q^A_\alpha, q^B_\beta \} = 2(P_0 - \mathrm{Im} \, Z)\epsilon_{\alpha\beta} \epsilon^{AB} \equiv 2 H \epsilon_{\alpha\beta} \epsilon^{AB} \,,
\end{equation}
giving a supersymmetric quantum mechanical system with four super-charges with energy $H$ above the BPS bound.  

The broken supercharges 
\begin{equation}
    \tilde q^A_\alpha \equiv \frac12 \left(Q_{\alpha}^A - \bar{Q}^A_{\alpha}\right)
\end{equation}
satisfy 
\begin{equation}
    \{ \tilde q^A_\alpha, q^B_\beta \} = P_{\alpha\beta} \epsilon^{AB}
\end{equation}
and are thus super-partners of the broken translations. They also satisfy 
\begin{equation}
\{ \tilde q^A_\alpha, \tilde  q^B_\beta \} = \left(\mathrm{Im} \, Z+ \frac12 H \right) \epsilon_{\alpha\beta} \epsilon^{AB} 
\,,
\end{equation}
which is approximately constant in a non-relativistic limit. 

We can recognize the non-relativistic hypermultiplet as being closely related to the dimensional reduction of a 4D ${\cal N}=1$ vectormultiplet, i.e. the $(2,2)$ SQM with gauge group $U(1)$ and no matter: the position $X^{\alpha \beta}$ arise from the three space components of the 4D gauge connection, the time-like component can be gauge-fixed away in a time-like gauge and the gauginos map to the Goldstinos $\psi^A_\alpha$ such that 
\begin{align}
q^A_\alpha &= \frac{P_\alpha^\beta}{\sqrt{\mathrm{Im} \, Z}} \, \psi^A_\beta\,, \qquad     \tilde q^A_\alpha = \sqrt{\mathrm{Im} \, Z}\, \psi^A_\alpha \,,
\end{align}
giving the conventional non-relativistic Hamiltonian. 

The Clifford algebra
\begin{equation}
    \{ \psi^A_\alpha, \psi^B_\beta \} = \epsilon_{\alpha\beta} \epsilon^{AB} 
\end{equation}
acts on four states, organized into two doublets $|\alpha\rangle$ and $|A\rangle$. The former are fermionic and represent the hypermultiplet fermions. The latter are bosonic and represent the hypermultiplet bosons. 

Every momentum mode comes with four polarization states. The zero-momentum (distributional) states are annihilated by all four $q^A_\alpha$. States of momentum $P_\alpha^\beta$ are annihilated by $q^A_\alpha- P_\alpha^\beta \tilde q^A_\beta$.

If we apply the HT twist, we take a cohomology with respect to 
\begin{equation}
Q = q^-_- = \frac{1}{\sqrt{\mathrm{Im} \, Z}}P_-^\beta \psi^-_\beta\,.
\end{equation}
The presence of a continuum of slow-moving particles, though, makes the cohomology problem somewhat poorly defined. 

Notice that time is one of the topological coordinates. 
Denoting by $x$ the space topological coordinate and by $z$ the holomorphic space coordinate, the HT super-charge $Q$ can be written as 
\begin{equation}
Q=dx \, \partial_x + d \bar z \, \partial_{\bar z}\,,
\end{equation}
i.e.\ we identify $dx = \psi^-_-$ and $d\bar z = \psi^-_+$. The $\psi^+_\beta$ are canonically conjugate variables to these. The states $|\alpha\rangle$ are 1-forms, while $|A\rangle$ are $0$- and $2$-forms. 

We are thus looking at the de Rham cohomology on $\mathbb{R}$ and Dolbeault cohomology on $\mathbb{C}$.

\subsection{SUSY traps}
We will now deform the problem in a manner which ``traps'' the particle at some location and thus gives a better notion of SUSY ground states and $Q$-cohomology.\footnote{For related discussion, see \cite{Cecotti:2011iy}} 
We are looking for deformations which are as ``canonical'' as possible, in the sense that they arise from some universal type of deformation in the original 4D theory.

In the context of the free BPS particle, there are some well-known ways to 
regularize the de Rham and Dolbeault differentials. We will describe them in detail and then recognize them as examples of a more general construction. 

\subsubsection{The Morse trap}
A natural deformation of the de Rham differential involves a Morse function $h(x)$ on $\mathbb{R}$:
\begin{align}
Q &= - i d - i dh \wedge = (P- i \partial h) \psi \\
\bar Q &= (P+ i \partial h) \bar \psi 
\,,
\end{align}
with $\{\psi, \bar \psi\} = 1$.

We can take a Morse function which is either positive at infinity or negative at infinity. Both choices result in a one-dimensional space of states. 
In the first case, the wavefunction is $e^{-h}$ and is bosonic. In the latter, it is $e^h dx$ and is fermionic. 

If $h(x)$ is slowly varying, there are approximate ground states at each zero of $\partial h$. The instanton effects lift pairs of approximate ground states to give a one-dimensional answer. 

The supersymmetric quantum mechanics with
\begin{align}
Q &=  \frac{P}{\sqrt{\mathrm{Im} \, Z}}\psi 
\,, \qquad 
\bar Q =  \frac{P}{\sqrt{\mathrm{Im} \, Z}}\bar \psi 
\end{align}
describes a non-relativistic BPS chiral multiplet in a $(2,2)$ 2D theory, with $\psi$ and $\bar \psi$ Goldstinos. 

\subsubsection{The equivariant trap}
A natural deformation of the Dolbeault differential employs the rotational symmetry of $\mathbb{C}$:
\begin{align}
Q &= - i \bar \partial - i z d\bar z \wedge = (\bar P- i z) \psi \\
\bar Q &= (P+ i \bar z) \bar \psi \,,
\end{align}
which gives 
\begin{equation}
\{Q, \bar Q\}  = |P|^2 + i (\bar z \bar P- z P)+ |z|^2 + 2 \psi \bar \psi H-J\,,
\end{equation}
where $J$ generates rotations of $\mathbb{C}$.
The cohomology consists of bosonic wavefunctions $z^n e^{-|z|^2}$. The opposite deformations would have given fermionic wave-functions $z^n e^{-|z|^2} d\bar z$. 
We will use the former option below. 

\subsubsection{Combining the traps}
An $\Omega$ deformation in the holomorphic plane is a well-understood way to trap the 4D dynamics near the origin of $\mathbb{C}$. The $\Omega$ deformation in a plane effectively preserves a $(2,2)$ 2D sub-algebra of the 4D SUSY algebra. The 2D $(2,2)$ super-algebra has central charges which receive contributions both from the 4D central charge and from the angular momentum $J$ in the holomorphic plane. The angular momentum contribution will appear as a shift of $H$: the Hamiltonian is not $Q$-exact but $H-J$ still is. Conversely, we can define the $\Omega$ deformation by presenting the 4D system as a 2D system with $(2,2)$ SUSY and turning on a ``twisted mass'' for the rotational symmetry in the holomorphic plane. 

We expect the $\Omega$ deformation of the bulk theory to give the above equivariant deformation of the Dolbeault differential for the BPS hypermultiplet. 

Based on a comparison to a similar 2d calculation in \cite{Gaiotto:2015aoa}, we expect that 
the deformation of the de Rham differential can be implemented universally by a small position-dependent rotation of the phase of the central charge of the particle. 

We can now consider a generic theory and assume we can find a locus in parameter space where all central charges are approximately aligned with the imaginary axis. If we turn on both the $\Omega$ deformation and some slowly-varying non-trivial profile for the phases $\vartheta_\gamma$, BPS particles of charge $\gamma$ will get trapped at locations where $\vartheta_\gamma$ passes through $\pi/2$. 

This will trap particles near the origin in the complex plane, with a tower of angular momentum modes, and near the loci where the central charges are purely imaginary. There will be ``fermionic'' traps which would trap fermion modes for a hypermultiplet and ``bosonic'' traps which would trap bosonic modes. 

\subsubsection{Algebras of (trapped) BPS particles}
We will now sketch how one may be able to employ intricate configurations of traps in order to define an algebra of BPS states. Our proposal has some flaws, but we hope they can be addressed in future work. 

Consider a situation where the central charges are constant and slightly away from the imaginary axis at large positive and negative $x$, but vary in between and possibly cross the imaginary axis at various locations.

The system will thus have approximate ground states consisting of collections of trapped BPS particles. Due to the angular momentum contribution from electric and magnetic fields, the quantum numbers of the resulting ground states are given by the twisted tensor product. True ground states will be described as the cohomology of some differential acting on approximate ground states. 

If we deform the $\vartheta_\gamma$ profile in a compact region, the approximate ground states may change but the true ground states will not. Accordingly, we must have families of quasi-isomorphisms relating the complexes of approximate ground states before and after the deformation. See \cite{Gaiotto:2015aoa} for a discussion of such Janus-like configurations and their deformations in a closely related 2d setup. 

We first consider the case where all phases are equal. Then we denote the states trapped at ``fermionic'' or ``bosonic'' traps  as $A$ and $A^!$ respectively. As the notation indicates, we hope that these coincide with the algebra of BPS states and its Koszul dual. The ``empty trap'' configurations where no particles are trapped provide candidate units for $A$ and $A^!$.\footnote{As discussed in the next footnote, it is more likely that the space of states are $A$ and $(A^!)^\vee$ and that empty configurations should give an unit and a co-unit. The matter requires a more careful analysis.}

In order to motivate the identification, observe first that if the phase profile has both a bosonic and a fermionic trap, it corresponds to a Morse function that is positive on one direction and negative on the other, and hence can be deformed to a configuration with no traps. 
We thus learn that there are compatible quasi-isomorphisms
\begin{equation}
\mathbb{C} \simeq A \otimes A^! \simeq A^! \otimes A \simeq A \otimes A^!\otimes A \otimes A^! \simeq \cdots
\end{equation}
Other trap configurations predict 
\begin{equation}
A \simeq A \otimes A^! \otimes A \simeq  A \otimes A^! \otimes  A \otimes A^! \otimes A \simeq \cdots
\end{equation}
and 
\begin{equation}
A^! \simeq A^! \otimes A \otimes A^! \simeq  A^! \otimes  A \otimes A^! \otimes A \otimes A^! \simeq \cdots
\end{equation}

In particular, if some no-exotic statement implies that $A$ is supported in ghost number $0$, we can place an empty trap in the middle of $A \otimes A^! \otimes A$, we get a map $A \otimes A \to A$ which should be associative, making $A$ into an algebra. Then the remaining relations do resemble axioms satisfied by $A^!$.\footnote{In examples of Koszul duality, $A \otimes A^!$ equipped with a canonical differential is quasi-isomorphic to $\bC$ only up to a degree shift. $A \otimes (A^!)^\vee$ would be a better candidate.}

\medskip

If we use different profiles for the $\vartheta_\gamma$, so that different particles are trapped at different locations, we obtain algebras $S_\gamma$ associated to each BPS ray. Deforming back to a single fermionic trap, we obtain an isomorphism between the ordered twisted tensor product of the $S_\gamma$ and $A$ and derive a PBW factorization of $A$. We conclude that the candidate algebra of BPS states admits a PBW factorization for any possible order of the BPS rays, which is one of the main properties of the CoHA which we employ in the main text. 

The analysis can be extended to configurations involving BPS line defects as well. In the presence of a single defect $L$ and in the absence of traps, the space of states coincides with the framed BPS degeneracies of the defect. If we add a fermionic trap we get an $A$-bimodule $B[L]$ equipped with all the PBW factorizations we employ in the main text. This map should be a functor, as local operators on the line defect will act on the space of ground states for the system. 

In the presence of two line defects, comparing the various trap configurations and various separations between the defects will relate the fusion of line defects and the tensor product of the associated bimodules over $A$. 

In conclusion, we expect a study of trapped configurations together with a series of no-exotic conjectures to imply the full collection of algebraic and categorical structures discussed in the main text. 

\subsection{Quiver quantum mechanics, equivariant cohomology and CoHA}

The conventional route to the CoHA is to consider the SQM associated to a quiver
\cite{Denef:2002ru}. This is a dimensional reduction of a 4d non-Abelian SUSY gauge theory: the fermions which used to be chiral or anti-chiral before the dimensional reduction now couple differently to the electric field $P_{\alpha \beta}$ and the magnetic field $[X^{\alpha \gamma},X^{\delta \beta}] \epsilon_{\gamma \delta}$. Physically, this is due to the fact that the quiver arises from a D-brane description of the BPS particles and string theory/supergravity does not have the $SU(2)_R$ symmetry. The discussion above will have to be adjusted accordingly. 

It should be possible to to turn on trapping deformations in the SQM as discussed above. We expect that the resulting algebra $A$ of BPS states trapped at a fermionic trap will coincide with the CoHA for the quiver. We leave the analysis to future work.

\bibliographystyle{JHEP}
\bibliography{biblio}

\end{document}